\newif\ifarxiv
\arxivtrue   

\documentclass[]{JFM-FLM_Au}

\usepackage{xcolor}
\usepackage{comment}
\usepackage{makecell}
\usepackage{subcaption}
\usepackage[normalem]{ulem}
\usepackage[export]{adjustbox}

\ifarxiv
\makeatletter

\gdef\@righttitle{}
\gdef\@lefttitle{}

\def\ps@titlepage{%
  \let\@oddhead\@empty
  \let\@evenhead\@empty
  \let\@oddfoot\@empty
  \let\@evenfoot\@empty
}

\renewcommand{\email}[1]{}
\renewcommand{\pagelimitfooter}{}

\def\ps@titlepage{%
  \let\@oddhead\@empty
  \let\@evenhead\@empty
  \def\@oddfoot{\hfil\thepage\hfil}%
  \let\@evenfoot\@oddfoot
}

\def\@titlefont{\centering}

\makeatother
\fi

\ifarxiv
\makeatletter
\gdef\@righttitle{}
\gdef\@lefttitle{}
\def\ps@titlepage{%
  \let\@oddhead\@empty
  \let\@evenhead\@empty
  \let\@oddfoot\@empty
  \let\@evenfoot\@empty
}
\renewcommand{\email}[1]{}
\renewcommand{\pagelimitfooter}{}

\def\ps@titlepage{%
  \let\@oddhead\@empty
  \let\@evenhead\@empty
  \def\@oddfoot{\hfil\thepage\hfil}%
  \let\@evenfoot\@oddfoot
}
\def\@maketitle{%
 \newpage
 \vspace*{10\p@}\addvspace{4.6pc}%
 {\centering
  {\titlefont
   \@title \par}%
  \vspace*{23.5\p@}%
  {\normalfont\authorfont\fontswitch\bfseries\baselineskip=12\p@
     \lowercase{\@author}\par}%
  \vspace*{2\p@}%
  {\normalfont\small
   \@affiliation \par\vspace*{2pt}}%
  \ifx\@corresau\empty\else{\normalfont\small\@corresau\vskip7pt}\fi
 \par}%
}
\makeatother
\fi

\title{Influence of wall thermal boundary condition on mean-flow characteristics of a Mach 2.5 fully rough turbulent boundary layer}
\author{Mateus A. R. Braga\aff{1} \and Robyn L. Macdonald\aff{1}}

\affiliation{\aff{1}Ann \& H.J. Smead Department of Aerospace Engineering Sciences, University of Colorado Boulder\\3775 Discovery Dr., Boulder, CO 80303, USA}

\corresau{mateus.braga@colorado.edu}

\newcommand{\pp}[2]{\frac{\partial #1}{\partial #2}}
\newcommand{\ol}[1]{\overline{#1}}
\newcommand{\wt}[1]{\widetilde{#1}}

\newlength{\rpad}
\allowdisplaybreaks 

\begin{document}
\maketitle

\begin{abstract}
We investigate turbulent boundary layers at Mach 2.5 over sinusoidal roughness at matched $Re_\tau=784$. We considered three wall temperatures, $T_w/T_r=$ 1.0, 0.7, and 0.4. Each wall temperature was repeated with a smooth and rough surface, the latter following a three-dimensional sinusoidal profile with effective slope 0.5 and matched $k^+=79.1$. In total six direct numerical simulations were completed. Analysis of mean wall shear and heat transfer detailed the augmentation in skin friction and heat transfer coefficient between smooth and rough cases; resulting in the classical failure of the Reynolds analogy for rough walls. We also show that differences in shear stress across wall temperatures are driven by the viscous component even though the pressure component dominates. The roughness sublayer was found to be $R_{RSL}=3k-6k$, consistent with the literature. Moreover, the wall offset $d=d_\Theta\approx0.5k$ for both momentum and thermal boundary layers, signifying the offset represents the half peak-to-valley height rather than the ``mean roughness height.'' For the mean momentum boundary layer, we find the present conditions recover the incompressible roughness function $\Delta u_1^+$ when matched via the semi-local roughness Reynolds number $k^*$. For this reason, an equivalent sand-grain roughness of $k_s^*\approx3.7k^*$ is proposed. The existing compressible transformations hold regardless of wall temperature or roughness -- contradicting recent claims otherwise. The generalized Reynolds analogy works for smooth walls but fails for rough walls; a roughness correction term is proposed and validated. Compressible mean temperature transformations fail for cold walls, especially when $(T_w-T_e)/(T_r-T_e)<0$. The thermal roughness function, $\Delta\Theta^+$, is reported with caution given transformation singularities. 
\end{abstract}

\ifarxiv
\fi

\newpage
\section{Introduction}
\label{sec:introduction}

Incompressible turbulent flow over rough walls has been the subject of extensive research as summarized by the reviews from \cite{Jimenez2004}, \cite{Chung2021}, \cite{Kadivar2021}, and \cite{Kadivar2025}. However, turbulence in compressible flows over distributed roughness, where the thermal boundary layer is important, has received fewer focused studies. \cite{Kadivar2025} in their review of turbulent heat transfer over roughness identified that the impact of boundary conditions (including roughness topography and thermal conditions), wall origin, and turbulent statistics remains comparatively under-explored for compressible turbulent boundary layers over surface roughness.

Recent experimental and numerical investigations have assessed rough-walled supersonic turbulent boundary layers (TBL) in an effort to explain how compressibility effects alter turbulence physics, wall interactions, and engineering predictions. The following studies are listed because of their relevance either to supersonic TBLs or sinusoidal surface roughness. \cite{Latin2000} performed experiments of Mach 2.9 TBLs over random sand-grain and uniformly machined rectangular and cubic elements. Similarly, the experiments from \cite{Kocher2022} assessed Mach 2 boundary layers with diamond-shaped patterned and random distributed surface roughnesses. \cite{Modesti2022} performed direct numerical simulations (DNS) of supersonic turbulent channel flow over cubical roughness elements, spanning bulk Mach numbers $M_b = 0.3-4$, finding that existing compressibility transformations are valid provided a roughness Reynolds number that accounts for the viscosity variations at the roughness crest is used. The DNS from \cite{Ma2023} also looked at turbulent channel flow, but over three-dimensional sinusoidal rough walls. Their focus on outer-layer similarity and energy transfer identified that for friction Reynolds numbers $Re_\tau \approx 1080$, the length scale and intensity of large-scale coherent structures increases for the small roughness ($k^+ = 10$), but decreases for large roughness ($k^+ = 60$), as compared with the smooth-wall case, providing a justification for an observed failure of outer-layer similarity (Townsend's wall-similarity hypothesis \citep{Townsend1976}) for the $k^+=10$ case. The DNS from \cite{Aghaei2023} were performed to characterize fully developed supersonic turbulent channel flows over isothermal rough walls. The key finding from their work was that the effects of roughness extend beyond the near-wall layer due to the shocks and suggest that outer-layer similarity may not fully apply to supersonic turbulent flow over rough walls. \cite{Chan2023} studied a spatially developing TBL over three-dimensional sinusoidal roughness with DNS, highlighting evidence for outer-layer similarity and the importance of the dispersive component, associated with the spatial inhomogeneity induced by the varying surface elevations, to characterise the energy balance in a TBL. However, the investigation from \cite{Chan2023} was limited to incompressible flow and should be extended to compressible flows. \cite{Cogo2025a} presented DNS of supersonic, zero-pressure-gradient (ZPG), adiabatic turbulent boundary layers at Mach 2 freestream, over cubical roughness elements. In addition to their focus on quantifying the development of an internal boundary layer, they found that the classical van Driest II transformation \citep{vanDriest1956} can also be applied to rough walls and that the quadratic velocity-temperature relation established for smooth walls \citep{Busemann1931,Crocco1932,Zhang2014} are also valid over rough surfaces; however, these findings have not been confirmed for diabatic walls. Similarly, for the same freestream conditions, \cite{Cogo2025b} investigated cubic- and diamond-shaped elements in aligned and staggered configurations, explaining the relative drag induced by each roughness shape by examining both the viscous and pressure drag contributions. Using the data from \citep{Cogo2025a,Cogo2025b}, \cite{Cogo2026} assess the validity of the Reynolds analogy \citep{Busemann1931, Crocco1932, walz69, Huang1994, Zhang2014}, building on their findings to propose a wall model for compressible rough-wall flows.

\cite{Wang2024} have recently and explicitly explored the effects of three-dimensional sinusoidal roughness on compressible TBLs under different Mach numbers and wall temperature conditions. \cite{Wang2024} found that the compressible velocity transformations they considered (\cite{vanDriest1951}, \cite{Zhang2012}, \cite{Trettel2016}, and \cite{Griffin2021}) did not make the logarithmic region of velocity profiles independent of the wall-to-adiabatic-wall temperature ratio $T_w/T_{aw}$ for the rough-walled cases due to strong wall heat transfer effects below the roughness peak. The roughness function, $\Delta u_1^+$, was found to decrease with decreasing $T_w/T_{aw}$, with the variation assigned to the variation in the non-dimensional mean shear, $S^+$, at wall-normal locations below the roughness crest. Aside from \cite{Wang2024}, and their subsequently derived supersonic rough-wall Reynolds-averaged Navier-Stokes (RANS) model \citep{Wang2026}, few studies have systematically examined the combined effects of surface roughness and wall temperature on the logarithmic mean velocity profile of compressible turbulent boundary layers. 

Common across the aforementioned investigations is the finding that the mean-flow structure of incompressible rough-wall boundary layers generally remains applicable to compressible, supersonic flows when an appropriate scaling is employed. Similarly, temperature–velocity relations, such as the Reynolds analogy established for smooth walls, have been shown to remain valid over rough surfaces, at least for adiabatic-wall conditions. Roughness, however, has a more pronounced influence on the thermal statistics, including the mean and fluctuating temperature fields, which generally do not exhibit the same outer-layer similarity observed in the velocity field. The latter appears to persist provided that shocks do not emanate from the roughness elements, although some studies have reported contradictory findings. Despite these advances, the existing literature does not provide a comprehensive understanding of the combined effects of wall temperature and sinusoidal surface roughness in compressible turbulent boundary layers. Previous studies have generally focused on only one or a subset of these effects: some have examined the influence of wall temperature over smooth surfaces, while others have investigated roughness effects under primarily adiabatic or isothermal conditions. Moreover, many of the studies considering roughness have focused on internal flows or on geometries such as cubic or diamond-shaped elements, rather than streamwise-developing boundary layers over sinusoidal roughness. Thus, while the individual effects of compressibility, wall temperature, and surface roughness have been considered to varying degrees, their combined influence on the mean velocity and thermal structure of a developing compressible turbulent boundary layer remains insufficiently understood. Furthermore, the reported trends from \cite{Wang2024} should be verified using an independent DNS dataset to extend the analysis through additional diagnostics and flow conditions, providing further insight into the mechanisms responsible for the observed lack of universal collapse. The thermal boundary layer, including temperature transformations, thermal roughness function, and velocity-temperature relationships for the mean flow should also be formally addressed. 

The primary objective of this work is to present a database of perfect gas air Mach 2.5 ZPG TBL DNSs which systematically varied wall surface geometry and wall temperature to enable direct comparison of their coupled effects on mean turbulence structures and heat transfer. The present work focuses on the effects of surface temperature and roughness on the boundary layer response, including surface characteristics, mean velocity and temperature scaling, the applicability of existing compressible similarity transformations, and velocity-temperature relationships. We detail the extent of the roughness sublayer and suggest metrics for the wall offset for both momentum and thermal boundary layers which are often overlooked~\citep{Kadivar2021,Kadivar2025}. This paper is organized as follows: In \S \ref{sec:computational-approach} we outline the computational approach, including the governing equations, numerical approach and setup, flow conditions and surface geometry, computational domain, grid and timestep resolution, and averaging procedure. In \S \ref{sec:surface} we present results and discuss the mean surface quantities of wall shear stress and heat transfer. \S \ref{sec:wall-origin-and-roughness-characterization} addresses the impact of the surface on the boundary layer by quantifying the roughness sublayer as well as providing a detailed exploration of the wall offset for both the momentum and thermal boundary layer. In \S \ref{sec:momentum-BL} we analyse the inner-scaled momentum boundary layer. In \S \ref{sec:vel-temp} we explore the validity of the quadratic velocity-temperature relationships for the mean flow, proposing a new roughness correction. In \S \ref{sec:thermal-BL} we analyse the inner-scaled thermal boundary layer. We conclude our findings and suggest future work in \S \ref{sec:Conclusion}.

\section{Computational Approach}\label{sec:computational-approach}

\subsection{Governing Equations}
Under the assumption of continuum flow, the flow of a thermally and calorically perfect gas can be described by the conservation of mass, momentum, and energy equations:

\begin{align}
    &\pp{\rho}{t}+\pp{\left(\rho u_j\right)}{x_j} = 0 \label{eq:mass-conservation}\\
    &\pp{\left(\rho u_i\right)}{t} + \pp{\left(\rho u_i u_j\right)}{x_j} + \pp{p}{x_i} - \pp{t_{ij}}{x_j} = 0 \label{eq:momentum-conservation}\\
    &\pp{E}{t} + \pp{\left(u_j\left[E+p\right]\right)}{x_j} - \pp{(u_i t_{ij})}{x_j} + \pp{q_j}{x_j} = 0\label{eq:energy-conservation}
\end{align}
where $t$ is time, $\rho$ is density, $p$ is static pressure, $x_i$ is the Cartesian direction, and $u_i$ is the velocity vector along the Cartesian direction $x_i$. Einstein summation convention is used and implied summation is given on repeated indices unless otherwise noted. Spatial coordinates $x_1$, $x_2$, $x_3$, and velocities $u_1$, $u_2$, $u_3$, correspond to the streamwise, wall-normal, and spanwise directions, respectively. Stokes' hypothesis assumes the the bulk viscosity is zero, making the viscous stress tensor, $t_{ij}$, as follows:
\begin{equation}
    t_{ij} = \mu\left(\pp{u_i}{x_j} + \pp{u_j}{x_i}\right) -\frac{2}{3}\mu\pp{u_k}{x_k}\delta_{ij}
    \label{eq:viscous-stress-tensor}
\end{equation}
where $\delta_{ij}$ is the Kronecker delta and $\mu$ is the dynamic viscosity from Sutherland's Law, $\mu(T) = \mu_{ref}\left(\frac{T}{T_{ref}}\right)^{3/2}\left(\frac{T_{ref}+S}{T+S}\right)$, where $T$ is the static temperature, and taking the reference values to be $\mu_{ref}=1.684\times10^{-5}$ kg~m$^{-1}$s$^{-1}$, $T_{ref}=273.15$~K, and $S=110.4$~K. From the energy equation, $E$ is the sum of the internal and kinetic energies:
\begin{equation}
    E = \rho\left(e+\frac{u_i u_i}{2}\right)
    \label{eq:total-energy}
\end{equation}

For calorically perfect gas, the specific heat at constant pressure, $c_p=1004.7$ J kg$^{-1}$K$^{-1}$, and constant volume, $c_v=717.6$ J kg$^{-1}$K$^{-1}$, are constant. The specific internal energy and specific enthalpy are defined as $e = c_vT$ and $h = e + \frac{p}{\rho} = c_pT$. The pressure, density, and temperature are related via the ideal gas equation of state and the perfect gas constant $R_{gas}=c_p-c_v=287.1$ J kg$^{-1}$K$^{-1}$ such that $p = \rho R_{gas} T$. The heat flux vector $q_j$ comes from Fourier's Law:

\begin{equation}
    q_j = -\lambda_\Theta\pp{T}{x_j}
    \label{eq:fourier-law}
\end{equation}
where the thermal conductivity $\lambda_\Theta = \mu c_p/Pr$, with $Pr=0.73$ the laminar Prandtl number.

\subsection{Numerical Approach}\label{sec:numerical-approach}
All DNSs were performed with US3D, the unstructured grid, finite volume, Navier-Stokes solver \citep{candler15_US3D,Nompelis05_US3D}. For spatial discretization, the fourth order kinetic energy consistent scheme from \cite{subbareddy09_US3D} was selected for the symmetric, non-dissipative portion of the inviscid fluxes. In this low-dissipation scheme, numerical dissipation is controlled by a switch $\alpha$, also known as a shock sensor, multiplied to the dissipative (upwind biased) component, where dissipation is only added in regions necessary to maintain stability of the solution at strong shocks and discontinuities. The present shock-sensor follows the dilatation based form from \cite{larsson11}. A weighted-least-squares reconstruction of primitive variables is used for cell-centred gradients, and viscous fluxes are computed with a deferred-correction approach.  Lastly, second order implicit Euler time integration with line relaxation is used. For complete descriptions of the numerical schemes, refer to \cite{kim03}, \cite{Nompelis05_US3D}, \cite{subbareddy09_US3D}, \cite{larsson11}, \cite{bartkowicz12_thesis}, and \cite{knutson20_thesis}.
    
\subsection{Synthetic Turbulence Generation}\label{sec:stg}
Synthetic turbulence generation (STG) is used to introduce ``realistic'' turbulence at the inlet of the simulations. The original idea from \cite{Kraichnan1970} to superimpose random Fourier modes has been extended by \cite{Shur2014} and it is the basis for the present implementation. The present work follows closely to the STG formulation from \cite{Shur2014}, with some alterations noted in the following paragraphs. Additionally, the concept is extended for compressible flow via velocity-thermodynamic property fluctuation correlations. At the inlet plane, velocity fluctuations are synthetically created through a superimposition of spatiotemporal Fourier modes with random amplitudes and phases, where additional constraints on the random number field are provided through a target energy spectrum, and anisotropy is introduced through a Cholesky decomposition of the target Reynolds stress tensor. The output of the STG is to obtain: $u_i(x_i,t) = \overline{u_i}(x_i) + u'_i(x_i, t)$ and $T (x_i, t) = \overline{T}(x_i) + T'(x_i, t)$ at the inflow boundary, where $u_i$ is the instantaneous velocity vector as a function of spatial location time, $\overline{u_i}$ is the mean velocity vector, and $u_i'$ is the velocity fluctuation vector; likewise for temperature $T$. The overbar and prime notation are for the mean and fluctuation components, respectively.

Notable deviations from the method presented by \cite{Shur2014} are as follows: The $f_\eta$ and $f_{cut}$ empirical functions to decay the energy spectrum after the Kolmogorov wavenumber and the Nyquist value are neglected. The macro-scale (advective) velocity, $U_0$, is taken in the present implementation to be the local mean velocity from a precursor RANS simulation $\overline{u_1}(x_2)$. To enforce periodicity in the spanwise direction, the wavenumber and random number assignment deviates from that of \cite{Shur2014}; rather, it follows \cite{Martinez-Sanchis2021} where wavenumbers are specified according to the minimum wavenumber (fundamental harmonic) set from the spanwise domain width and assigning the spanwise component of the random unit vector $\mathbf{d}^n$ to be an integer multiple of the fundamental wavenumber. The standard STG formulation only provides velocity fluctuations; however, the thermodynamic fluctuations should also be considered for the compressible flows of interest. This is accomplished by relating the temperature and velocity fluctuations by the Strong Reynolds Analogy (SRA) \citep{Morkovin1962}. Furthermore, the pressure fluctuations are neglected, letting the density be computed from the instantaneous temperature and mean flow pressure from the ideal gas equation of state. The SRA is defined:

\begin{equation}
    \frac{T'/\overline{T}}{(\gamma-1)\overline{M}^2(u_1'/\overline{u_1})}\approx-1
\end{equation}
where $\gamma$ is the ratio of specific heats and the mean Mach number, $\overline{M}$, is:

\begin{equation}
    \overline{M}^2 = \frac{\overline{u_1}^2}{\gamma R_\text{gas}\overline{T}}
\end{equation}
The thermodynamic fluctuations are therefore:

\begin{align}
    T &= \overline{T} + T'= \overline{T} - \frac{(\gamma-1)\overline{M}^2u_1'\overline{T}}{\overline{u_1}}\\
    p &= \overline{p} +p' = \overline{p}\\
    \rho &= \frac{\overline{p}}{R_\text{gas}T}
\end{align}

In practice, a sufficiently long development region is provided downstream of the inflow to minimize the effect of any errors introduced by the assumptions inherent in the SRA, such as adiabaticity. The present method leads to a turbulence adjustment region (TAR) approximately $20\delta_i$, where $\delta_i$ is the inlet boundary layer height. A requirement is that a precursor RANS simulation has been performed, subsequently permitting knowledge of the mean flow, Reynolds stress tensor, and relevant length, velocity, and time scales such as the Kolmogorov length scale. Therefore, for all scale-resolving simulations, a precursor RANS was performed at matching conditions (wall temperature, Reynolds number, etc.) necessary to inform the inlet STG. The RANS were also performed with US3D; but with a Modified Steger-Warming (MSW) flux scheme and first order implicit Euler time integration. The turbulence model was the Menter SST two-equation model with vorticity source term (SST-V) \citep{Menter92}.

\subsection{Flow Conditions and Surface Geometry}\label{sec:flow-conds-and-surf-geom}
Six DNSs were performed to include three wall temperatures, $T_w/T_r=1.0$, 0.7, and 0.4 each repeated with a smooth and rough-walled surface. The cases are denoted TwTr\_p\_\_, where the first \_p\_ indicates the wall-to-recovery temperature ratio, for example $T_w/T_r=1.0$ would lead to TwTr1p0. The final character in the naming convention is whether the surface is rough (r) or smooth (s). The inlet to all simulations is supersonic with mean Mach number $M_{in}=2.5$, streamwise velocity $U_{in}=823.4$ m s$^{-1}$, density $\rho_{in}=0.1$ kg m$^{-3}$, and temperature $T_{in}=270.0$ K, following the STG procedure outlined in \S \ref{sec:stg}. The inlet values are denoted with subscript \textit{in} to clearly distinguish between the inlet values and the freestream values at the analysis locations (subscript $\infty$). The inlet and freestream conditions are indistinguishable for the smooth wall cases. However, for the rough-walled cases, a weak oblique shock occurs at the onset of the roughness, modifying the freestream conditions slightly. The shock angle is approximately $24.1$\textdegree{}, 2\% larger than the Mach angle $\alpha=\sin^{-1}(1/M_{in})=23.6$\textdegree{}. For consistency, the recovery temperature is defined as $T_r = T_{in}(1+0.883((\gamma-1)/2)M_{in}^2) = 568.0$ K. The flow conditions, including freestream values at the analysis locations, wall temperature, inlet boundary layer heights coming from the STG inflow ($\theta_i$ is momentum thickness at inlet), and roughness parameters are listed in Table \ref{tab:freestream-and-wall-conditions}.

\begin{figure}[h!]
    \centering
    \begin{subfigure}[b]{0.325\textwidth}
        \centering
        \includegraphics[width=\textwidth]{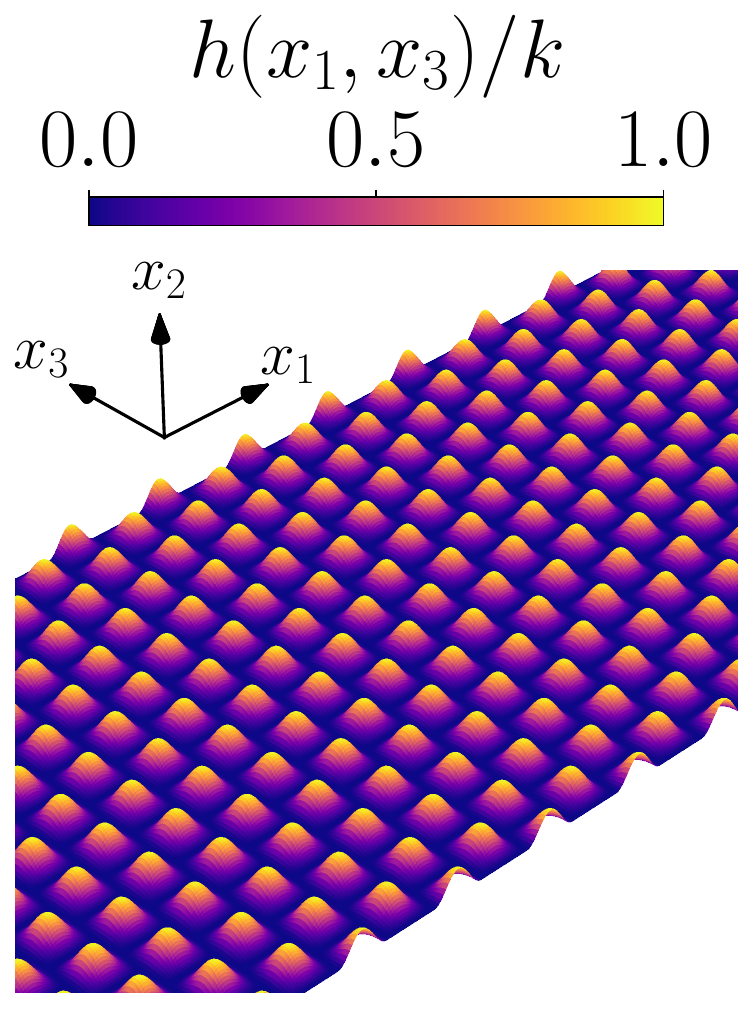}
        \vspace{0.1cm}
        \caption{3D perspective view}
        \label{fig:surface-profile-3D}
    \end{subfigure}
    \begin{subfigure}[b]{0.325\textwidth}
        \centering
        \includegraphics[width=\textwidth]{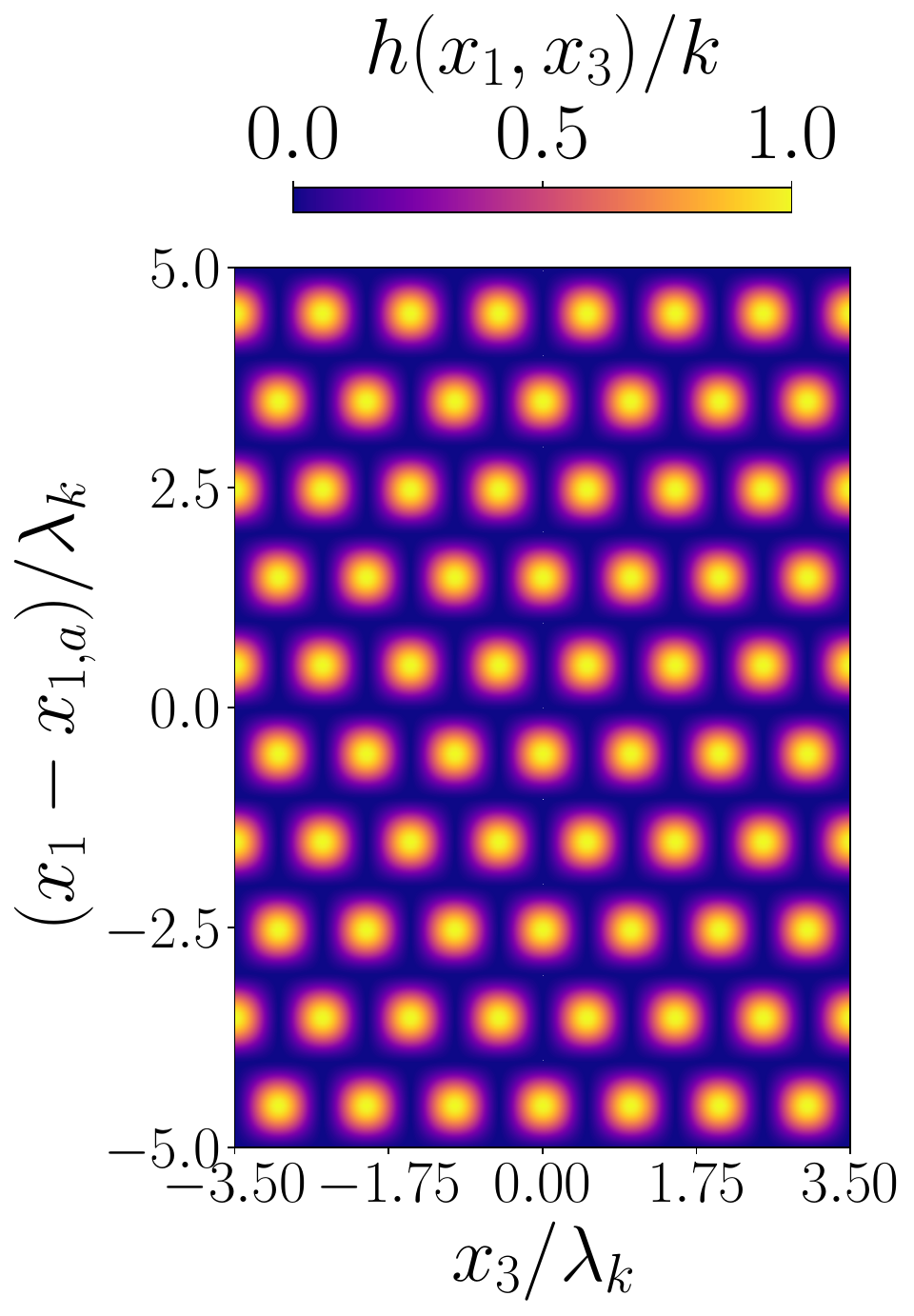}
        \caption{Top view}
        \label{fig:surface-profile-top-view}
    \end{subfigure}
    \begin{subfigure}[b]{0.325\textwidth}
        \centering
        \includegraphics[width=\textwidth]{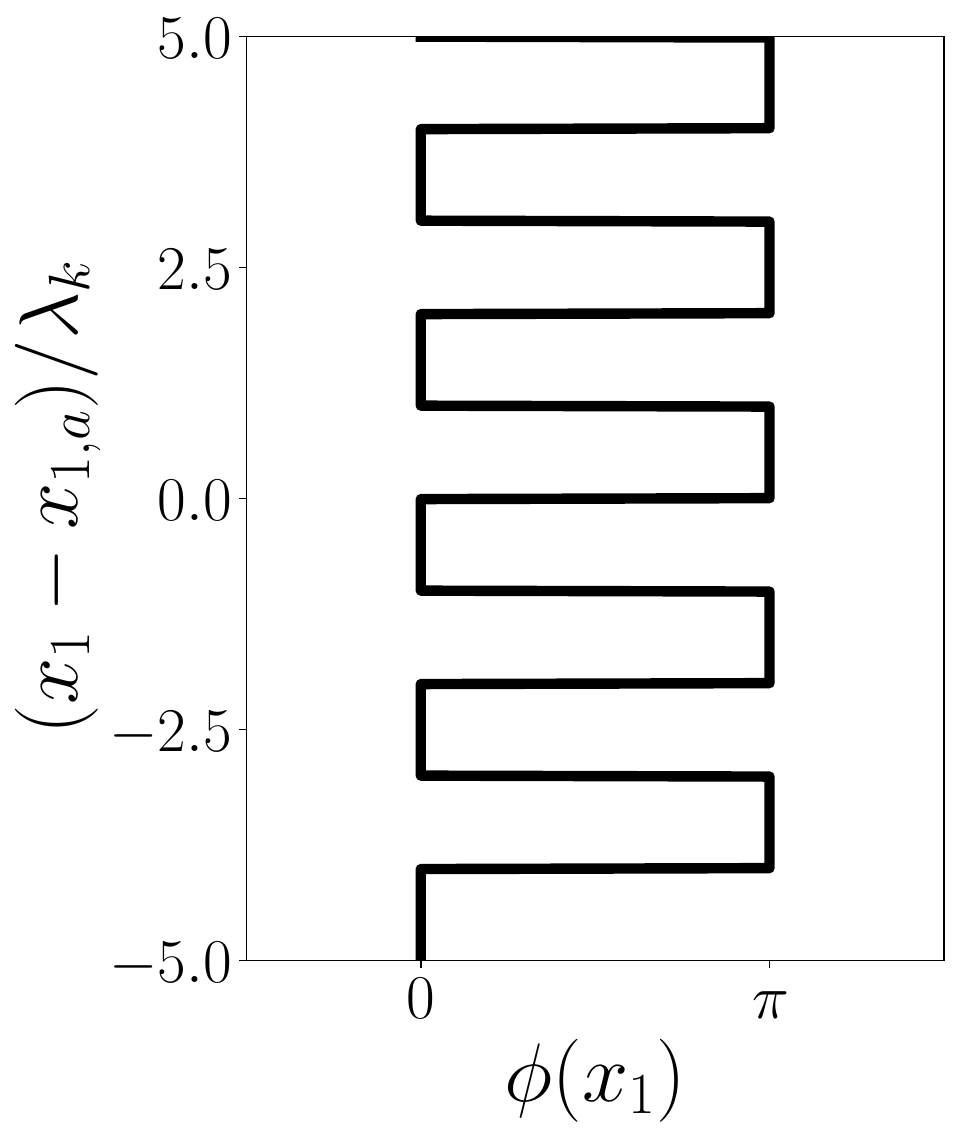}
        \caption{Phasing}
        \label{fig:surface-profile-phasing}
    \end{subfigure}
    \caption{Surface profile for sinusoidal roughness defined in Eq. \ref{eq:idealized_sin_rough}. The 3D perspective view in (a) is for the TwTr0p4r case, and the top view in (b) is from the TwTr1p0r case. All rough-wall cases follow analogous profiles, regardless of wall temperature; however, the wavelength and amplitude are sized according to the desired $k^+$ value. (c) is included to illustrate the staggered phasing from Eq. \ref{eq:span-phase-staggered}. The streamwise coordinate $x_1$ is centred about some analysis location $x_{1,a}$, and all spatial coordinates are normalised by the amplitude $k$ (equivalent to the roughness peak-to-valley-height).}
    \label{fig:surface-profile}
\end{figure}

The distributed surface roughness follows an idealised sinusoidal profile with streamwise rows of roughness elements staggered 180\textdegree{} out of phase. Equation \ref{eq:idealized_sin_rough} describes the surface profile. The roughness elements exist entirely on top of the originally smooth surface ($x_2\ge0$), and are implemented by perturbing the surface grid points of the computational grid. In Eq. \ref{eq:idealized_sin_rough} $h$ is the distance in the wall-normal ($x_2$) direction to perturb the computational surface grid, $k = \lambda_k/2$ is the amplitude, $\kappa_k = 2\pi/\lambda_k$ is the roughness wavenumber, $\lambda_k$ is the roughness wavelength, $x_{1,r}$ is the streamwise coordinate at the onset of the roughness, and $\phi$ is the spanwise phasing. The phase is specified according to Eq. \ref{eq:span-phase-staggered}. Roughness is introduced at the downstream location $x_{1,r}=20\delta_i$ to allow for the TAR from the STG. This guarantees that it is possible to observe the completion of the TAR, ensuring realistic and physical turbulence over the flat plate before it is altered by the roughened portion of the domain. All rough wall cases are designed to have the same effective slope $ES=0.5$ and viscous scaled peak-to-valley height $k^+=k u_\tau\rho_w/\mu_w\approx80$. Additional details for the roughness parameters are tabulated in Table \ref{tab:roughness-parameters}, and visualizations of the surface roughness profile are shown in Figure \ref{fig:surface-profile}.

\begin{table}
    \centering
    \begin{tabular}{lccccccccccc}
       Case  &  $M_\infty$ & $U_\infty$, m/s & $\rho_\infty$, kg m$^{-3}$ & $T_\infty$, K & $T_w$, K & $T_w/T_r$ & $\delta_i$, mm & $\theta_i$, mm & $\lambda_k$, mm & $k/\delta_i$ & $k^+$ \\\hline
       TwTr1p0s & 2.50 & 823.8 & 0.100 & 269.7 & 568.0 & 1.0 & 12.0 & 0.825 & -- & -- & -- \\
       TwTr1p0r & 2.48 & 819.9 & 0.102 & 272.8 & 568.0 & 1.0 & 4.0 & 0.296 & 1.600 & 0.200 & 79 \\
       TwTr0p7s & 2.50 & 823.6 & 0.100 & 269.8 & 397.6 & 0.7 & 7.3 & 0.584 & -- & -- & -- \\
       TwTr0p7r & 2.45 & 816.5 & 0.105 & 275.6 & 397.6 & 0.7 & 3.0 & 0.243 & 1.000 & 0.167 & 87 \\
       TwTr0p4s & 2.50 & 823.7 & 0.100 & 269.8 & 227.2 & 0.4 & 3.0 & 0.289 & -- & -- & -- \\
       TwTr0p4r & 2.47 & 819.1 & 0.103 & 273.4 & 227.2 & 0.4 & 1.5 & 0.151 & 0.375 & 0.125 & 71 
    \end{tabular}
    \caption{Freestream, wall-temperature, inlet, and roughness sizing parameters. Freestream values and $k^+$ are reported at the analysis location as defined in \S \ref{sec:comp-domain-grid-res-time-res}.}
    \label{tab:freestream-and-wall-conditions}
\end{table}

\begin{align}
	h(x_1,x_3) &= \frac{k}{4}\left[1-\cos\left(\kappa_kx_3+\phi\right)\right]\left[1-\cos\left(\kappa_k(x_1-x_{1,r})\right)\right]\label{eq:idealized_sin_rough}\\ 
    \phi(x_1) &=
    \begin{cases}
        \pi, & \text{if } \text{\textbf{mod}}\left(\left\lfloor \frac{(x_1-x_{1,r})}{\lambda_k}\right\rfloor,2\right)=0\\
        0, & \text{otherwise}
    \end{cases} \label{eq:span-phase-staggered}
\end{align}

\begin{table}
    \centering
    \begin{tabular}{llc}
        Parameter & Mathematical Description & Value \\ \hline
        Surface mean height & $R_m=\frac{1}{Lx_3Lx_1}\iint h\;dx_3dx_1$ & $k/4.0$\\
        Peak-to-valley height & $R_{pv}=\max(h-R_m)-\min(h-R_m)$ & $k$   \\
        Arithmetic mean deviation & $R_a = \frac{1}{Lx_3Lx_1}\iint|h-R_m|\;dx_3dx_1$ &  $k/4.3$  \\
        Root-mean-square & $R_{RMS}=\sqrt{\frac{1}{Lx_3Lx_1}\iint|h-R_m|^2\;dx_3dx_1}$ &  $k/3.6$ \\
        Skewness & $s_k=\frac{1}{(R_{RMS})^3}\frac{1}{Lx_3Lx_1}\iint|h-R_m|^3\;dx_3dx_1$ & 1.6\\
        Kurtosis & $k_u=\frac{1}{(R_{RMS})^4}\frac{1}{Lx_3Lx_1}\iint|h-R_m|^4\;dx_3dx_1$ & 3.0  \\
        Effective slope & $ES=\frac{1}{Lx_3Lx_1}\iint\left|\frac{\partial h}{\partial x_1}\right|\;dx_3dx_1$ & 0.5  \\
    \end{tabular}
    \caption{Roughness parameters. Physical heights are shown in normalised form with respect to the amplitude $k=\lambda_k/2$ and measured from the originally smooth surface $x_2=0$. Parameter definitions are according to \cite{Kadivar2021}.}
    \label{tab:roughness-parameters}
\end{table}

The flow conditions were chosen to have a consistent Reynolds number and roughness height over all the cases, to isolate the wall temperature as the only variable. Moreover, the flow conditions are selected to be in the fully rough regime, where the flow should essentially be Reynolds number independent \citep{Chung2021}. In the present work, the friction Reynolds number, $Re_\tau=\delta u_\tau \rho_w/\mu_w$, and roughness Reynolds number $k^+=k u_\tau \rho_w/\mu_w$ were set to match across all cases ($\delta$ is the boundary layer height, $u_\tau=\sqrt{\tau_w/\rho_w}$ is the friction velocity, $\rho_w$ is the density at the wall, and $\mu_w$ is the dynamic viscosity at the wall). To set up the subsequent results and discussion, Table \ref{tab:bl-parameters} shows the friction and roughness Reynolds numbers at the analysis location $x_{1,a}$, where the location is normalised by the inlet momentum thickness $\theta_i$. The average friction Reynolds number is $Re_\tau=784^{+4\%}_{-5\%}$ and the roughness Reynolds number is matched to be on average $k^+=79.1\pm10\%$.

\begin{table}
    \centering
    \begin{tabular}{lccccccccccc}
        Case & $x_{1,a}/\theta_i$ & $Re_\tau$ & $k^+$ \\\hline
        TwTr1p0s & 394 & 797 & -- \\
        TwTr1p0r & 687 & 790 & 79.2 \\
        TwTr0p7s & 313 & 812 & -- \\
        TwTr0p7r & 395 & 743 & 87.2 \\
        TwTr0p4s & 277 & 796 & -- \\
        TwTr0p4r & 316 & 769 & 70.8  
    \end{tabular}
    \caption{Matched friction Reynolds number and roughness Reynolds number at the analysis location $(x_{1,a})$. }
    \label{tab:bl-parameters}
\end{table}

\subsection{Computational Domain, Grid and Timestep Resolution} \label{sec:comp-domain-grid-res-time-res}
\begin{figure}[h!]
    \centering
    \includegraphics[width=0.9\textwidth]{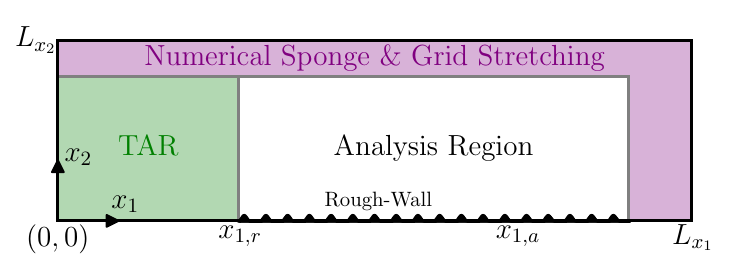}
    \caption{Annotated schematic of computational domain. Flow direction is from left to right. Streamwise direction is along $x_1$, wall-normal direction is along $x_2$, and spanwise direction is along $x_3$ (direction out of page, not shown in figure). Roughness onset at $x_{1,r}$ and analysis location at $x_{1,a}$. Smooth wall cases omit roughness.}
    \label{fig:computational-domain}
\end{figure}

The computational domain for the DNS follows a simple rectangular flat plate setup. Figure \ref{fig:computational-domain} shows a schematic of the computational domain. In Figure \ref{fig:computational-domain}, the flow direction is from left to right, with the left boundary being a supersonic inflow with synthetic turbulence as described in \S \ref{sec:stg}; the right and top boundaries are supersonic outflows. The spanwise boundaries are periodic in the $x_3$ direction. The wall is no-slip with isothermal wall temperature assigned based on the case in Table \ref{tab:freestream-and-wall-conditions}. The total domain length, height, and width are denoted by $L_{x_1}$, $L_{x_2}$, and $L_{x_3}$, respectively. Grid stretching and a numerical sponge are used for the last 10\% of the $x_1$ direction and after the last 20\%-25\% in the $x_2$ direction. The numerical sponge is added to increase dissipation near the outflow by gradually transitioning the spatial discretization from a centred to an upwind-biased scheme using the shock sensor described in \S \ref{sec:numerical-approach}. The numerical sponge in conjunction with grid stretching ramps from ``fully off'' to ``fully on'' following a hyperbolic tangent blending. Consistent with \S \ref{sec:flow-conds-and-surf-geom},  $x_{1,r}$ is the streamwise coordinate at the onset of the roughness after the TAR. $x_{1,a}$ is the streamwise coordinate at the analysis location and varies case-by-case to ensure matched friction Reynolds number, $Re_\tau$, at the analysis location. 

Grid details, including domain sizing and grid resolution are tabulated in Table \ref{tab:grid-info}. The number of grid points in the streamwise, wall-normal, and spanwise directions is given by $N_{x_1}$, $N_{x_2}$, and $N_{x_3}$, with the total number of cells on the order of 230,000,000 to 320,000,000 cells. The corresponding grid spacings are $\Delta x_1$, $\Delta x_2$, and $\Delta x_3$. In the refined regions away from the sponges near the outflow, the streamwise and spanwise spacings are constant; while the wall-normal grid spacing is stretched using a hyperbolic tangent away from the wall. The inner scaled (plus units) grid spacing are approximately $\Delta x_1^+\approx8.5$ in the streamwise direction and $\Delta x_3^+\approx5.0$ in the spanwise direction, with the largest grid spacing across all the cases never exceeding $\Delta x_1^+=9.6$ and $\Delta x_3^+=5.8$. In the wall-normal direction, the grid spacing in the first cell off the wall is given by $\Delta x_{2,w}^+$, with the average across all cases $\Delta x_{2,w}^+\approx0.56$ and no grid exceeding $\Delta x_{2,w}^+=0.79$. At the boundary layer edge the grid spacing $\Delta x_{2,\delta}$ is similar for all cases and rounds to 2\% of the boundary layer height. In terms of roughness wavelengths $\lambda_k$, the spanwise domain widths range from $7\lambda_k$ to $11\lambda_k$. There are 20 grid cells per wavelength in the streamwise direction and 30 to 33 grid cells in the spanwise direction for all rough-walled cases.

\begin{table}
    \centering
    \begin{tabular}{lcccccccccccc}
        Case & $N_{x_1}\!\times\!N_{x_2}\!\times\! N_{x_3}$ & $L{x_1}/\theta_i$ & $L_{x_2}/\theta_i$ & $L_{x_3}/\theta_i$ & $L_{x_3}/\lambda_k$ & $\Delta x_1^+$ & $\Delta x_{2,w}^+$ & $\Delta x_{2,\delta}/\delta$ & $\Delta x_3^+$  & $\lambda_k/\Delta x_1$ & $\lambda_k/\Delta x_3$ \\\hline
        TwTr1p0s & $3077\!\times\!310\!\times\!304$ & 640 & 364 & 29 & -- & 8.8 & 0.45 & 0.024 & 4.4 & -- & -- \\
        TwTr1p0r & $3180\!\times\!310\!\times\! 232$ & 947 & 270 & 38 & 7 & 7.9 & 0.79 & 0.021 & 4.8  & 20 & 33 \\
        TwTr0p7s & $2752\!\times\!300\!\times\!362$ & 481 & 250 & 31 & -- & 8.9 & 0.47 & 0.024 & 4.8  & -- & -- \\
        TwTr0p7r & $3020\!\times\!300\!\times\!270$ & 678 & 247 & 37 & 9 & 8.7 & 0.58 & 0.023 & 5.8  & 20 & 30 \\
        TwTr0p4s & $2484\!\times\!300\!\times\!392$ & 400 & 208 & 31 & -- & 9.6 & 0.48 & 0.024 & 5.2 & -- & -- \\
        TwTr0p4r & $3240\!\times\!300\!\times\!330$ & 441 & 199 & 27 & 11 & 7.1 & 0.60 & 0.023 & 4.7 & 20 & 30 
    \end{tabular}
    \caption{Grid resolution and domain sizes. Cell spacing reported in inner scaled units are based on the turbulence statistics from the analysis location. Domain widths are normalised by the turbulent inflow momentum thickness and the sinusoidal roughness wavelength.}
    \label{tab:grid-info}
\end{table}

Simulation time step and statistics sampling temporal information is provided in Table \ref{tab:sim-tim-info}. The time step is denoted $\Delta t$ and normalised by the friction velocity, $u_\tau$, and wall kinematic viscosity, $\nu_w=\mu_w/\rho_w$, to get an inner scaled non-dimensional time step $\Delta t^+$. During statistics collection, the time step is held constant and set well below the $\Delta t^+ \le 0.4$ temporal resolution necessary to capture accurate turbulence statistics per \cite{Choi1994} -- with no case exceeding $\Delta t^+ = 0.154$. In terms of statistics collection (time averaging), $T_s$ is the statistics collection time and $T_{int}$ is the integral time scale based on the streamwise velocity fluctuations, $u_1'$. The integral time scale $T_{int}$ is computed from the temporal autocorrelation of a data probe (time series) at $x_2/\delta\approx0.2$ and $x_2/\delta\approx0.5$, reporting data from the location with the smaller of the two $T_s/T_{int}$. Statistics collection began after the initial transients passed and the simulations reached a statistically stationary state. Subsequent statistics collection was sufficiently long for $T_s u_\tau/\delta\ge8$ in terms of outer time scaling and $T_s/T_{int}>200$ in terms of integral time scales, with many simulations exceeding this minimum by at least a factor of two.

\begin{table}
    \centering
    \begin{tabular}{lcccc}
        Case & $\Delta tu_\tau^2/\nu_w$ & $T_su_\tau/\delta$ & $T_s/T_{int}$ \\\hline
        TwTr1p0s & 0.077 & 8.8 & 627\\
        TwTr1p0r & 0.143 & 11.7 & 328\\
        TwTr0p7s & 0.089 & 9.6 & 658\\
        TwTr0p7r & 0.154 & 11.0 & 321\\
        TwTr0p4s & 0.124 & 8.7 & 380\\
        TwTr0p4r & 0.118 & 8.8 & 219
    \end{tabular}
    \caption{Simulation time step and statistics sampling -- temporal information.}
    \label{tab:sim-tim-info}
\end{table}

\subsection{Density Weighting and Double Averaging}
Compressible flows exhibit significant density variations; therefore, Favre (density-weighted) averaging is employed to account for density and temperature fluctuations and to recover a turbulence formulation analogous to the well-known incompressible theory. The time average follows the standard Reynolds or Favre decomposition as follows for a given variable $\psi$:

\begin{align}
    \psi(x_i,t) &= \ol{\psi}(x_i) + \psi'(x_i,t)\;\;\;\text{(Reynolds decomposition)}\\
    \psi(x_i,t) &= \wt{\psi}(x_i) + \psi''(x_i,t)\;\;\;\text{(Favre decomposition)}\\
    \ol{\psi'} &= 0\\
    \wt{\psi} &= \frac{\ol{\rho\psi}}{\ol{\rho}}=\overline{\psi}+\frac{\overline{\rho'\psi'}}{\overline{\rho}}\\
    \ol{\psi''} &= \frac{-\ol{\rho'\psi'}}{\ol{\rho}} \neq 0\\
    \ol{\rho\psi''} &= 0 
\end{align}

A double-average is defined as both the temporal and spatial average in the spanwise periodic direction. The spatial averaging is denoted with angle brackets $\left<\psi\right>$ and is performed in the spanwise $x_3$-direction. Due to the presence of grid-resolved roughness, the computational mesh is deformed in the vicinity near the roughness elements. Therefore, the spatial averaging is defined in Eq. \ref{eq:bin-averaging} for the discrete computational grid uniformly spaced in the spanwise direction. The spanwise mean profile is obtained by first interpolating the time-averaged field onto a common wall-normal coordinate. The spatial average is then performed in the periodic spanwise direction using only the fluid locations available at each wall-normal position:

\begin{equation}
    \left<\wt{\psi}\right>(x_2) = \frac{1}{N_{f}(x_2)}\sum_{I\in\mathcal{F}({x_2})} \wt{\psi}_{I}(x_2)
    \label{eq:bin-averaging}
\end{equation}
where $\wt{\psi}_{I}(x_2)$ denotes the Favre time-averaged value of $\psi$ interpolated onto the common wall-normal grid, $\mathcal{F}(x_2)$ denotes set the spanwise $x_3$ locations containing fluid at the wall-normal location $x_2$, and $N_{f}(x_2)$ is the corresponding number of spanwise fluid locations. Due to the presence of the roughness, the time-averaged signal may still possess a spatially varying profile -- a spatial fluctuation, $\wt{\psi}^\dagger$ . The spatial decomposition of the time average is as follows:

\begin{equation}
    \wt{\psi}(x_i) = \left<\wt{\psi}\right>(x_2) + \wt{\psi}^\dagger(x_i)
\end{equation}
Including the span-averaged mean, the spatial (dispersive deviation), and the turbulent fluctuation, the complete decomposition of an instantaneous value is:

\begin{equation}
    \psi(x_i,t) = \left<\wt{\psi}\right>(x_2) + \wt{\psi}^\dagger(x_i) + \psi''(x_i,t)
\end{equation}
Outside of \S \ref{sec:R_RSL}, the double average angle bracket notation, $\left<\cdot\right>$, is dropped and $\ol{(\cdot)}$ or $\wt{(\cdot)}$ implies a double average.

\section{Surface Response Characteristics} \label{sec:surface}
The results and discussion begin with an analysis of the surface quantities. The wall shear stress and surface heat transfer are investigated in detail, describing the relative contributions of the viscous and pressure components to the total shear stress at the wall. Likewise, similar attention is given to surface heating. These two parameters are important for the subsequent momentum and thermal boundary layer analyses, because they dictate the slope of the inner scaled velocity or temperature profiles. 

\subsection{Wall Shear Stress and Heat Transfer}
The wall shear stress, $\tau_w$, is obtained by summing both the viscous shear and pressure drag effects, integrated across the entire span and over one roughness wavelength in the streamwise direction. The viscous component is given by $\frac{1}{A_P}\int_S\tau_vdS$, where S is the wetted surface area, $A_P$ is the projected planform area, $\tau_v=\left.\mu\frac{\partial u_i}{\partial n}\right|_w\cdot\hat{e}_{x_1}$ is the local tangential viscous wall shear stress (specifically the streamwise component), $\mu$ is the dynamic viscosity, $u_i$ is the Cartesian velocity vector, $n$ is the wall-normal direction (distance), and $\hat{e}_{x_1}$ is the streamwise unit vector. The pressure component is given by $\frac{1}{A_P}\int_S-p_w(\hat{n}\cdot\hat{e}_{x_1})dS$, where $p_w$ is the surface pressure, and $\hat{n}$ is the wall-normal unit vector. The wall heat flux $q_w$ is defined per Eq. \ref{eq:fourier-law} and $\left.\pp{T}{n}\right|_w$ is the temperature gradient in the wall-normal direction at the wall. The sign convention is such that negative $q_w$ indicates heat transfer out of the fluid into (toward) the boundary. Equations \ref{eq:tauw-definition} and \ref{eq:qw-definition} summarize the wall shear stress and heat transfer.

\begin{align}
    \tau_w &= \frac{1}{A_P}\int_S\left[\left.\mu\frac{\partial u_i}{\partial n}\right|_w\cdot\hat{e}_{x_1}-p_w(\hat{n}\cdot\hat{e}_{x_1})\right]dS \label{eq:tauw-definition}\\
    q_w &= \frac{1}{A_P}\int_S \left[\left.-\lambda_\Theta\pp{T}{n}\right|_w\right] dS \label{eq:qw-definition}
\end{align}

We further define skin friction coefficient $C_f$, and a heat transfer coefficient $C_h$ (in Stanton-number form) based on recovery temperature, boundary layer edge velocity $u_e$, and edge density $\rho_e$ as:

\begin{align}
    C_f &= \frac{2\ol{\tau_w}}{\rho_e u_e^2} \label{eq:skin-fric-coeff}\\
    C_h &= \frac{\ol{q_w}}{\rho_e u_e c_p(T_w-T_r)} \label{eq:heat-trans-coeff}
\end{align}

\begin{figure}[h!]
    \centering
    \begin{subfigure}[b]{0.329\textwidth}
        \centering
        \includegraphics[width=\textwidth]{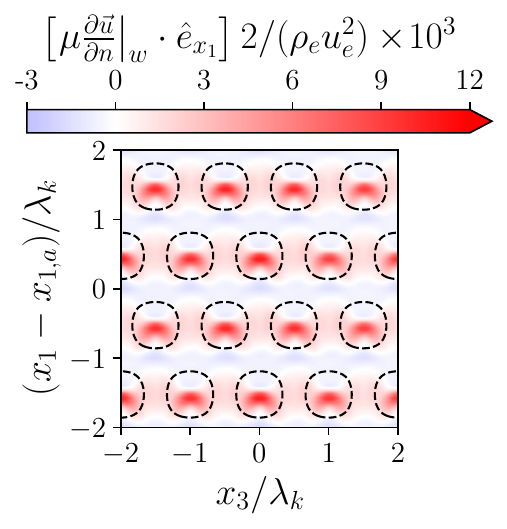}
        \caption{TwTr1p0r}
        \label{fig:tauv1p0}
    \end{subfigure}
    \begin{subfigure}[b]{0.329\textwidth}
        \centering
        \includegraphics[width=\textwidth]{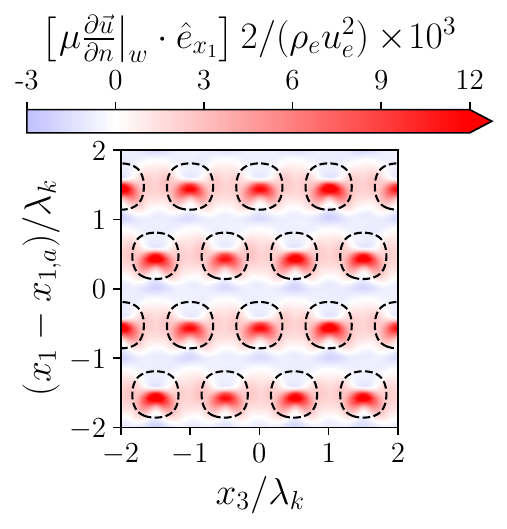}
        \caption{TwTr0p7r}
        \label{fig:tauv0p7}
    \end{subfigure}
    \begin{subfigure}[b]{0.329\textwidth}
        \centering
        \includegraphics[width=\textwidth]{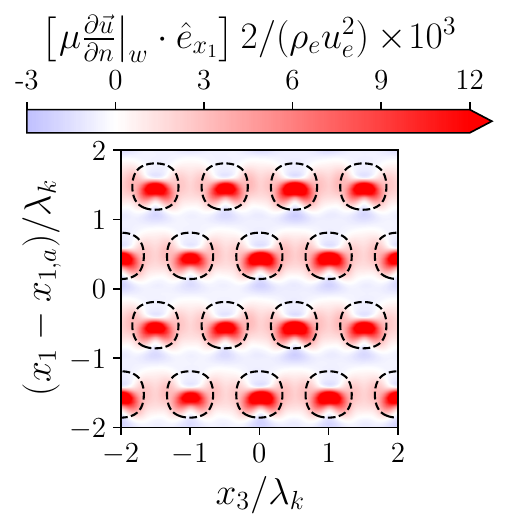}
        \caption{TwTr0p4r}
        \label{fig:tauv0p4}
    \end{subfigure}
    \caption{Contours of the time-averaged, viscous shear component of the wall shear stress on the surface. The contour represents the first term of the integrand in Eq. \ref{eq:tauw-definition}, computed locally over the surface and then normalised by edge conditions consistent with the skin friction coefficient normalisation. The spanwise, $x_3$, and streamwise $x_1$, spatial locations are normalised by the roughness wavelength $\lambda_k$. The streamwise location is centred about the analysis location $x_{1,a}$. Black dashed circles indicate the individual roughness elements, with the circles corresponding to the wall-normal locations $x_2=R_m$, the surface mean height.}
    \label{fig:tauv-surf-contours}
\end{figure}

\begin{figure}[h!]
    \centering
    \begin{subfigure}[b]{0.329\textwidth}
        \centering
        \includegraphics[width=\textwidth]{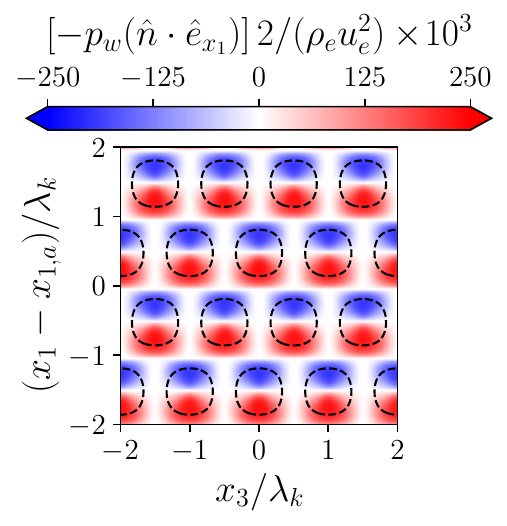}
        \caption{TwTr1p0r}
        \label{fig:pwndotv1p0}
    \end{subfigure}
    \hfill
    \begin{subfigure}[b]{0.329\textwidth}
        \centering
        \includegraphics[width=\textwidth]{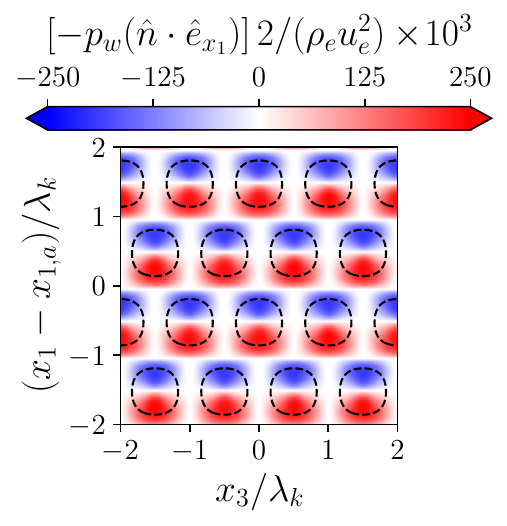}
        \caption{TwTr0p7r}
        \label{fig:pwndotv0p7}
    \end{subfigure}
    \hfill
    \begin{subfigure}[b]{0.329\textwidth}
        \centering
        \includegraphics[width=\textwidth]{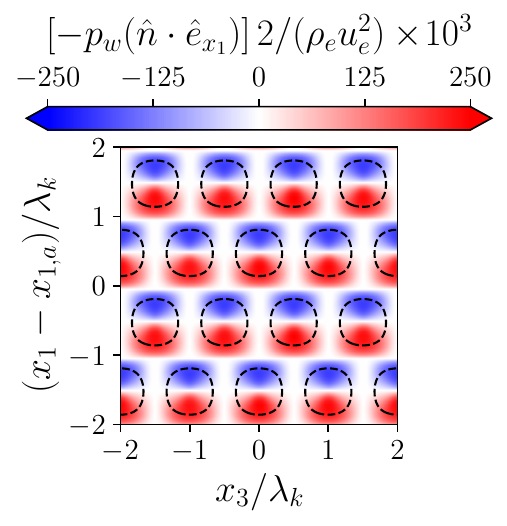}
        \caption{TwTr0p4r}
        \label{fig:pwndotv0p4}
    \end{subfigure}
    \caption{Contours of the time-averaged, pressure component of the wall shear stress on the surface. The contour represents the second term of the integrand in Eq. \ref{eq:tauw-definition}, computed locally over the surface and then normalised by edge conditions consistent with the skin friction coefficient normalisation. The spanwise, $x_3$, and streamwise $x_1$, spatial locations are normalised by the roughness wavelength $\lambda_k$. The streamwise location is centred about the analysis location $x_{1,a}$. Black dashed circles indicate the individual roughness elements, with the circles corresponding to the wall-normal locations $x_2=R_m$, the surface mean height.}
    \label{fig:pwndotv-surf-contours}
\end{figure}

\begin{figure}[h!]
    \centering
    \begin{subfigure}[b]{0.329\textwidth}
        \centering
        \includegraphics[width=\textwidth]{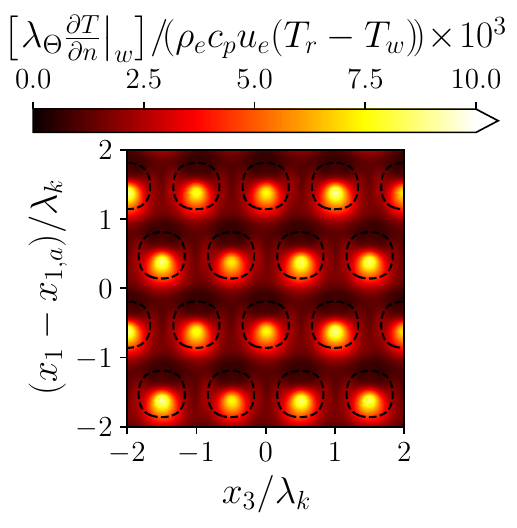}
        \caption{TwTr0p7r}
        \label{fig:qw0p7}
    \end{subfigure}
    \hspace{1cm}
    \begin{subfigure}[b]{0.329\textwidth}
        \centering
        \includegraphics[width=\textwidth]{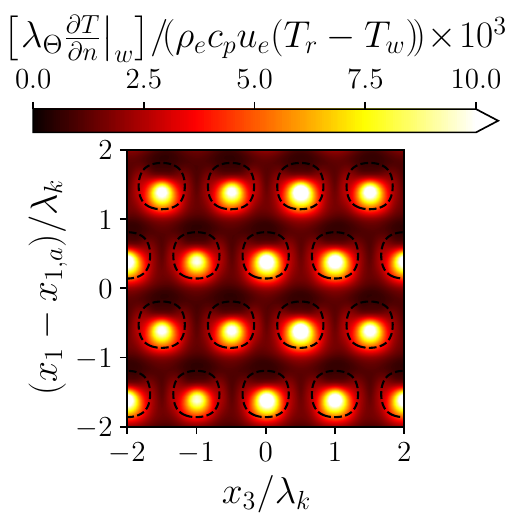}
        \caption{TwTr0p4r}
        \label{fig:qw0p4}
    \end{subfigure}
    \caption{Contours of the time-averaged wall heat flux (positive into the surface). The contour represents the negative of the integrand in Eq. \ref{eq:qw-definition}, computed locally over the surface and then normalised by edge conditions consistent with the heat transfer coefficient (Stanton number) normalisation. The spanwise, $x_3$, and streamwise $x_1$, spatial locations are normalised by the roughness wavelength $\lambda_k$. The streamwise location is centred about the analysis location $x_{1,a}$. Black dashed circles indicate the individual roughness elements, with the circles corresponding to the wall-normal locations $x_2=R_m$, the surface mean height. The adiabatic TwTr1p0r case is not shown because the heat transfer coefficient normalisation is undefined when the wall heat flux is zero and the wall temperature equals the recovery temperature.}
    \label{fig:qw-surf-contours}
\end{figure}

Figures \ref{fig:tauv-surf-contours} to \ref{fig:qw-surf-contours} show contours of the time-averaged skin friction and heat transfer coefficients on the surface over four roughness wavelengths near the analysis location $x_{1,a}$. To orient the view, black dashed circles indicate the individual roughness elements by placing them at the surface mean height wall-normal location $x_2=R_m$. The contours in Figure \ref{fig:tauv-surf-contours} represent the first term of the integrand in Eq. \ref{eq:tauw-definition} (viscous shear component), computed locally over the surface and then normalised by edge conditions consistent with the skin friction coefficient normalisation. Qualitatively, the maximum value of the viscous shear is larger with colder wall temperature. In the valleys behind the roughness peaks, the viscous shear drops to zero with small negative viscous shear indicating re-circulation or flow separation behind the roughness elements. In line with this, on the lee-side of the roughness elements, the viscous shear is almost zero. The peak shearing is near the crests and takes an inverted u-shape on the windward side of the roughness elements. The positive viscous shear contour on the windward side of the roughness elements takes the inverted u-shape because those are the areas exposed to the oncoming flow; the centre of the roughness element is shielded by the previous row of roughness with the same phase.

The contours in Figure \ref{fig:pwndotv-surf-contours} represent the second term of the integrand in Eq. \ref{eq:tauw-definition} (pressure component), computed locally over the surface and then normalised by edge conditions consistent with the skin friction coefficient normalisation. The pressure configuration normal to the surface is simpler than for the viscous shear. There is high-pressure covering most of the windward side of the the roughness elements and low pressure on the lee-side of the roughness elements. Across the peaks and in the valleys behind the roughness elements the surface pressure component becomes small. Visually, the low and high-pressure regions on the lee and windward sides, respectively, appear to be similar in magnitude and almost an order of magnitude larger than the viscous shear component from Figure \ref{fig:tauv-surf-contours}. Though the local pressure contribution is high, when integrating across the roughness element the low and high regions cancel and result in a net positive force (leading to drag) that is similar in magnitude to the viscous shear component.

The contours in Figure \ref{fig:qw-surf-contours} represent the negative of the integrand in Eq. \ref{eq:qw-definition} (wall heat flux), computed locally over the surface and then normalised by edge conditions consistent with the heat transfer coefficient normalisation from Eq. \ref{eq:heat-trans-coeff}. In Figure \ref{fig:qw-surf-contours} only the two diabatic wall cases, TwTr0p7r and TwTr0p4r, are shown. Qualitatively, the peak heating occurs on the windward side of the roughness elements and near the crests. To the sides of the roughness elements, slightly increased heating can be seen wrapping along the sides of the elements in a u-shape around the base of the element. The lowest heating appears on the lee side and valleys behind the roughness elements. The contours from the viscous shear component and the heating appear to be inverses of each other. Recalling Figure \ref{fig:tauv-surf-contours}, there is an inverted u-shape on the windward side of the roughness elements of high, positive viscous shear near the peak and sides of the windward side of the roughness element -- leading to a pocket of blue/white in the middle of the windward side of the roughness elements with low shear. In Figure \ref{fig:qw-surf-contours}, it appears that the peak in heating is occurring in that low shear region in the middle of the windward side roughness elements. When comparing the heating across wall temperature cases, the lower wall temperature naturally results in higher heat flux into the surface due to the larger gradient in temperature between the fluid phase and the cold wall.

\begin{figure}[h!]
    \centering
    \includegraphics[width=0.75\linewidth]{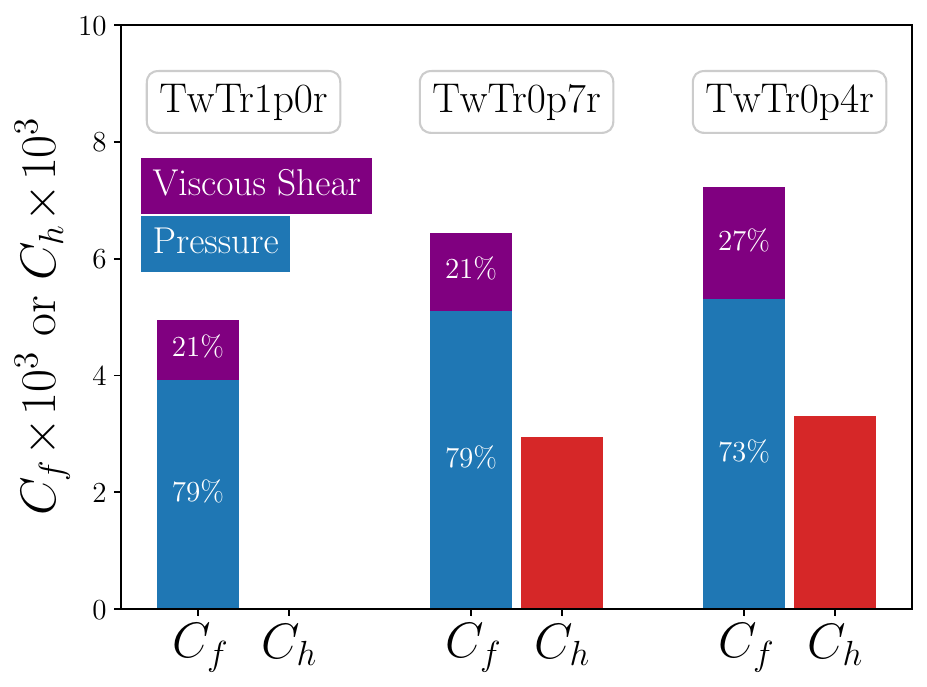}
    \caption{Integrated skin friction, $C_f$, and heat transfer, $C_h$, coefficients over the roughness elements located at analysis location $x_{1,a}$. Integration is performed over one roughness wavelength, $\lambda_k$, in the streamwise direction and also over the entire spanwise extent. The skin friction coefficient is shown as composition of pressure and viscous shear components, consistent with Eq. \ref{eq:tauw-definition}, and is defined per Eq. \ref{eq:skin-fric-coeff}. The heat transfer coefficient is defined per Eq. \ref{eq:heat-trans-coeff}.}
    \label{fig:cf-ch-bar-chart}
\end{figure}

The total, integrated wall shear stress and surface heat flux are reported in Figure \ref{fig:cf-ch-bar-chart} for all three wall temperature conditions for one row of roughness elements. At the analysis location $x_{1,a}$, integration is performed over one roughness wavelength, $\lambda_k$ , in the streamwise direction and also over the entire spanwise extent. In the bar chart, the total wall shear stress (skin friction coefficient) is separated into the contributions from the pressure and the viscous shear components. From Figure \ref{fig:cf-ch-bar-chart} the skin friction coefficient increases with decreasing wall temperature. This is driven by both an increase in contribution to the viscous shear and pressure components, with the latter being a much smaller relative increase. Due to the variation in skin friction with wall temperature for otherwise matched rough surface conditions, this suggests that the shear stress augmentation depends on both wall thermal condition and surface roughness boundary conditions.

In addition to the visualisations from Figure \ref{fig:cf-ch-bar-chart} for the three rough wall simulations, Table \ref{tab:surface-analysis} tabulates the skin friction and heat transfer coefficient values for all six conditions presently considered, including comparisons to the smooth-wall counterparts. For both the smooth ($C_{f,sw}$) and rough walls ($C_{f,rw}$), the skin friction coefficient increases with decreasing wall temperature conditions. The average skin friction augmentation due to the roughness is $C_{f,rw}/C_{f,sw}=2.96\pm0.28$ and has no clear trend on wall temperature dependence; however, it does map linearly with roughness Reynolds number $k^+$ (though additional data points beyond the three considered are necessary to confirm the precise trend). For the two diabatic cases, decreased wall temperature leads to increased wall heating, and adding roughness ($C_{h,rw}$) increases the heating from the smooth wall ($C_{h,sw}$) by a factor of 2.11 to 2.50, with less heating augmentation for the colder wall. A compelling insight is that the increase in skin friction due to the presence of the roughness is larger than the increase in heat transfer due to the roughness. When comparing the relative increase, the skin friction augmentation is 128\% of the augmentation in heat transfer (average of the two diabatic cases). Consequently, shear stress-heat transfer relationships like the classical Reynolds analogy factor, $s=2C_h/C_f$ would not be constant across smooth and rough cases. In terms of the Chilton-Colburn form \citep{CC-RA-1934}, $2Pr^{2/3}C_h/C_f\approx1$, the rough-walled cases result in a value approximately 0.75 for both cases as indicated by Table \ref{tab:surface-analysis}. The literature has accredited this result to the fact that the drag augmentation increase is primarily due to the increased pressure contribution in addition to the viscous component, whereas heat transfer does not have an additional pressure-like mechanism \citep{Modesti2022,Kadivar2025}. Our $2Pr^{2/3}C_h/C_f\approx0.75$ for the rough-wall cases at matched $k^+$ is consistent with \cite{Modesti2022} who also found that the Reynolds analogy factor mainly depends on the roughness Reynolds number.

\begin{table}
    \centering
    \begin{tabular}{lcccccccc}
        Case & $C_f\!\times\!10^3$ & $C_h\!\times\!10^3$ & $2\frac{C_h}{C_f}$ & $2Pr^{2/3}\frac{C_h}{C_f}$ & $\frac{C_{f,rw}}{C_{f,sw}}$ & $\frac{C_{h,rw}}{C_{h,sw}}$ & $\frac{C_{f,rw}}{C_{f,sw}}\!/\!\frac{C_{h,rw}}{C_{h,sw}}$ & $k^+$\\ \hline
        TwTr1p0s & 1.6625 & -- & -- & -- & -- &  -- &  -- & --\\ 
        TwTr1p0r & 4.9534 & -- & -- & -- & 2.9751 & -- & -- & 79\\ 
        TwTr0p7s & 1.9892 & 1.1830 & 1.1894 & 0.9703 & -- & -- & -- & --\\
        TwTr0p7r & 6.4473 & 2.9555 & 0.9168 & 0.7479 & 3.2412 & 2.4983 & 1.2973 & 87\\
        TwTr0p4s & 2.7025 & 1.5677 & 1.1602 & 0.9465 & -- & -- & -- & --\\
        TwTr0p4r & 7.2280 & 3.3142 & 0.9171 & 0.7481 & 2.6746 & 2.1140 & 1.2651 & 71
    \end{tabular}
    \caption{Skin friction and heat transfer coefficient surface analysis, integrated over the row of roughness elements at $x_{1,a}$. Reynolds analogy factor comparison and roughness heating-to-drag augmentation ratios are included. Roughness Reynolds number $k^+$ included for reference.}
    \label{tab:surface-analysis}
\end{table}

The implication of the data in this section is that future (or tuning of existing) roughness augmentation models may obtain accurate heating augmentation directly from drag augmentation without needing wall-temperature-dependent terms. Moreover, for rough-walled, turbulent flows, the skin friction augmentation is often said to be “dominated” by the pressure and a function of the roughness height $k^+$. From the present results, we see that the pressure is the dominant component to the skin friction in the presence of roughness, but that the viscous shear can make more than 20\% of the total skin friction for the present conditions. Moreover, differences between wall-temperature cases with similar roughness heights (surface geometries) are driven by the viscous shear component and is a result of the larger wall-normal velocity gradients at the wall with decreasing wall temperature. Additionally, heat transfer (energy) does not have an additional component (like pressure to shear stress / momentum), so its augmentation due to roughness is less than the augmentation in drag due to roughness.

\section{Wall Origin and Roughness Characterization} \label{sec:wall-origin-and-roughness-characterization}

This section will discuss how surface properties are integrated to obtain estimates of the wall offset. In addition, we discuss the extent of the roughness sublayer through the spatial dispersion. This analysis aims to quantify the effect of the roughness and wall thermal boundary condition on the boundary layer.

\subsection{Roughness Sublayer Extent}\label{sec:R_RSL}

As described by \cite{Florens2013}, the spatial dispersion of a time-averaged quantity $\ol{\psi}$ or $\wt{\psi}$ based on the double-average decomposition is defined as the standard deviation or error of the spatial variation of the various $\psi$ quantities:

\begin{equation}
    D\left(\wt{\psi}(x_2)\right) = \sqrt{\left< {\wt{\psi}^\dagger}{\wt{\psi}^\dagger}\right>(x_2)}
    \label{eq:spatial-dispersion}
\end{equation}
where presently $\wt{\psi}$ is $\wt{u_1}$, $\wt{u_2}$, and $\wt{u_3}$. Above the roughness sublayer (RSL), the flow should become spatially homogeneous in the spanwise direction and the dispersive terms should vanish. For the present work, the extent of the roughness sublayer height ($R_{RSL}$) is found based on the range of true geometric wall-normal locations $(x_2-0)$ where the normalised spatial dispersion reaches the threshold values of 2.5\% and 5.0\%. The spatial dispersion is taken as twice that defined in Eq. \ref{eq:spatial-dispersion}, and normalisation is by the corresponding local double average (for $\wt{u_2}$, $\wt{u_3}$ which are close to zero, the normalisation is by $\left<\wt{u_1}\right>$). The factor of two is applied to the root-mean-square (RMS) spatial fluctuation to represent a two-standard-deviation envelope, capturing approximately 95\% of the spatial variability.

\begin{equation}
    R_{RSL} = x_2\;\;\;\text{s.t.}\;\;\;\frac{2D(\wt{\psi})}{\left<\wt{\psi}\right>} \le 0.025 \text{ to } 0.050
    \label{eq:spat-disp-Rrsl-threshold}
\end{equation}

\begin{figure}[h!]
    \centering
    \begin{subfigure}[b]{0.3\textwidth}
        \centering
        \includegraphics[width=\textwidth]{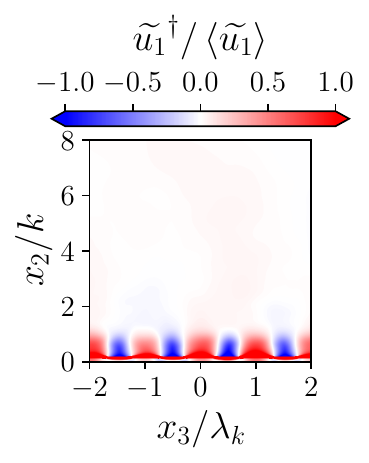}
        \caption{Str. velocity}
        \label{fig:spat-fluct-ufav-0p4}
    \end{subfigure}
    \hfill
    \begin{subfigure}[b]{0.3\textwidth}
        \centering
        \includegraphics[width=\textwidth]{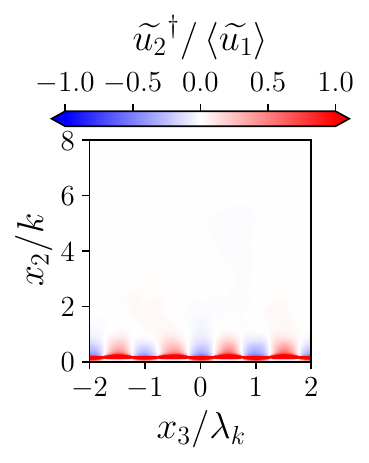}
        \caption{W-N velocity}
        \label{fig:spat-fluct-vfav-0p4}
    \end{subfigure}
    \hfill
    \begin{subfigure}[b]{0.3\textwidth}
        \centering
        \includegraphics[width=\textwidth]{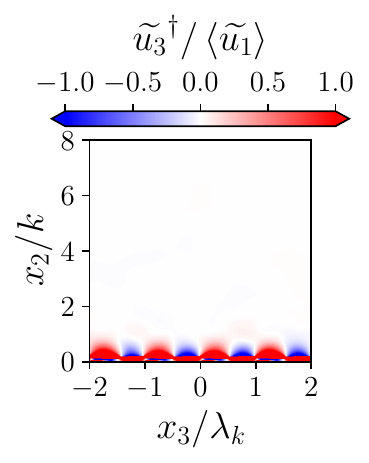}
        \caption{Span. velocity}
        \label{fig:spat-fluct-wfav-0p4}
    \end{subfigure}
    \hfill
    \caption{Spatial fluctuations of the mean velocities for the TwTr0p4r case. View zoomed on four roughness wavelengths in the span and eight roughness heights (four wavelengths) in the wall-normal directions.}
    \label{fig:spatial-fluctuations-TwTr0p4r}
\end{figure}

\begin{figure}[h!]
    \centering
    \begin{subfigure}[b]{0.245\textwidth}
        \centering
        \includegraphics[width=\textwidth]{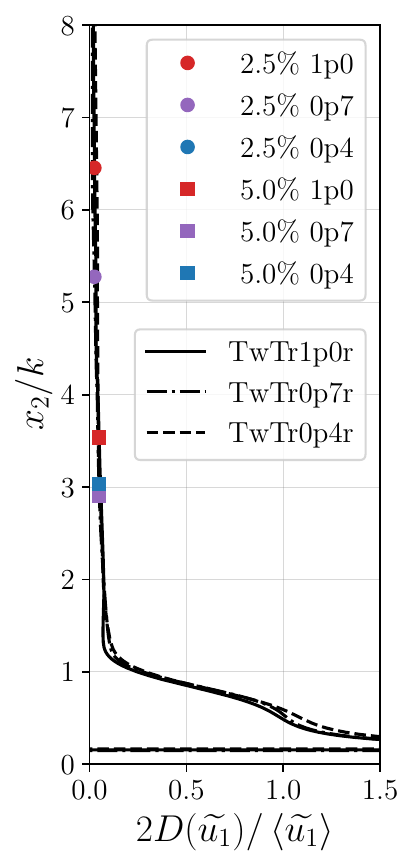}
        \caption{Str. velocity}
        \label{fig:RSL-ufav}
    \end{subfigure}
    \hfill
    \begin{subfigure}[b]{0.245\textwidth}
        \centering
        \includegraphics[width=\textwidth]{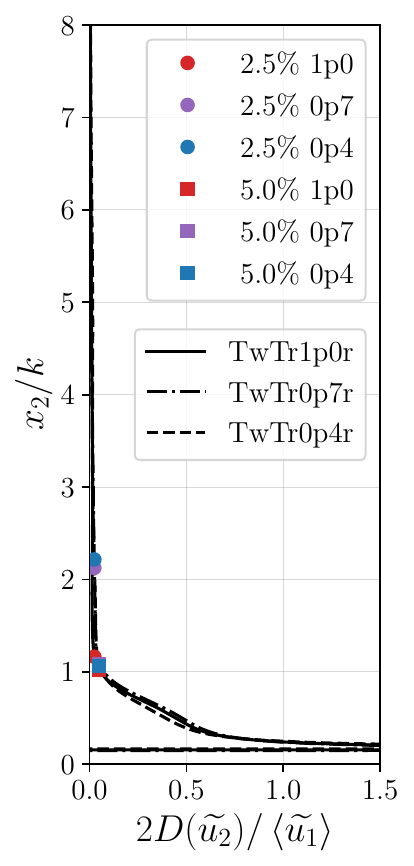}
        \caption{W-N velocity}
        \label{fig:RSL-vfav}
    \end{subfigure}
    \hfill
    \begin{subfigure}[b]{0.245\textwidth}
        \centering
        \includegraphics[width=\textwidth]{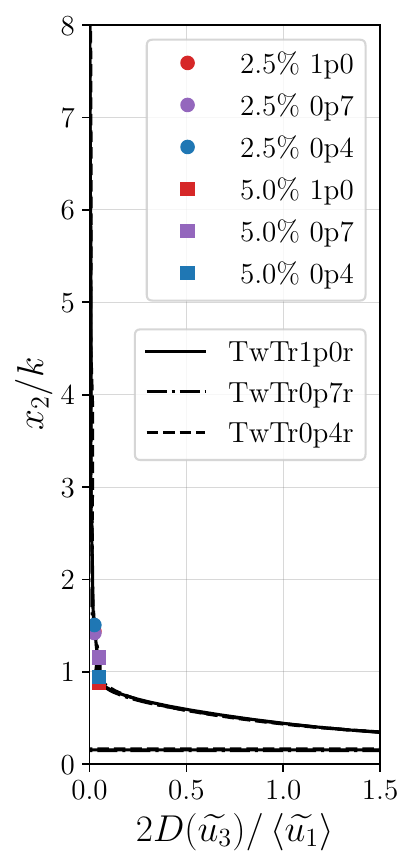}
        \caption{Span. velocity}
        \label{fig:RSL-wfav}
    \end{subfigure}
    \caption{Spatial dispersion for the mean velocities at one plane located at streamwise station $x_{1,a}$. Roughness sublayer height $R_{RSL}$ defined based on the the wall-normal location $x_2$ at which $2D(\wt{\psi})/\left<\wt{\psi}\right> \le 0.025 \text{ or } 0.050$. 2.5\% represented with circles and 5.0\% represented with squares.}
    \label{fig:spatial-dispersion}
\end{figure}

Figure \ref{fig:spatial-fluctuations-TwTr0p4r} shows contours of the normalised spatial fluctuations of the time-averaged quantities for the TwTr0p4r case at a single spanwise-wall-normal plane for one streamwise position (at the analysis location). The other wall temperature conditions are omitted for brevity since they are qualitatively similar. The streamwise and spanwise velocities have the largest spatial fluctuations, followed by the wall-normal velocity. Visually, the largest distortion to the mean flow remains confined to below the roughness crests. The streamwise and spanwise velocity components respond strongest to roughness geometry because the no-slip condition is oriented in those wall-parallel directions. To qualitatively report the wall-normal extent of the spatial variations seen in Figure \ref{fig:spatial-fluctuations-TwTr0p4r}, Figure \ref{fig:spatial-dispersion} shows the normalised spatial dispersion per Eq. \ref{eq:spat-disp-Rrsl-threshold} for the three velocity components at all wall temperature conditions. Table \ref{tab:RSL-extent} tabulates the wall-normal locations at which the 2.5\% and 5.0\% thresholds are achieved. Multiple streamwise locations over one roughness wavelength near the analysis location were considered to asses any roughness phasing effects. Figure \ref{fig:spatial-dispersion} only shows one representative location; however, the range of values from all streamwise locations are reported in Table \ref{tab:RSL-extent}. For all rough-walled cases, based on streamwise velocity, the roughness sublayer height is generally found to be approximately $R_{RSL}\approx 3k-6k$, where $k$ is the amplitude (peak-to-valley height) of the sinusoidal roughness. The coldest wall, $T_w/T_r=0.4$ case, suggests the RSL extends just beyond $8k$ -- the most significant outlier. It is known that cold-wall conditions can enhance compressibility effects \citep{Hadjadj2015, Yu2021}, which may extend the influence of the roughness further from the wall. However, a 2.5\% threshold is very stringent and the outlier $R_{RSL}$ values are regarded with caution. After accounting for the wall offset, the range $R_{RSL}\approx 3k-6k$ is consistent with the commonly cited $2k-5k$ from \cite{Raupach1991}. Other conclusions are that the the streamwise phase has limited effect on $R_{RSL}$, wall temperature conditions have limited effect on $R_{rsl}$, and the extent of any spatial variation in the spanwise and wall-normal mean velocities are limited close to the roughness crest at $x_2=1k-2k$.

\begin{table}
    \centering
    \begin{tabular}{cccccccc}
        & \multicolumn{3}{c}{$R_{RSL}/k$ at 5.0\%}
        & \multicolumn{4}{c}{$R_{RSL}/k$ at 2.5\%} \\ \cline{2-4} \cline{6-8}
        $\wt{\psi}$ & \makecell{TwTr1p0r\\min$|$max} & \makecell{TwTr0p7r\\min$|$max} & \makecell{TwTr0p4r\\min$|$max} & & \makecell{TwTr1p0r\\min$|$max} & \makecell{TwTr0p7r\\min$|$max} & \makecell{TwTr0p4r\\min$|$max} \\ \hline
        $\wt{u_1}$ & $3.54|3.54$ & $2.67|2.91$ & $3.04|3.11$ & & $6.46|6.46$ & $5.28|5.29$ & $8.45|8.65$\\ 
        $\wt{u_2}$ & $0.90|1.22$ & $0.97|1.25$ & $0.90|1.26$ & & $1.05|1.33$ & $1.86|2.17$ & $1.94|2.22$\\ 
        $\wt{u_3}$ & $0.88|1.25$ & $0.89|1.27$ & $0.94|1.32$ & & $1.44|1.48$ & $1.36|1.46$ & $1.38|1.51$\\ 
    \end{tabular}
    \caption{Roughness sublayer extent based on spatial dispersion thresholds at the analysis location. Heights are normalised by roughness amplitude $k$. min and max indicate minimum and maximum $R_{rsl}/k$ values over one roughness wavelength in the streamwise direction.}
    \label{tab:RSL-extent} 
\end{table}

\subsection{Wall Offset}\label{sec:wall-offset}
\subsubsection{Momentum}
The reviews by \cite{Kadivar2021} and \cite{Chung2021} summarize the body of literature which describe how the presence of surface roughness has the effect of shifting the mean flow profiles away from the wall such that the origin of the outer turbulent flow is at some wall-normal location $x_2=d$ rather than the $x_2=0$ smooth-wall reference plane on which the roughness elements lie. This is known as the wall offset (other names include the zero-plane displacement, wall origin, or virtual origin). The identification and selection of an appropriate wall offset is critical to maintain physical consistency and to collapse all profiles in the outer layer \citep{Chung2021}. The simplest choice is to set the offset as the mean height of the roughness elements. In two dimensions, over the whole surface, the mean height is $R_m=k/4$, per Table \ref{tab:roughness-parameters}. Integrating only in the streamwise direction, over one wavelength, down the centreline of the roughness element, the mean height is $R_{m,oe}=k/2$. Pure geometric measures are appealing due to their simplicity, but do not reflect the fact that the virtual origin is a dynamic parameter associated with the origin of the logarithmic region of the flow \citep{MacDonald2018}. Another option, as suggested by \cite{Jackson1981}, is to set $d$ as the centre of the drag profile on the roughness element. In other words, the drag-centroid defined as $d=M_d/\tau_w$, where $M_d$ is the moment per unit area (about $x_2=0$) due to the horizontal drag forces acting on the roughness element, and $\tau_w$ is the wall shear stress comprised of both viscous and pressure components. The full equation for the drag moment is found in Eq. \ref{eq:drag-moment}, where $h$ is the wall-normal height of the surface profile above $x_2=0$, per Eq \ref{eq:idealized_sin_rough}. Though commonly cited, the integrated resultant force approach (drag centroid) does not always produce the correct wall offset as shown by \cite{Chan2015_JFM} for sinusoidal roughness in pipe flow, and for high aspect ratio spanwise-aligned bars by \cite{MacDonald2018}. If mean flow data beyond the surface is available, it is possible to define a ``no-slip'' origin as the furthest wall-normal location where the the streamwise velocity crosses zero $d=x_2(\widetilde{u_1}=0)$ as was done by \cite{Chan2023}. The statement the \textit{furthest wall-normal location} is crucial to the definition since the zero-crossing may occur at multiple wall-normal locations in the wakes, separated, and recirculating regions behind large roughness\footnote{In this paper, when reporting zero-crossing heights, the mean flow data is taken from the streamwise location in between rows of roughness elements where the phasing is such that the surface height is $h(x_{1,a},x_3)=0$ across the entire span.}. Rather than assigning the wall offset from surface quantities like shear stress or geometry, other approaches determine the wall offset from the mean flow. \cite{Modesti2022} tested different wall offsets until the difference between the transformed smooth and rough mean velocity profiles in the expected logarithmic region returned a constant roughness function $\Delta u_1^+=u^+_{1,sw} - u^+_{1,rw}$. They empirically found a wall offset of $d=0.9k$ adequate for their conditions. With a more rigorous optimisation, \cite{Su2026} sweep through different values of $d$ searching for the optimal zero-plane displacement determined by identifying the value that maximizes the longest continuous region in which the diagnostic function, $\Xi=du_1^+/d\ln(x_2^+-d^+)$, remains within 10\% of the theoretical logarithmic-law slope (theoretical inverse von K\'arm\'an constant, $1/\kappa$) under the chosen compressibility transformation. Equation \ref{eq:wall-offset-cases} summarizes the presently listed drag-centroid definitions.

\begin{equation}
    M_d = \frac{1}{A_P}\int_S h\left[\left.\mu\frac{\partial u_i}{\partial n}\right|_w\cdot\hat{e}_{x_1}-p_w(\hat{n}\cdot\hat{e}_{x_1})\right]dS \label{eq:drag-moment}
\end{equation}
\begin{equation}
    d = \begin{cases}
        R_m = k/4, & \text{mean height} \\
        R_{m,oe} = k/2, & \text{mean height (one element)} \\
        M_d/\tau_w, & \text{drag-centroid} \\
        x_2(\widetilde{u_1}=0), & \text{zero-crossing} \\
        d \text{ s.t. } \Xi = \frac{1}{\kappa}\pm 10\%,   & \text{diagnostic function}\\
        d \text{ s.t. } \Delta u_1^+\neq\Delta u_1^+(x_2), & \text{constant } \Delta u_1^+ \text{(in log-layer)}
    \end{cases} \label{eq:wall-offset-cases}
\end{equation}

\begin{figure}[h!]
    \centering
    \begin{subfigure}[b]{0.325\textwidth}
        \centering
        \includegraphics[width=\textwidth]{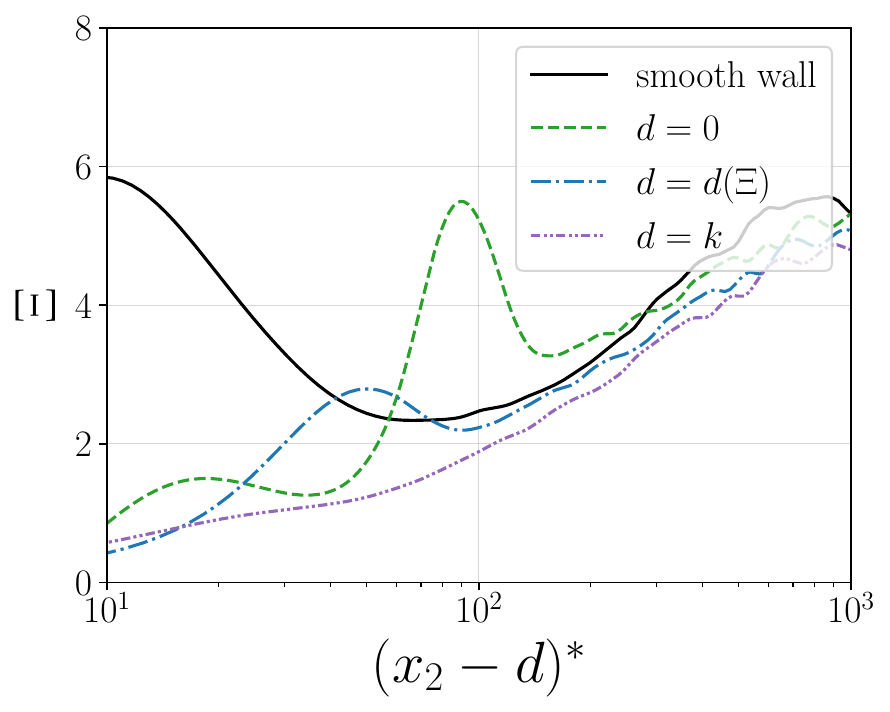}
        \caption{TwTr1p0}
        \label{fig:d_df_1p0}
    \end{subfigure}
    \hfill
    \begin{subfigure}[b]{0.325\textwidth}
        \centering
        \includegraphics[width=\textwidth]{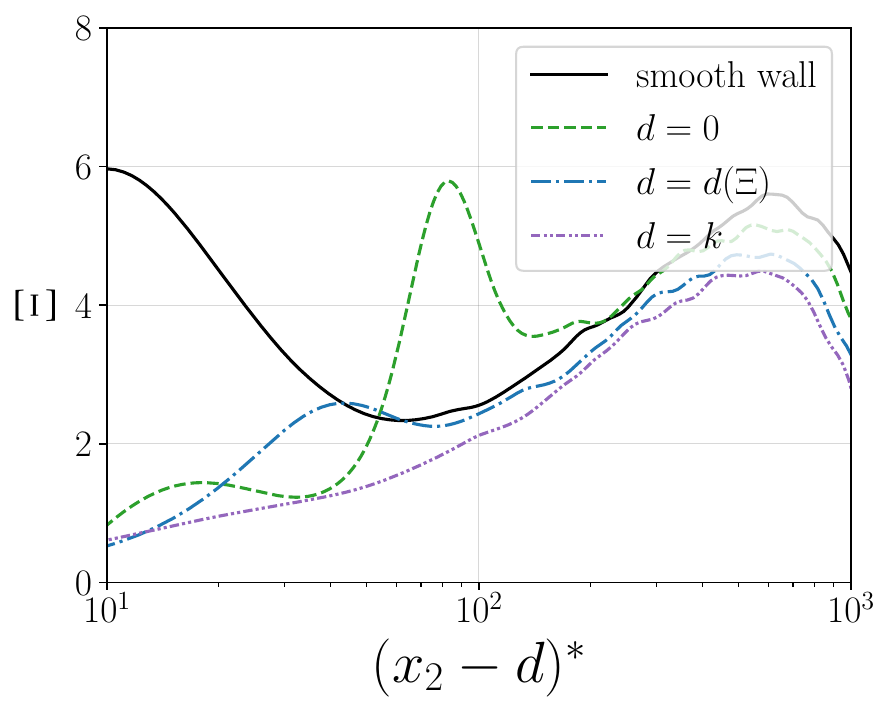}
        \caption{TwTr0p7}
        \label{fig:d_df_0p7}
    \end{subfigure}
    \hfill
    \begin{subfigure}[b]{0.325\textwidth}
        \centering
        \includegraphics[width=\textwidth]{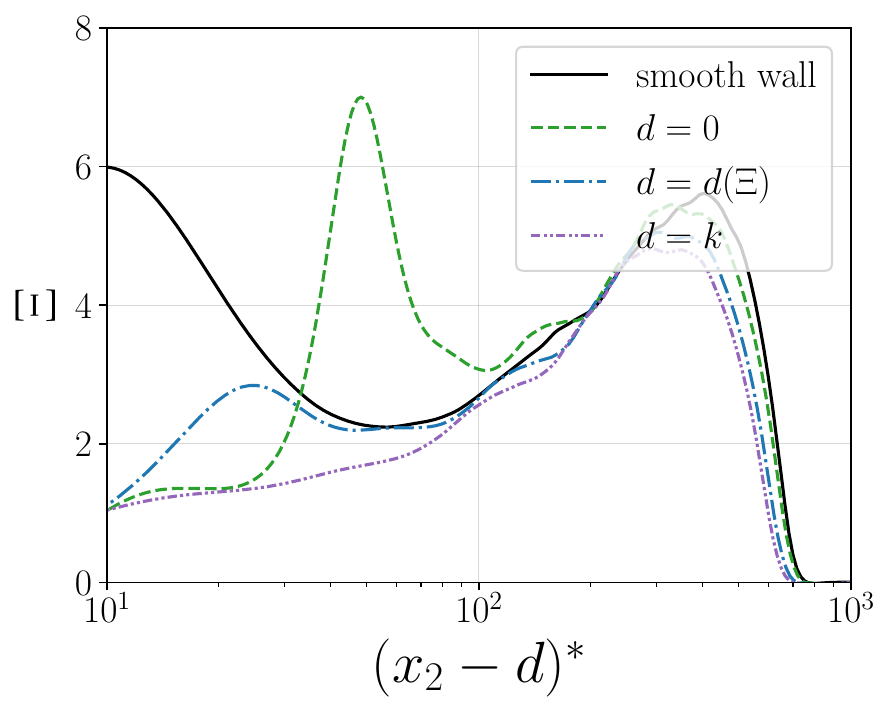}
        \caption{TwTr0p4}
        \label{fig:d_df_0p4}
    \end{subfigure}
    \caption{Sensitivity of the diagnostic function $\Xi=du_1^+/d\ln(x_2-d)^*$ to wall offset $d$. The optimal $d=d(\Xi)$ is identified as the value that maximizes the longest continuous region in which the diagnostic function remains within 10\% of the theoretical logarithmic-law slope (theoretical inverse von K\'arm\'an constant, $1/\kappa$, $\kappa=0.41$). The mean velocity is scaled following the \cite{Griffin2021} compressibility transformation. No wall offset and the sinusoidal peak-to-valley height $k$ are also shown for reference.}
    \label{fig:wall-offset-df}
\end{figure}

\begin{figure}[h!]
    \centering
    \begin{subfigure}[b]{0.325\textwidth}
        \centering
        \includegraphics[width=\textwidth]{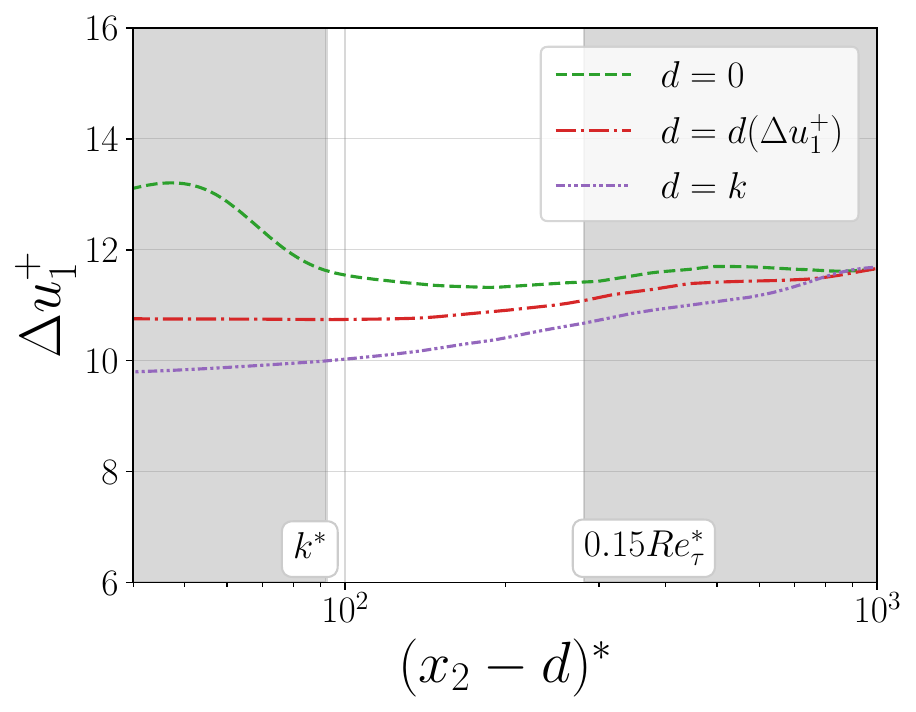}
        \caption{TwTr1p0}
        \label{fig:d_dU_1p0}
    \end{subfigure}
    \hfill
    \begin{subfigure}[b]{0.325\textwidth}
        \centering
        \includegraphics[width=\textwidth]{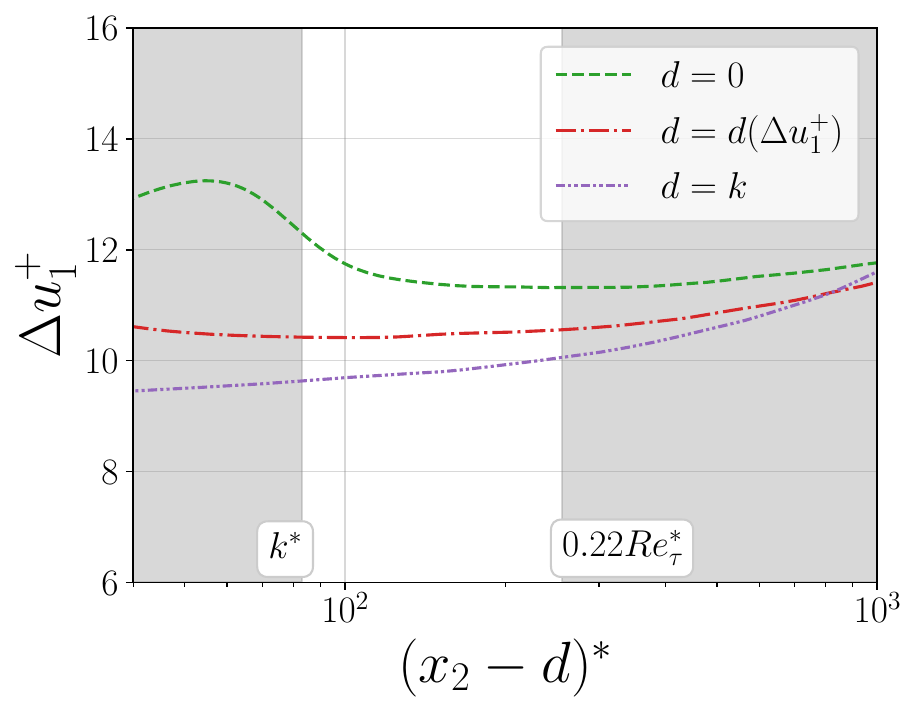}
        \caption{TwTr0p7}
        \label{fig:d_dU_0p7}
    \end{subfigure}
    \hfill
    \begin{subfigure}[b]{0.325\textwidth}
        \centering
        \includegraphics[width=\textwidth]{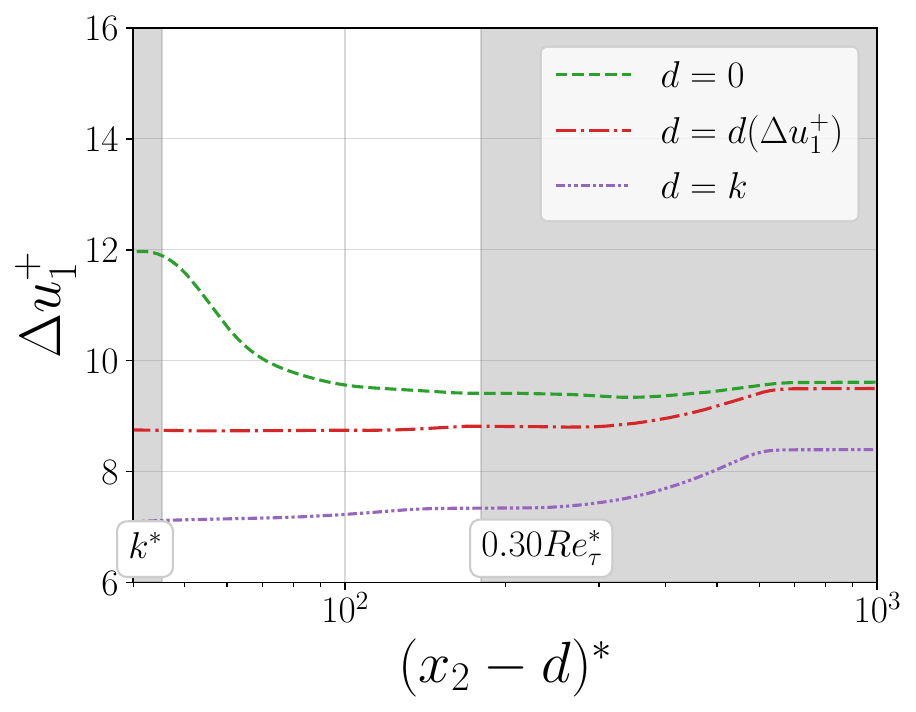}
        \caption{TwTr0p4}
        \label{fig:d_dU_0p4}
    \end{subfigure}
    \caption{Sensitivity of the roughness function $\Delta u_1^+$ to wall offset $d$. The optimal $d=d(\Delta u_1^+)$ is identified as the value that returns the flattest slope in the expected logarithmic region, shown as the non-grey portions of the plot. The mean velocity is scaled following the \cite{Griffin2021} compressibility transformation. No wall offset and the sinusoidal peak-to-valley height $k$ are also shown for reference.}
    \label{fig:wall-offset-Du}
\end{figure}

The geometric, drag centroid, no-slip, and mean velocity profile methods are reported for comparison in Table \ref{tab:wall-offsets}. For clarity, the notation for the wall offset $d$ is as follows: $R_m$ for the mean height; $R_{m,oe}$ for the one-dimensional, one-element mean height; $M_d/\tau_w$ for the drag-centroid; $x_2(\widetilde{u_1}=0)$ for the zero-crossing; $d(\Xi)$ for the diagnostic function; and $d(\Delta u_1^+)$ for the constant roughness function. Notably, all non-geometric measures report wall offsets larger than the true mean height $R_m$, with the exception of the zero-crossing. The zero-crossing suggests a value of approximately $d\approx0.16k$ and no clear trend in the wall offset with wall temperature condition. Moreover, this method significantly depends on the streamwise phasing location of the mean velocity profile over the roughness location and was inconsistent with the other methods presently considered. Across all three wall temperature conditions, the drag centroid suggests a wall offset on the order of $d=0.5k$, with the offset increasing slightly with decreasing wall temperature condition. Similarly, the diagnostic function approach led to a wall offset on the order of $d=0.6k$. Figure \ref{fig:wall-offset-df} shows the diagnostic function defined using the semi-local wall normal coordinate $(x_2-d)^*=(x_2-d)\sqrt{\tau_w/\rho(x_2)}/\nu(x_2)$ and the \cite{Griffin2021} velocity transformation for wall offsets of $d=0$, $d=d(\Xi)$, and $d=k$. Figure \ref{fig:wall-offset-df} highlights the ability of the offset $d=d(\Xi)$ to capture a region such that $1/\Xi\approx\kappa=0.41$. This logarithmic region, ideally a flat segment around $x_2^*\approx100$, is limited in width for the rough-walled cases due to the size of the roughness $k^+\approx80$ and the limitations on Reynolds number from the computational resources -- the rough wall diagnostic plots indicate more inflectional profiles, rather than a flat segment. Nevertheless, with the correct wall offset the same inverse von K\'arm\'an constant as the smooth wall is reached before eventually recovering to the smooth wall in the outer layer. Lastly, the roughness function method in the expected logarithmic region is shown in Figure \ref{fig:wall-offset-Du} based again on the \cite{Griffin2021} transformed velocity profiles. Table \ref{tab:wall-offsets} reports the roughness function wall offset of $d=0.39k$ for the $T_w/T_r=1.0$ case, $d=0.38k$ for the $T_w/T_r=0.7$ case, and $d=0.53k$ for the $T_w/T_r=0.4$ case. The flatness of the roughness function over the experted logarithmic region is highlighted as the non-greyed regions in Figure \ref{fig:wall-offset-Du}. The flatness is measured from $x_2^*(k)\le x_2^* \le cRe_\tau^*$, where $c$ is taken from the smooth-wall velocity such that $cRe_\tau^*$ occurs at the location where the velocity profiles deviate from the log-law. This was found to be $c=0.15$, 0.22, and 0.3 for the $T_w/T_r=1.0$, 0.7, and 0.4 cases, respectively. $Re_\tau^*=\delta\sqrt{\tau_w/\rho_e}/\nu_e$ is a semi-local friction Reynolds number based on edge conditions.

From Table~\ref{tab:wall-offsets}, there is no clear trend on the effect of wall temperature on the ideal offset. However, we find that the true mean height $R_m$ is an inaccurate wall offset because none of the other methods produce a similar offset of $d = 0.25k$. Most methods suggest a wall offset of $d\approx0.5k$, so if a simple geometric measure is desired, the one-element mean height or half the peak-to-valley height is better suited for the present sinusoidal roughness. The results in \S \ref{sec:momentum-BL} are centred around the roughness function and so the wall offset $d(\Delta u_1^+)$ will be used. Unless otherwise stated, all momentum boundary layer parameters reported are with respect to this wall offset and all wall-normal coordinates denoted as $x_2$ are $x_2-d$.

\begin{table}
    \centering
    \begin{tabular}{lcccccc}
        Case & \makecell{$R_m/k$\\(mean height)} & \makecell{$R_{m,oe}/k$\\(mean height)} & \makecell{$(M_d/\tau_w)/ k$\\(drag-centroid)} & \makecell{$x_2(\widetilde{u_1}=0)/k$\\(zero-crossing)} & \makecell{$d(\Xi)/k$\\(diag. func.)} & \makecell{$d(\Delta u_1^+)/k$\\(constant $\Delta u_1^+$)} \\ \hline
        TwTr1p0r & 0.250 & 0.500 & 0.497 & 0.156 & 0.550 & 0.389 \\ 
        TwTr0p7r & 0.250 & 0.500 & 0.507 & 0.152 & 0.602 & 0.378 \\
        TwTr0p4r & 0.250 & 0.500 & 0.549 & 0.168 & 0.602 & 0.533
    \end{tabular}
    \caption{Wall offsets as computed from the mean surface height, drag-centroid, zero crossing, diagnostic function, and roughness function. Heights are normalised with respect to the sinusoid peak-to-valley height $k$. }
    \label{tab:wall-offsets}
\end{table}

\subsubsection{Thermal}
\begin{figure}[h!]
    \centering
    \begin{subfigure}[b]{0.325\textwidth}
        \centering
        \includegraphics[width=\textwidth]{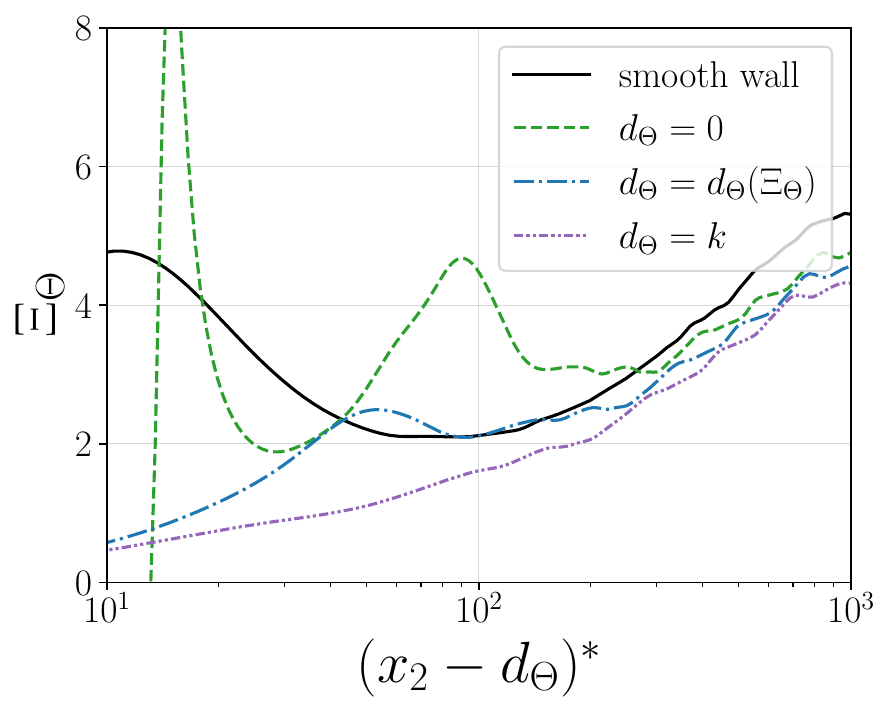}
        \caption{TwTr1p0}
        \label{fig:dT_df_1p0}
    \end{subfigure}
    \begin{subfigure}[b]{0.325\textwidth}
        \centering
        \includegraphics[width=\textwidth]{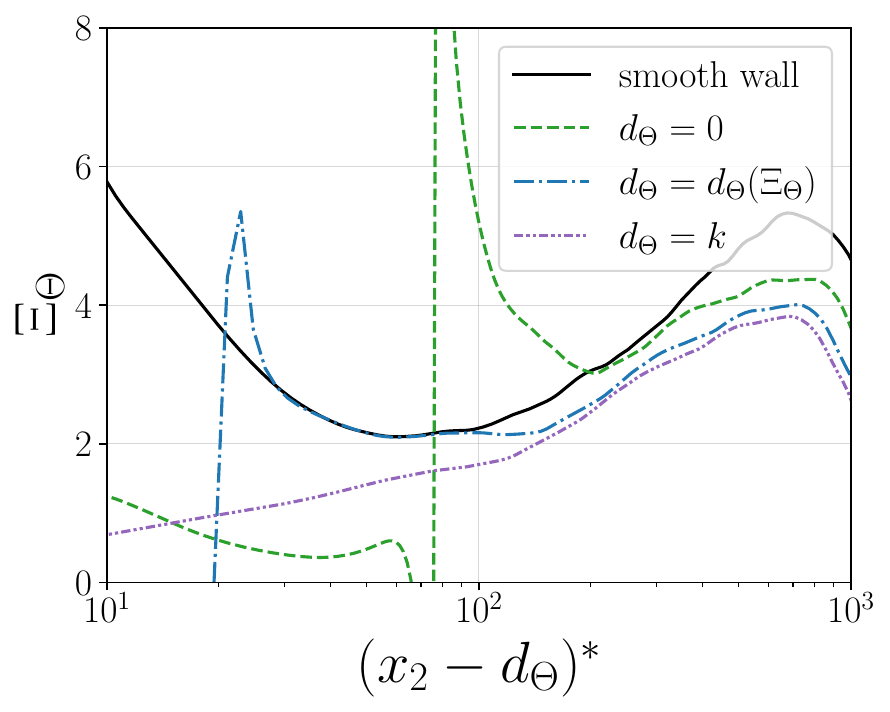}
        \caption{TwTr0p7}
        \label{fig:dT_df_0p7}
    \end{subfigure}
    \caption{Sensitivity of the diagnostic function $\Xi_\Theta=d\Theta^+/d\ln(x_2-d_\Theta)^*$ to wall offset $d_\Theta$. The optimal $d_\Theta=d_\Theta(\Xi_\Theta)$ is identified as the value that maximizes the longest continuous region in which the diagnostic function remains within 10\% of the equivalent wall-temperature smooth wall theoretical logarithmic-law slope (theoretical inverse von K\'arm\'an constant for temperature, $1/\kappa_{\Theta,sw}$). The mean temperature is scaled following the \cite{Liang2026} compressibility transformation. No wall offset and the sinusoidal peak-to-valley height $k$ are also shown for reference.}
    \label{fig:thermal-wall-offset-df}
\end{figure}

\begin{figure}[h!]
    \centering
    \begin{subfigure}[b]{0.325\textwidth}
        \centering
        \includegraphics[width=\textwidth]{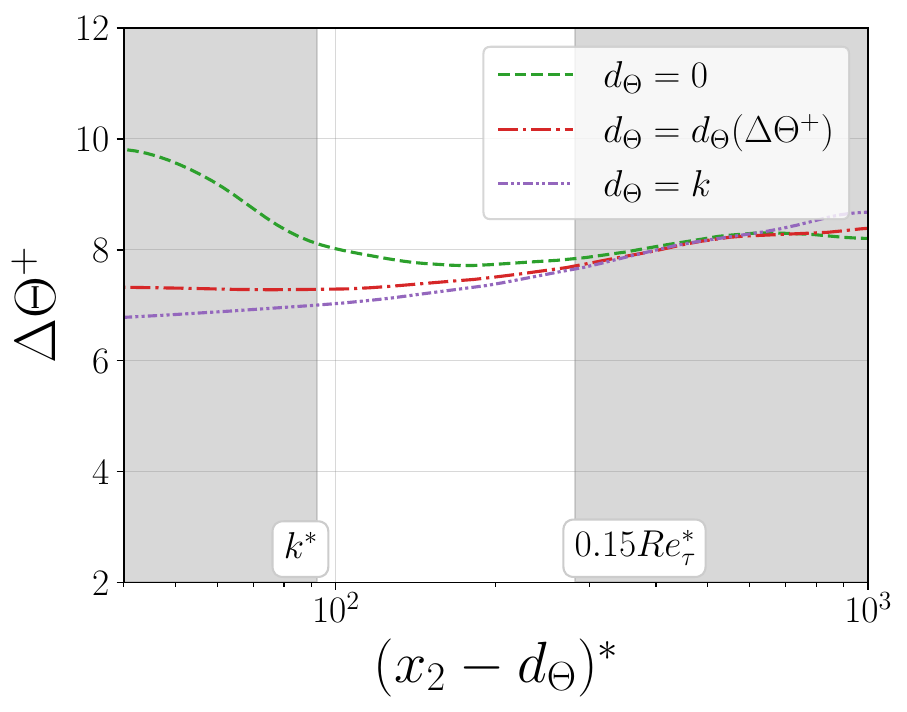}
        \caption{TwTr1p0}
        \label{fig:dT_DT_1p0}
    \end{subfigure}
    \begin{subfigure}[b]{0.325\textwidth}
        \centering
        \includegraphics[width=\textwidth]{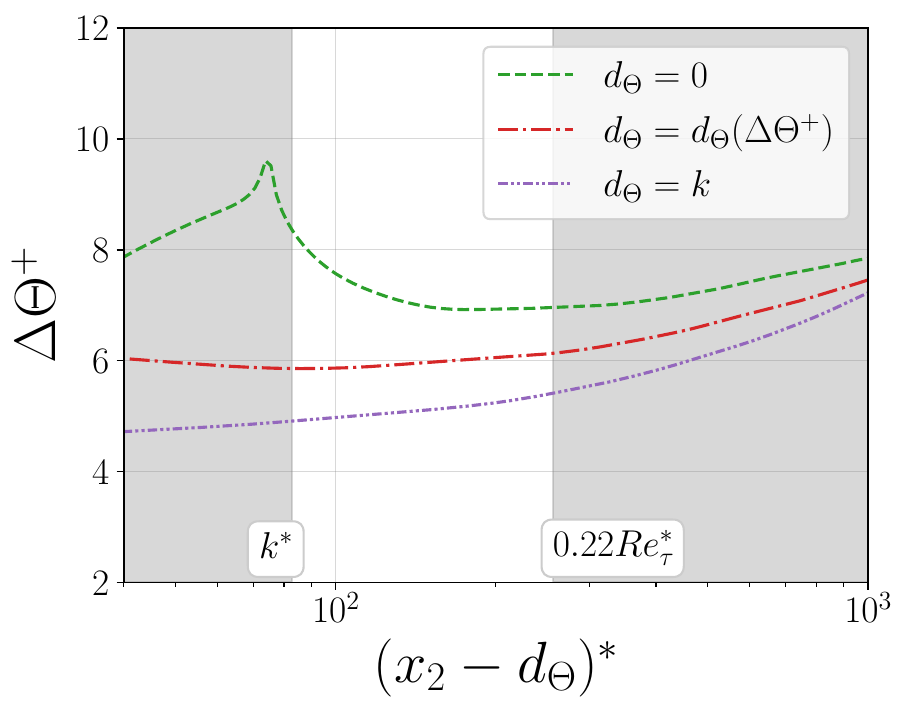}
        \caption{TwTr0p7}
        \label{fig:dT_DT_0p7}
    \end{subfigure}
    \caption{Sensitivity of the thermal roughness function $\Delta \Theta^+$ to wall offset $d_\Theta$. The optimal $d_\Theta=d_\Theta(\Delta \Theta^+)$ is identified as the value that returns the flattest slope in the expected logarithmic region, shown as the non-grey portions of the plot. The mean temperature is scaled following the \cite{Liang2026} compressibility transformation. No wall offset and the sinusoidal peak-to-valley height $k$ are also shown for reference.}
    \label{fig:thermal-wall-offset-dT}
\end{figure}

Extending the concept from the momentum boundary layer, analogous to the wall offset $d$, a thermal wall offset $d_\Theta$ can be defined. Presently we compute a thermal wall offset for each wall temperature condition by extending the momentum parameters to their thermal counterparts. We contrast these values to the equivalent velocity origin and assess whether they are wall temperature boundary condition dependent. A similar metric to the drag-centroid can be defined based on temperature rather than momentum. The thermal wall offset $d_\Theta$ in terms of a centre-of-heat is defined analogously to the drag-centroid, now taking into consideration the wall heat flux $q_w$ and a heat flux moment $M_{d_\Theta}$, as listed in Eq. \ref{eq:qw-definition} and \ref{eq:heat-moment}, respectively. Regarding the zero-crossing parameter, this is omitted because the temperature profile is non-monotonic for all wall boundary conditions and a mean temperature zero crossing location due to the roughness is not well defined like it is for the velocity. Assuming the mean temperature profile follows a logarithmic profile in the overlap layer, both a diagnostic function and log-layer shift (thermal roughness function) wall offset are computed. Figures \ref{fig:thermal-wall-offset-df} and \ref{fig:thermal-wall-offset-dT} display the diagnostic function and log-layer shift based on the thermal boundary layer for the TwTr1p0r and TwTr0p7r cases. For the diagnostic function, the von K\'arm\'an constant for temperature depends on the flow dependent turbulent Prandtl number, $\kappa_\Theta= \kappa/Pr_t$. Since $\kappa_\Theta$ may change with the flow, the target $1/\kappa_{\Theta,sw}$ for the diagnostic function is based on the smooth-wall simulation at matched wall temperature, rather than a constant von K\'arm\'an constant as was done for the velocity analysis. Equation \ref{eq:thermal-wall-offset-cases} lists the thermal wall offset definitions presently considered and Table \ref{tab:wall-offsets-thermal} tabulates the various wall offsets based on the thermal boundary layers. Due to the limitations of the various metrics there are three omissions in Table \ref{tab:wall-offsets-thermal}: the centre-of-heat is omitted for the $T_w/T_r=1.0$ case because of lack of wall heat flux, and both the diagnostic function and roughness function cases are omitted for the $T_w/T_r=0.4$ case due to the singularity present after the temperature transformation for this cold wall case, which will be discussed in \S \ref{sec:thermal-BL}. From Table \ref{tab:wall-offsets-thermal}, we find the thermal wall offset approximately the same as the wall offset from the momentum boundary layer, $d_\Theta\approx0.5k\approx d$, regardless of wall temperature condition. This finding is consistent with \cite{Kadivar2025}, citing that most studies recommend using the velocity wall origin as the reference for the temperature wall origin $(d_\Theta = d)$. For the remainder of this paper, the thermal wall offset is taken to be the same as the momentum wall offset, $d_\Theta=d$.

\begin{equation}  
    M_{d_\Theta} = \frac{1}{A_P}\int_S h\left[\left.-\lambda_\Theta\pp{T}{n}\right|_w\right] dS \label{eq:heat-moment}
\end{equation}

\begin{align}
    d_\Theta &= \begin{cases}
        R_m = k/4, & \text{mean height}\\
        R_{m,oe} = k/2, & \text{mean height (one element)} \\
        M_{d_\Theta}/q_w, & \text{centre-of-heat}\\
        d_\Theta \text{ s.t. } \Xi_\Theta = \frac{1}{\kappa_{\Theta,sw}}\pm 10\%,   & \text{diagnostic function}\\
        d_\Theta \text{ s.t. } \Delta \Theta^+\neq\Delta \Theta^+(x_2), & \text{constant } \Delta \Theta^+ \text{(in log-layer)}
    \end{cases} \label{eq:thermal-wall-offset-cases}
\end{align}

\begin{table}
    \centering
    \begin{tabular}{lcccccc}
        Case & \makecell{$R_m/k$\\(mean height)} & \makecell{$R_{m,oe}/k$\\(mean height)} & \makecell{$(M_{d_\Theta}/q_w)/ k$\\(centre-of-heat)} & \makecell{$d_\Theta(\Xi_\Theta)/k$\\(diag. func.)} & \makecell{$d_\Theta(\Delta \Theta^+)/k$\\(constant $\Delta \Theta^+$)} \\ \hline
        TwTr1p0r & 0.250 & 0.500 & -- & 0.445 & 0.506 \\ 
        TwTr0p7r & 0.250 & 0.500 & 0.384 & 0.689 & 0.460 \\
        TwTr0p4r & 0.250 & 0.500 & 0.434 & -- & --
    \end{tabular}
    \caption{Wall offsets as computed from the mean surface height, centre-of-heat, diagnostic function, and roughness function based on the thermal boundary layer. Heights are normalised with respect to the sinusoid peak-to-valley height $k$. }
    \label{tab:wall-offsets-thermal}
\end{table}

\section{Momentum Boundary Layer Scaling} \label{sec:momentum-BL}
Table \ref{tab:momentum-bl-parameters} tabulates a number of boundary layer parameters at the analysis location relevant to the momentum (streamwise velocity) mean flow. In Table \ref{tab:momentum-bl-parameters} $x_{1,a}/\theta_i$ is the streamwise analysis location normalised by the inlet momentum thickness, $Re_\theta=\rho_e u_e\theta/\mu_e$ is the momentum thickness Reynolds number based on edge parameters, $Re_\tau=\rho_w u_\tau \delta/\mu_w$ is the friction Reynolds number, $Re_{\delta_2}=\rho_e u_e\theta/\mu_w$ is the momentum thickness Reynolds number with wall viscosity, $Re_\tau^*=\rho_e\sqrt{\tau_w/\rho_e}\delta/\mu_e$ is the semi-local Reynolds number based on edge parameters, $\theta$ is the momentum thickness, $H=\delta^*/\theta$ is the shape factor defined as the ratio of the displacement to the momentum thickness, $\delta=\delta_{99.5}$ is the boundary layer thickness based on the 99.5\% freestream velocity, $\delta_\nu=\mu_w/(u_\tau \rho_w)$ is the wall viscous length scale, $M_\tau=u_\tau/\sqrt{\gamma R_{gas} T_w}$ is the friction Mach number, $k^+=\rho_w u_\tau k/\mu_w$ is the roughness Reynolds number based on wall quantities, and $k^*=\rho(k) \sqrt{\tau_w/\rho(k)} k/\mu(k)$ is the semi-local roughness Reynolds number based on local quantities.

\begin{table}
    \centering
    \begin{tabular}{lccccccccccccc}
        Case & $x_{1,a}/\theta_i$ & $Re_\theta$ & $Re_\tau$ & $Re_{\delta_2}$ & $Re_\tau^*$ & $\theta$, mm & $H$ & $\delta$, mm & $\delta_\nu$, $\mu$m & $u_\tau$, m s$^{-1}$ & $M_\tau$ & $k^+$ & $k^*$ \\\hline
        TwTr1p0s & 394 & 5191 & 797 & 3067 & 1944 & 1.09 & 4.13 & 14.15 & 17.7 & 34.02 & 0.071 & -- & --  \\
        TwTr1p0r & 687 & 3617 & 790 & 2169 & 1876 & 0.765 & 5.26 & 7.98 & 10.1 & 57.73 & 0.121 & 79.2 & 92.3 \\
        TwTr0p7s & 313 & 3645 & 812 & 2745 & 1303 & 0.757 & 3.23 & 8.57 & 10.6 & 31.26 & 0.078 & -- & --\\
        TwTr0p7r & 395 & 2186 & 743 & 1669 & 1163 & 0.454 & 4.42 & 4.26 & 5.73 & 55.16 &  0.138 & 87.2 & 82.9 \\
        TwTr0p4s & 277 & 1906 & 796 & 2202 & 631 & 0.393 & 2.29 & 3.54 & 4.45 & 27.53 &  0.091 & -- & --\\
        TwTr0p4r & 316 & 1287 & 769 & 1498 & 601 & 0.263 & 3.14 & 2.04 & 2.65 & 44.67 &  0.148 & 70.8 & 45.2 
    \end{tabular}
    \caption{Momentum boundary layer parameters at analysis location $(x_{1,a})$. All boundary layer height metrics have the wall offset pre-subtracted.}
    \label{tab:momentum-bl-parameters}
\end{table}

\subsection{Inner Layer and Roughness Function Analysis}
Six velocity transformations from the literature are considered in the present analysis: \cite{vanDriest1951} VD, \cite{Zhang2012} ZBHLS, \cite{Trettel2016} TL, \cite{Volpiani2020} VIPL, \cite{Griffin2021} GFM, and \cite{Hasan2023} HLPP. The form of the transformations is provided in Appendix \ref{app:vel-trans}. Figure \ref{fig:BL-inner} shows the smooth-wall and rough-wall mean streamwise velocity profiles in both the wall and semi-local scalings from Appendix \ref{app:vel-trans}. The appropriate wall offset $d$ has already been applied in the wall-normal coordinate and transformations and the $x_2-d$ notation is dropped for simplicity of notation to $x_2$.

\begin{figure}[h!]
    \centering
    \begin{subfigure}[b]{0.495\textwidth}
        \centering
        \includegraphics[width=\textwidth]{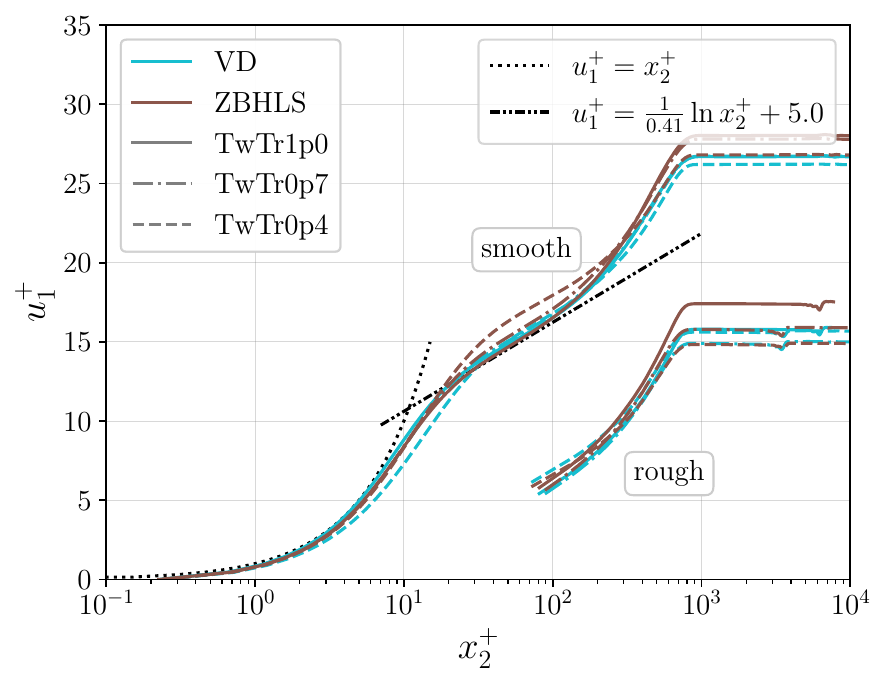}
        \caption{Wall scaling}
        \label{fig:BL-local}
    \end{subfigure}
    \hfill
    \begin{subfigure}[b]{0.495\textwidth}
        \centering
        \includegraphics[width=\textwidth]{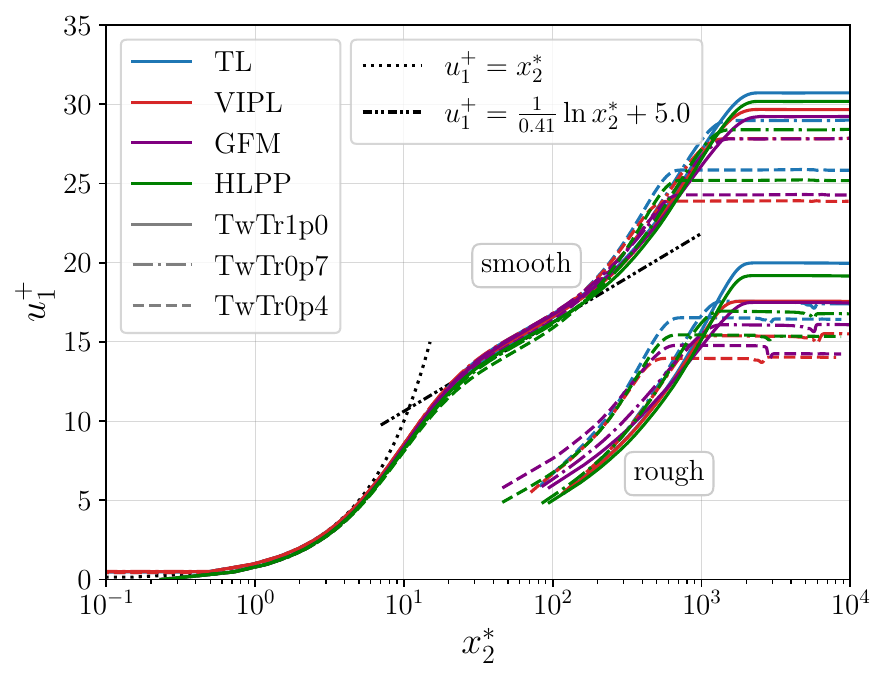}
        \caption{Semi-local scaling}
        \label{fig:BL-semi-local}
    \end{subfigure}
    \caption{Smooth- and rough-wall mean streamwise velocity profiles in inner scaling at analysis location $(x_{1,a})$. The transformed smooth-wall profiles follow closely to the classical law of the wall, and the transformed rough-wall profiles are shifted down by the roughness function $\Delta u_1^+$. Rough-walled profiles are truncated below the roughness Reynolds number $k^+$.}
    \label{fig:BL-inner}
\end{figure}

From Figure \ref{fig:BL-local} the VD transformation struggles to collapse the profiles in the buffer layer for the cold wall cases, but does well for the adiabatic cases as expected from its original formulation. The ZBHLS transformation does quite well in the viscous sublayer and buffer layer regardless of wall temperature, but over predicts in the log-layer for the cold wall case. Interestingly, for all rough-wall cases, both the VD and ZBHLS transformations do relatively well at collapsing all profiles regardless of wall temperature condition, resulting in a $T_w/T_r$ independent roughness function. Moving to Figure \ref{fig:BL-semi-local}, all the semi-local transformations perform well for the smooth wall cases, with the VIPL and GFM recovering to the log-layer slightly better than the TL and HLPP. However, for the rough-wall cases the transformations appear to be wall temperature dependent, for the most part collapsing the velocity profiles within a given wall-temperature grouping, but dispersing across different $T_w/T_r$ conditions. \cite{Wang2024} also reported this discrepancy and claim that none of the existing velocity transformations proposed for compressible turbulent boundary layers over smooth walls are able to make the logarithmic region of velocity profile independent of the wall-to-adiabatic wall temperature ratio, $T_w/T_{aw}$, for their rough cases. They credit the difference in roughness function between $T_w/T_{aw}$ conditions to dispersion of the non-dimensional mean shear $S^+_{VD}=\sqrt{\rho^+}\frac{du_1^+}{dx_2^+}$ (from the integrand of the \cite{vanDriest1951} transformation) and $S^+_t$ (from the integrand of the \cite{Griffin2021} transformation) below the roughness peak under different $T_w/T_{aw}$. For the full $S^+$ definition see Appendix \ref{app:vel-trans}. They further note that the the region below the roughness peak under different $T_w/T_{aw}$ conditions reveals that the dispersion of $S^+_{VD}$ is smaller than that of $S^+_t$ because $\sqrt{\rho^+}$ has a certain influence to characterize the wall heat transfer effect within the roughness region, going on to suggest a new non-dimensional mean shear $S^+_\rho=\left(\sqrt{\rho^+}\right)^{1/(T_w/T_{aw})}\frac{d u_1^+}{dx_2^+}$ applicable to the region below the roughness peak. Ultimately they propose combining their new mean shear definition $S^+_\rho$ below the roughness peak and then reverting to the well performing GFM total stress based mean shear $S^+_t$ from \cite{Griffin2021} above the roughness peak, with the switching occur exactly at $x_2 = k$.

We propose an alternate explanation that does not require a new velocity transformation. The apparent dependence of the roughness function on $T_w/T_r$  can be explained entirely based on existing fully rough asymptotic relations from the  incompressible literature and variation in the semi-local roughness Reynolds number $k^*=\rho(k) \sqrt{\tau_w/\rho(k)} k/\mu(k)$ rather than a wall only roughness Reynolds number $k^+=\rho_w u_\tau k/\mu_w$. In the fully rough regime, for similar effective slopes, the roughness function is a function of the roughness Reynolds number alone, $\Delta u_1^+(k^+)=\frac{1}{\kappa}\ln(k^+)$ + C, \citep{Nikuradse1933, Nikuradse1933_NACA, Clauser1954, Hama1954, Jimenez2004, Flack2010}.  The variation in $\Delta u_1^+$ can therefore be found from the variation in $k^+$ according to Eq \ref{eq:rough-func-vary-w/-k} as follows:

\begin{align}
\left(\Delta u_1^+\right)_1-\left(\Delta u_1^+\right)_2&=\left(\frac{1}{\kappa}\ln(k_1^+) + C\right) - \left(\frac{1}{\kappa}\ln(k_2^+) + C\right)\\
\left(\Delta u_1^+\right)_1-\left(\Delta u_1^+\right)_2&=\frac{1}{\kappa}\ln(k_1^+)-\frac{1}{\kappa}\ln(k_2^+)\\
\left(\Delta u_1^+\right)_1-\left(\Delta u_1^+\right)_2&=\frac{1}{\kappa}\ln(k_1^+/k_2^+) \label{eq:rough-func-vary-w/-k}
\end{align}

\begin{figure}[h!]
    \centering
    \begin{subfigure}[b]{0.495\textwidth}
        \centering
        \includegraphics[width=\textwidth]{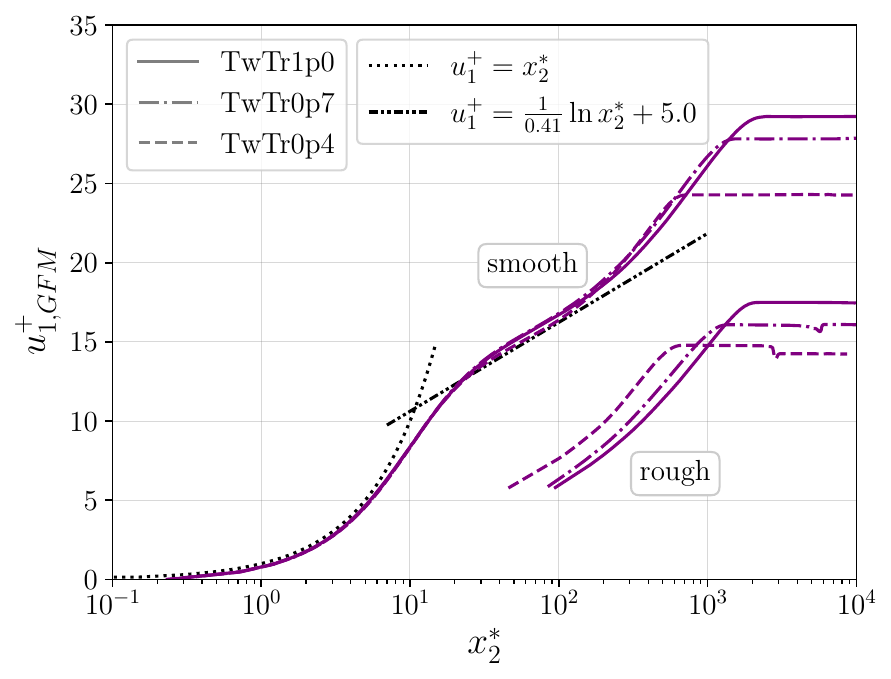}
        \caption{Standard}
        \label{fig:uGFM-only}
    \end{subfigure}
    \hfill
    \begin{subfigure}[b]{0.495\textwidth}
        \centering
        \includegraphics[width=\textwidth]{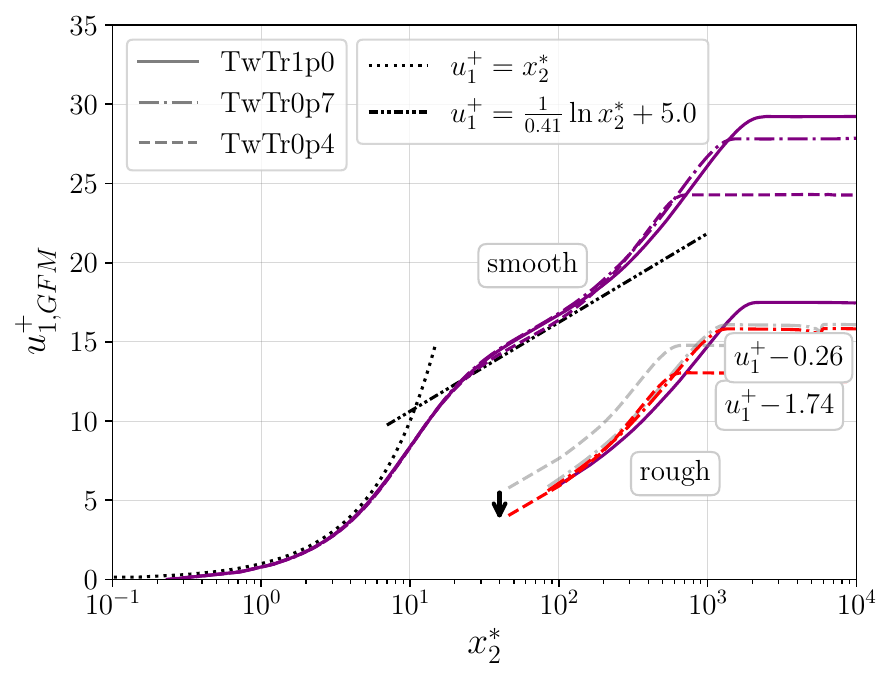}
        \caption{Corrected for $k^*$}
        \label{fig:uGFM-corrected}
    \end{subfigure}
    \caption{Smooth- and rough-wall mean streamwise velocity profiles at the analysis location $(x_{1,a})$ based on the \cite{Griffin2021} scaling. (a) Standard velocity profiles. (b) Profiles corrected (shifted) to account for $k^*$ variation.}
    \label{fig:uGFM-duplus-based-on-kplus}
\end{figure}

Using the wall-parameter only roughness Reynolds number $k^+$, the average of all three $T_w/T_r$ cases is $k^+=79.1$, varying only $\pm8.2$. Taking the largest difference to be $k^+_1=87.2$ and $k^+_2=70.8$ the expected difference in roughness function would only be $\left(\Delta u_1^+\right)_1-\left(\Delta u_1^+\right)_2=\frac{1}{0.41}\ln(87.2/70.8)=0.51$. This explains the relatively good collapse of the rough-walled velocity profiles in Figure \ref{fig:BL-local} based on wall scaling. If instead a semi-local roughness Reynolds number $k^*$ is used, the coldest wall TwTr0p4r case led to a $k^*$ of only 45.2 as compared to 82.9 and 92.3 for the TwTr0p7r and TwTr1p0r cases respectively. Using the semi-local roughness Reynolds number the expected difference in roughness function between the TwTr0p4r and TwTr1p0r cases would be $\left(\Delta u_1^+\right)_1-\left(\Delta u_1^+\right)_2=\frac{1}{0.41}\ln(92.3/45.2)=1.74$, and between the TwTr0p7r and TwTr1p0r cases would be $\left(\Delta u_1^+\right)_1-\left(\Delta u_1^+\right)_2=\frac{1}{0.41}\ln(92.3/82.9)=0.26$. Figure \ref{fig:uGFM-duplus-based-on-kplus} isolates only the GFM transformation, with Figure \ref{fig:uGFM-corrected} shifting the rough-walled profiles by the $k^*$ dependent correction factors $\left(\Delta u_1^+\right)_1-\left(\Delta u_1^+\right)_2=1.74$ and $\left(\Delta u_1^+\right)_1-\left(\Delta u_1^+\right)_2=0.26$, respectively. After applying the correction all profiles collapse and return the same roughness function. This result contradicts \cite{Wang2024} as we find that $T_w/T_r$ affect the local properties and modify the effective roughness Reynolds number which manifests as a variation in the momentum loss in the boundary layer (which can be captured by $\Delta u_1^+$) rather than the inability of the existing velocity transformations to account for $T_w/T_r$. \cite{Modesti2022} came to a similar conclusion with their DNS of supersonic turbulent channel flow over
cubical roughness elements at different Mach numbers, recommending a relevant roughness Reynolds number based on the viscosity and density at the roughness crest as a key aspect for compressible flows over roughness to agree with incompressible data. In brief, the present DNS results indicate that the supersonic turbulent boundary layer over fully rough sinusoidal roughness at varying wall temperatures recovers the incompressible theory so long as the semi-local roughness Reynolds number, $k^*$, and not just the roughness Reynolds number, $k^+$, is matched. 

\subsection{Equivalent Sand-Grain Roughness Height}
Originating from incompressible close-packed, uniform sand-grain rough-wall experiments by \cite{Nikuradse1933,Nikuradse1933_NACA}, and later formalised by \cite{Schlichting1937_NACA}, the equivalent sand-grain roughness height $k_s$ is defined as the diameter of uniform sand grains that would produce the same roughness function $\Delta u_1^+$ (log-law shift) as an arbitrary rough surface under dynamically equivalent flow conditions. In the fully rough regime, this leads to Eq. \ref{eq:ksplus} which represents the roughness Reynolds number $k_s^+=k_su_\tau/\nu_w$ \citep{Chung2021}. Per the classical smooth-wall, incompressible law of the wall $u_1^+=(1/\kappa)\ln x_2^+ + A$, $A$ is the log-law intercept set to $A=5.0$ and $\kappa=0.41$ is the von K\'arm\'an constant. In Eq. \ref{eq:ksplus}, $B_s(\infty)=8.5$ is the large roughness limit of the log-law intercept function over the uniform sand grains measured by \cite{Nikuradse1933,Nikuradse1933_NACA} in the fully rough regime. The log-law shift $\Delta u_1^+$ is found by subtracting the correctly scaled rough-wall velocity profile $u^+_{1,rw}$ with that of the smooth wall $u^+_{1,sw}$. When scaling the velocity profiles, an appropriate compressible velocity transformation should be used to collapse the data to the incompressible values. We use the VD transformation for wall scaling \citep{vanDriest1951} and the GFM transformation for semi-local scaling \citep{Griffin2021}. Based on the significance of using a semi-local roughness Reynolds number, as described in the previous section, the equivalent sand-grain roughness definition from Eq. \ref{eq:ksplus} is extended to be in terms of semi-local equivalent sand-grain roughness $k_s^*$ as shown in Eq. \ref{eq:ksstar}. Lastly, all wall-normal distance measurements should account for the wall offset $d$ in consideration.

\begin{align}
    k_s^+ &= \exp\left\{\kappa\left(\Delta u_1^+ + B_s(\infty)-A\right)\right\} \label{eq:ksplus}\\
    k_s^* &= \exp\left\{\kappa\left(\Delta u_1^+ + B_s(\infty)-A\right)\right\} \label{eq:ksstar}\\
    \Delta u_1^+ &= u^+_{1,sw} - u^+_{1,rw}\;\;\;\;\;(\text{at matched }x_2^*)
\end{align}

\begin{figure}[h!]
    \centering
    \begin{subfigure}[b]{0.495\textwidth}
        \centering
        \includegraphics[width=\textwidth]{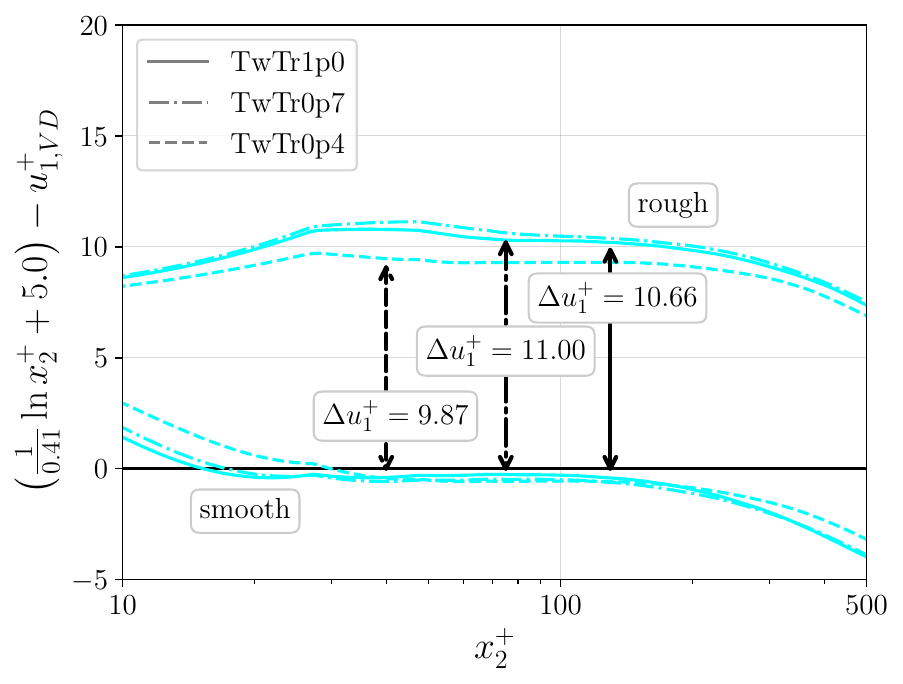}
        \caption{Wall scaling}
        \label{fig:loc-rough-func}
    \end{subfigure}
    \hfill
    \begin{subfigure}[b]{0.495\textwidth}
        \centering
        \includegraphics[width=\textwidth]{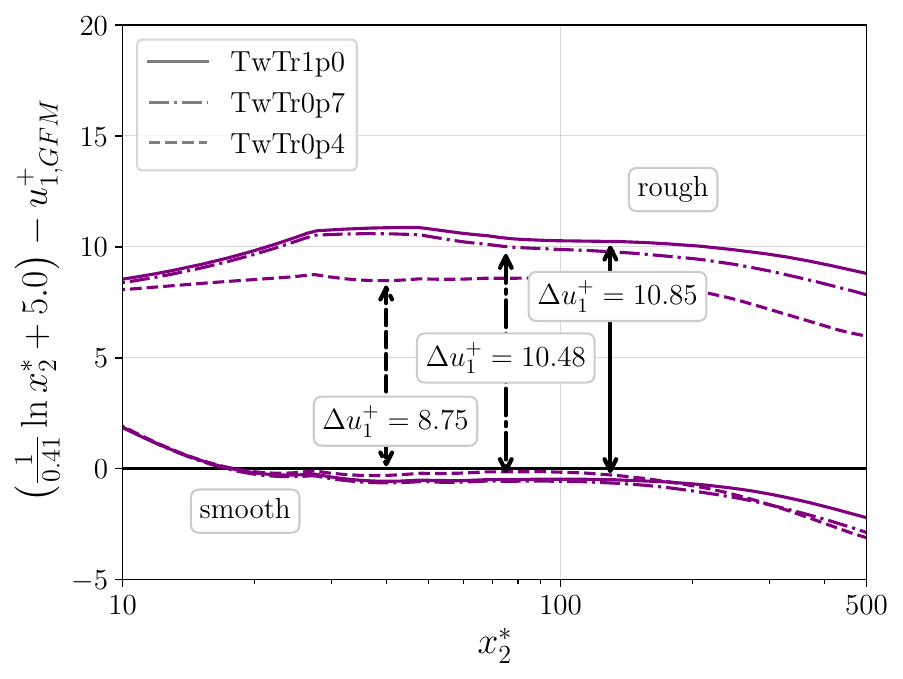}
        \caption{Semi-local scaling}
        \label{fig:sl-rough-func}
    \end{subfigure}
    \caption{$\Delta u_1^+=u^+_{1,sw} - u^+_{1,rw}$ in the logarithmic layer. Roughness function $\Delta u_1^+$ measured as average along viable logarithmic region, per \S \ref{sec:wall-offset}.}
    \label{fig:BL-rough-func}
\end{figure}

Figure \ref{fig:loc-rough-func} shows the roughness functions based on the VD transformation and Figure \ref{fig:sl-rough-func} shows the roughness function from the semi-local scaling GFM transformation. Table \ref{tab:equiv-sandgrain-height} tabulates the roughness functions and the corresponding equivalent sand-grain roughness heights. The roughness function is taken as the average $u^+_{1,sw} - u^+_{1,rw}$ from $k^+\le x_2^+ \le cRe_\tau$ (or $k^*\le x_2^* \le cRe_\tau^*$), where $c$ is taken from the smooth-wall velocity such that $cRe_\tau$ occurs at the location where the velocity profiles deviate from the log-law. The same as in \S \ref{sec:wall-offset}, this was found to be $c=0.15$ for all VD plots, and for the GFM plots $c=0.15$, 0.22, and 0.3 for the $T_w/T_r=1.0$, 0.7, and 0.4 cases, respectively.

\begin{table}
    \centering
    \begin{tabular}{lcccccc}
       Case  & $\Delta u^+_{VD}$ & $k_s^+$ & $k_s^+/k^+$ & $\Delta u^+_{GFM}$ & $k_s^*$& $k_s^*/k^*$ \\\hline
       TwTr1p0r  & 10.66 & 331.84 & 4.19 & 10.85 & 359.55 & 3.90\\
       TwTr0p7r  & 11.00 & 381.28 & 4.37 & 10.48 & 308.13 & 3.72\\
       TwTr0p4r  & 9.87 & 240.12 & 3.39 & 8.75 & 152.04 & 3.36
    \end{tabular}
    \caption{Roughness functions and equivalent sandgrain roughness heights. The equivalent sandgrain roughness heights are also reported relative to the original sinusoidal peak-to-valley height.}
    \label{tab:equiv-sandgrain-height}
\end{table}

From Table \ref{tab:equiv-sandgrain-height} we find $3.4k^+\le k_s^+\le4.4k^+$ with an average $k_s^+=3.98k^+$ and limited dependence on $T_w/T_r$. Likewise, we find $3.4k^*\le k_s^*\le3.9k^*$ with an average $k_s^*=3.66k^*$ and a slight decrease in the ratio with decreasing $T_w/T_r$. Moreover, the primary impact of the wall temperature condition is to modify $k^+$ or $k^*$, with variations in $k_s/k$ across wall temperature conditions limited to approximately 10\% of the respective means. The $k_s/k$ values in Table \ref{tab:equiv-sandgrain-height} are consistent and within the range of existing incompressible studies with similar three-dimensional sinusoidal roughness. \cite{Chan2015_JFM} report $k_s^+=4.1k^+$, \cite{Ma2020} report $k_s^+=3.7k^+$, and \cite{Chan2023} report $k_s^+=3.2k^+$. Given slight variations in choice of wall offset, von K\'arm\'an constant, and log-law intercept, and recalling the inability for the roughness Reynolds number $k^+$ to collapse the roughness functions, for the present type of sinusoidal roughness and Mach 2.5 flow conditions we recommend the average $k_s^*=3.7k^*$ -- extending the findings from incompressible theory to the present compressible, supersonic conditions.

\section{Velocity-Temperature Relationships}\label{sec:vel-temp}

Owing to the similarity between the transport mechanisms of momentum and thermal energy, it has been shown that the total enthalpy, $\ol{H} = c_p\ol{T}+\ol{u_1}^2/2$, is directly related to the velocity, $\ol{u_1}$, with extension to turbulent flows also assuming a strong analogy between the total enthalpy and velocity fluctuations $H' \propto u_1'$ \citep{Zhang2014}. Therefore, the mean temperature is proportional to the square of the mean streamwise velocity. In the context of laminar boundary layers \cite{Busemann1931} and \cite{Crocco1932} independently derived quadratic mean velocity-temperature relationships of the form:

\begin{equation}
    \frac{T}{T_e} = \frac{T_w}{T_e} + \frac{T_{c,e}-T_w}{T_e}\frac{u_1}{u_e} + \frac{T_e-T_{c,e}}{T_e}\left(\frac{u_1}{u_e}\right)^2 \label{eq:crocco-busemann}
\end{equation}
\begin{equation}
    T_{c,e} = T_e + c\frac{u_e^2}{2c_p}
\end{equation}
where $c=1$ in the original Crocco-Busemann relation, assuming a Prandtl number of unity. Later that factor was modified to the recovery factor $r$ by \cite{walz69}. Resulting in Eq. \ref{eq:crocco-busemann} being equivalent to Eq. \ref{eq:Walz} if $c$ is set to the recovery factor $r$. Edge velocity, temperature, and Mach number are denoted by $u_e$, $T_e$, and $M_e$ respectively, and $\gamma$ is the ratio of specific heats.

\begin{equation}
    \frac{T}{T_e} = 1+\frac{T_r-T_w}{T_e}\left(\frac{u_1}{u_e}-1\right)+r\frac{\gamma-1}{2}M_e^2\left(1-\left(\frac{u_1}{u_e}\right)^2\right) \label{eq:Walz}
\end{equation}
The recovery temperature is defined as follows, where $r=0.883$ for the present work:

\begin{align}
    T_r &= T_e\left(1+r\frac{\gamma-1}{2}M_e^2\right)\\
    T_r &= T_e + r\frac{u_e^2}{2c_p}
\end{align}

\begin{figure}[h!]
    \centering
    \begin{subfigure}[b]{0.495\textwidth}
        \centering
        \includegraphics[width=\textwidth]{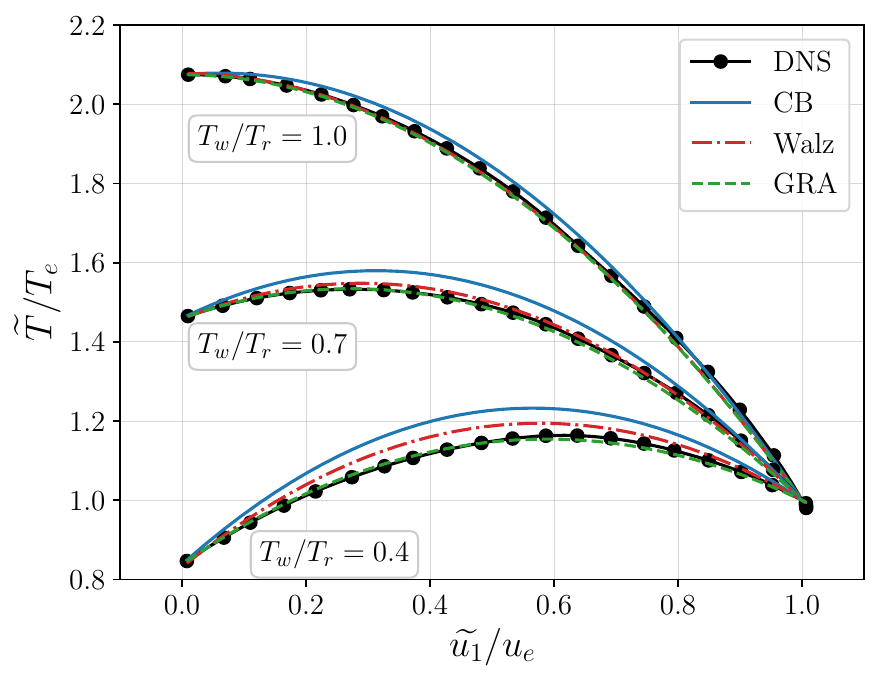}
        \caption{Smooth wall}
        \label{fig:vel-temp-sw}
    \end{subfigure}
    \hfill
    \begin{subfigure}[b]{0.495\textwidth}
        \centering
        \includegraphics[width=\textwidth]{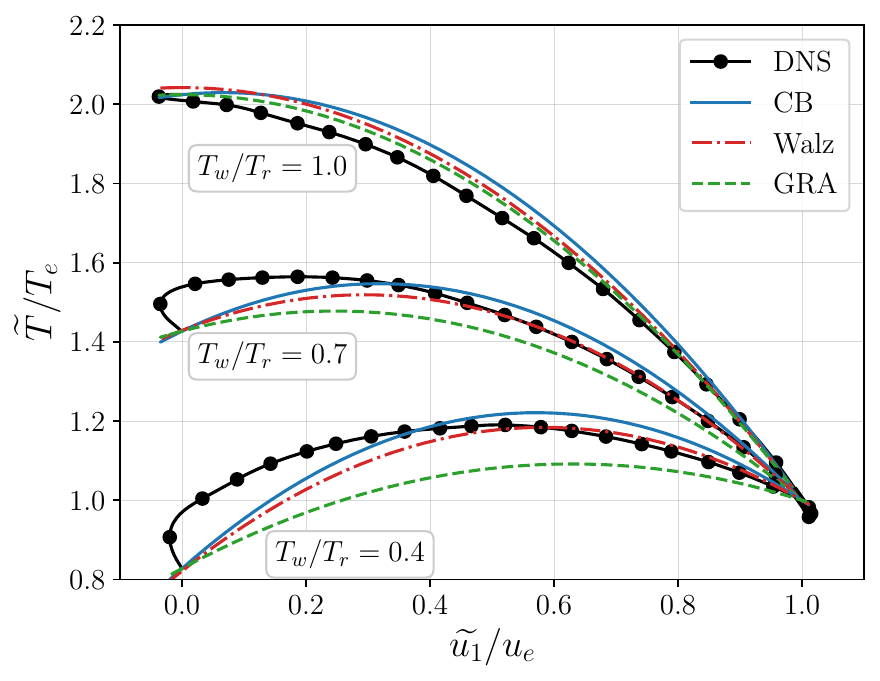}
        \caption{Rough wall}
        \label{fig:vel-temp-rw}
    \end{subfigure}
    \caption{Velocity-temperature quadratic relationship from the various Reynolds analogies.}
    \label{fig:vel-temp}
\end{figure}

Presently, these equations are written in terms of instantaneous values; however, they are applicable for mean (Reynolds or Favre) temperatures and velocities. Although the Walz relation improves upon the Crocco-Busemann relation for non-adiabatic flows, recent work has focused on further modifying these quadratic velocity-temperature relationships. To account for the effects where $Pr\ne1$ and diabatic walls, \cite{Zhang2014} developed a generalized Reynolds analogy (GRA) with the same quadratic form as the Crocco–Busemann relation and Walz’s equation, but adopting the general recovery factor $r_g$. They introduce a generalized analogy between the total enthalpy and streamwise velocity $H_g-H_w=U_w u_1$, where $H_g=c_pT+r_gu_1^2/2$ and $U_w=-Pr\ol{q_w}/\ol{\tau_w}$. A key assumption in their work is that the effective turbulent Prandtl number is constant and equal one, $\ol{Pr_e}\approx1$. The \cite{Zhang2014} velocity-temperature relationship is detailed in Eqs.~\ref{eq:GRA-mean}-\ref{eq:GRA-s}:

\begin{align}
        \frac{T}{T_e} &= \frac{T_w}{T_e} + \frac{T_{r}-T_w}{T_e}f\left(\frac{u_1}{u_e}\right) + \frac{T_e-T_{r}}{T_e}\left(\frac{u_1}{u_e}\right)^2 \label{eq:GRA-mean}\\
        f\left(\frac{u_1}{u_e}\right) &= (1-sPr)\left(\frac{u_1}{u_e}\right)^2+sPr\left(\frac{u_1}{u_e}\right)\\
        r_g &= \frac{T_w-T_\infty}{u_e^2/(2c_p)} - \frac{2Pr}{u_e}\frac{q_{w}}{\tau_w} = r\left[sPr+(1-sPr)\Theta\right]\;\;\;(\text{if } \ol{Pr_e}=1)\\
        \Theta &= \frac{T_w-T_e}{T_r-T_e}\\
        s &\equiv \frac{2C_h}{C_f} = \frac{q_w u_e}{\tau_w c_p(T_w-T_r)}\;\;\;(\text{Reynolds analogy factor})\label{eq:GRA-s}
\end{align}
where $C_h=\frac{q_w}{\rho_e u_e c_p(T_w-T_r)}$ is the Stanton number and $C_f=\frac{2\tau_w}{\rho_e u_e^2}$ is the skin friction coefficient, per Eq. \ref{eq:heat-trans-coeff} and Eq. \ref{eq:skin-fric-coeff}. Figure \ref{fig:vel-temp} shows the Crocco-Busemann (CB), Walz, and \cite{Zhang2014} GRA applied to our six DNS simulations. For the smooth wall data, the GRA effectively recovers the velocity-temperature relationship from the DNS, with the increasing number of assumptions for the Walz and CB failing for the colder wall cases. On the contrary, the rough wall destroys the classical velocity-temperature relationship and all three methods fail. The roughness increases both the skin friction and heat transfer, but by disparate amounts. The heat transfer does not possess an equivalent additional mechanism like pressure does for the skin friction. This disparity disrupts the traditional Reynolds analogies and can be seen in the surface analysis in Table \ref{tab:surface-analysis}. 

It is expected that the roughness destroys the quadratic velocity-temperature relationship in the near wall; however, in the outer layer, we postulate that the GRA, $\wt{H_g}(\wt{u_1}) - H_w =U_w\wt{u_1}$, still holds so long as a near-wall roughness correction is considered. The idea of modelling the near-wall roughness behaviour to relate the temperature to the velocity in the context of wall modelling for compressible turbulent boundary layers over prism-shaped roughness at Mach 2 and 4 was recently proposed by \cite{Cogo2026}. They use the \cite{Huang1994} quadratic velocity-temperature relationship with the wall modelling framework introduced by \cite{Yang2016}. Consistent with the outer-layer similarity arguments of \cite{Townsend1976}, \cite{Cogo2026} found the parabolic velocity-temperature relation remains valid above the roughness crest and the Reynolds analogy recovers outside the roughness sublayer at two to three times the roughness element height. \cite{Su2026} attempt a roughness modification to the GRA by introducing an equivalent slip-plane or reference-point boundary conditions to bypass the near-wall thermal heterogeneity. However, this slip velocity must be prescribed, and the corresponding reference height, virtual temperature, and the ratio of heat flux to skin friction are adjusted to best fit the outer-region velocity-temperature distribution.

\begin{figure}[h!]
    \centering
    \begin{subfigure}[b]{0.495\textwidth}
        \centering
        \includegraphics[width=\textwidth]{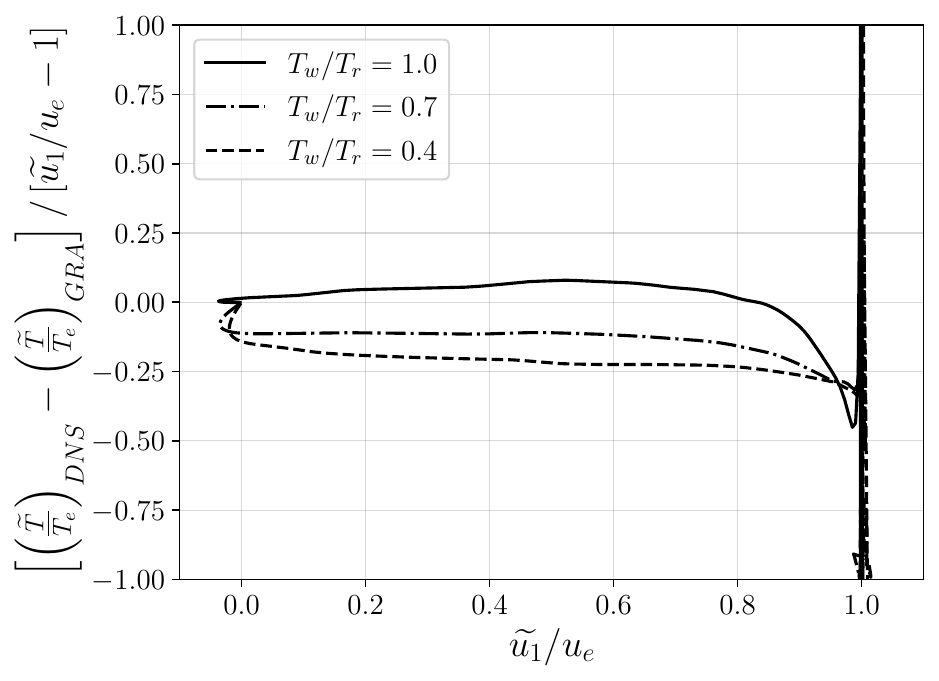}
        \caption{DNS to GRA difference}
        \label{fig:vel-temp-DNS-GRA_linear}
    \end{subfigure}
    \hfill
    \begin{subfigure}[b]{0.495\textwidth}
        \centering
        \includegraphics[width=\textwidth]{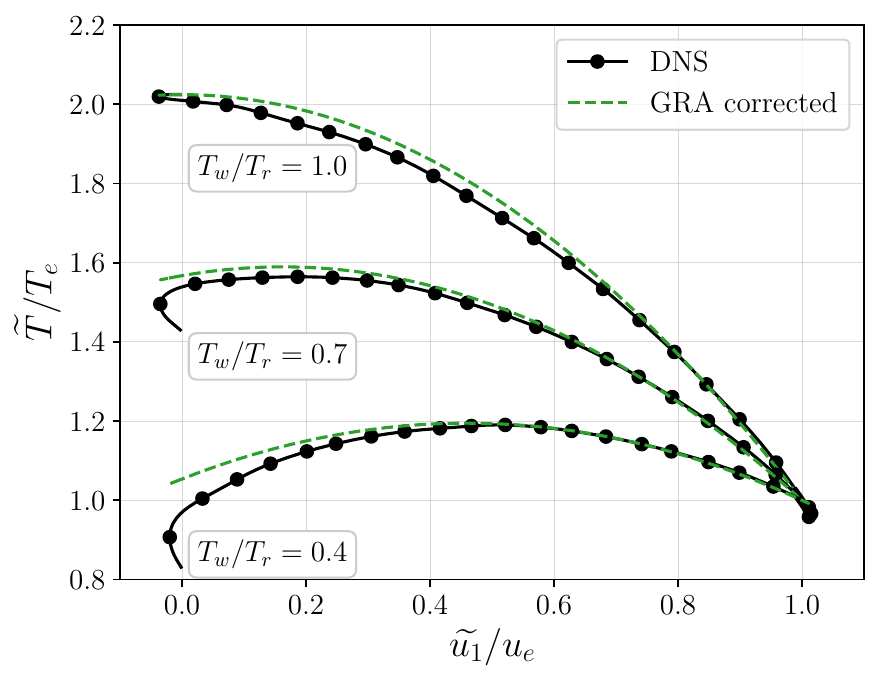}
        \caption{rcGRA}
        \label{fig:vel-temp-rcGRA}
    \end{subfigure}
    \caption{ (a) Difference in the predicted $\wt{T}/T_e$ from the GRA relationship and the actual $\wt{T}/T_e$ for the rough-walled DNS. The difference between the GRA an DNS is linear with respect to $(\wt{u_1}/u_e-1)$ as indicated by the mostly constant line. (b) Roughness corrected generalized Reynolds analogy (rcGRA) from Eq. \ref{eq:full-rcGRA} and \ref{eq:rcGRA}.}
    \label{fig:vel-temp-corrected}
\end{figure}

We would like a correction $\delta U_w(\wt{u_1})$ such that $\delta U_w(\wt{u_1}=u_e)=0$ at the edge and $\delta U_w(\wt{u_1}=0)=\Delta U_w$ at the wall. Figure \ref{fig:vel-temp-DNS-GRA_linear} shows that the difference between the DNS data and the basic GRA without correction for the rough wall cases is linear with respect to $\left(\wt{u_1}-u_e\right)$. Therefore, the simplest form that satisfies the boundary conditions is:

\begin{equation}
    \delta U_w(\wt{u_1}) = \Delta U_w\left(\wt{u_1}-u_e\right)
\end{equation}
Note that this satisfies the edge condition $\wt{H_g}(u_e) - H_w =U_w u_e$, consistent with the smooth wall GRA and outer layer similarity, but by necessity the wall boundary condition $\wt{H_g}(0) - H_w = 0$ is violated. To be clear, $H_w=c_p T_w$ in this equation is functioning as the smooth-wall reference enthalpy, not the rough-wall value at $\wt{u_1}=0$ which may have $\wt{T}\ne T_w$. The roughness corrected GRA then takes the form:

\begin{equation}
    \wt{H_g}(\wt{u_1}) - H_w =U_w\wt{u_1} + \Delta U_w\left(\wt{u_1}-u_e\right)
    \label{eq:rcGRA_enthalpy-form}
\end{equation}
From \cite{Zhang2014}, $U_w$ functions as an effective velocity scale matching momentum and thermal energy transport, taking the ratio $U_w=-Pr\ol{q_w}/\ol{\tau_w}$. Here $\Delta U_w=U_{w,sw}-U_{w,rw}$ is taken as the difference between the smooth wall and rough wall. As seen by Table \ref{tab:surface-analysis} the Reynolds analogy is broken over rough walls due to the wall heat flux lacking an equivalent mechanism to pressure for the wall shear stress. Extending the definition from Eq. \ref{eq:rcGRA_enthalpy-form} to the commonly plotted temperature form, where $H_g=c_pT + r_g u_1^2/2$ is the general recovery enthalpy based on the general recovery factor $r_g$, per \cite{Zhang2014} and $H_w=c_pT_w$:

\begin{align}
    \wt{T}(\wt{u_1}) &= T_w + \frac{U_w}{c_p}\wt{u_1} - \frac{r_g}{2c_p}\wt{u_1}^2 + \frac{\Delta U_w}{c_p}\left(\wt{u_1}-u_e\right)\\
    \frac{\wt{T}}{T_e} &= \underbrace{\frac{T_w}{T_e} + \frac{U_w u_e}{c_pT_e}\frac{\wt{u_1}}{u_e} - \frac{r_gu_e^2}{2c_pT_e}\left(\frac{\wt{u_1}}{u_e}\right)^2}_{\left(\frac{\wt{T}}{T_e}\right)_{GRA}} + \underbrace{\frac{\Delta U_w u_e}{c_pT_e}\left(\frac{\wt{u_1}}{u_e}-1\right)}_{\left(\frac{\wt{T}}{T_e}\right)_{rc}}
    \label{eq:full-rcGRA}
\end{align}
Using the heat transfer and skin friction coefficient form for $U_w$ from \cite{Zhang2014}, the roughness correction $\left(\frac{\wt{T}}{T_e}\right)_{rc}$ is as follows:

\begin{align}    
    \left(\frac{\wt{T}}{T_e}\right)_{rc} &= \frac{-Pr\left[(\ol{q_w}/\ol{\tau_w})_{sw}-(\ol{q_w}/\ol{\tau_w})_{rw}\right] u_e}{c_pT_e}\left(\frac{\wt{u_1}}{u_e}-1\right)\\
    \left(\frac{\wt{T}}{T_e}\right)_{rc} &= \frac{\Delta s Pr (T_w-T_r)}{T_e}\left(\frac{\wt{u_1}}{u_e}-1\right) \label{eq:rcGRA}
\end{align}
where $s=2C_h/C_f$ is the Reynolds analogy factor and $\Delta s= s_{sw}-s_{rw}$. For the present diabatic simulations $\Delta s\approx0.25$. From Eq. \ref{eq:rcGRA} there is no roughness correction when either $C_h=0$ or $T_w=T_r$. Figure \ref{fig:vel-temp-rcGRA} shows the roughness corrected GRA (rcGRA). The original uncorrected GRA achieved a percent error less than 2.5\% of the DNS values after $\wt{u_1}/u_e \ge 0.8$ ($x_2\ge5k$) for the TwTr0p7r case, $\wt{u_1}/u_e \ge 0.9$ ($x_2\ge8k$) for the TwTr0p4r case, and is always below 2.5\% error for the TwTr1p0r case. Including the correction, the TwTr1p0r case remains the same, the TwTr0p7r case error is less than 2.5\% for any $\wt{u_1}/u_e \ge 0$ (corresponding to $x_2 \ge 0.1k$), and the TwTr0p4r case error is less than 2.5\% for any $\wt{u_1}/u_e \ge 0.2$ (corresponding to $x_2 \ge 0.9k$). In brief, the correction reduces the percent error to below 2.5\% for all wall-normal locations outside the immediate vicinity of the roughness height, conservatively stated as $x_2\ge k$, while the original uncorrected GRA failed to recover the DNS $\wt{T}/T_e$ for most of the boundary layer.

\section{Thermal Boundary Layer Scaling}\label{sec:thermal-BL}
Table \ref{tab:thermal-bl-parameters} tabulates the thermal boundary layer parameters at the analysis location. In the table $x_{1,a}/\theta_i$ is the streamwise analysis location normalised by the inlet momentum thickness, $T_w$ is the wall temperature, $T_r$ is the recovery temperature, $T_e$ is the boundary layer edge temperature, $\delta_\Theta$ is the boundary layer thickness based on the 99.5\% freestream temperature ($x_2-d_\Theta$ where $T_w-\wt{T}=0.995(T_w-T_\infty)$), $\delta_\Theta/\delta$ is the thermal-to-momentum boundary layer height ratio, and $Pr=0.7368$ is the molecular Prandtl number. Based on $\delta_\Theta/\delta$, the thermal boundary layer is slightly larger than the momentum boundary layer. To relate these two heights with the Prandtl number, Table \ref{tab:thermal-bl-parameters} includes a column for $n_\Theta= -\ln(\delta_\Theta/\delta)/\ln(Pr)$ such that $Pr^{-n_\Theta}=\delta_\Theta/\delta$. Finally, the last two columns report the friction temperature, $\Theta_\tau=q_w/(\rho_w u_\tau c_p)$, where $q_w$ is the wall heat transfer and $c_p=1004.7$ J kg$^{-1}$K$^{-1}$ is the specific heat at constant pressure, and $B_q=\Theta_\tau/T_w$ the dimensionless wall heat transfer rate. Here the sign convention follows positive heat transfer from the wall into the fluid, so a cooled wall will have a negative heat transfer rate. For the temperature analysis, a mean temperature difference is defined as the difference with the wall temperature, $\Theta=T_w-T$, where both a Reynolds or Favre averaged temperature difference are possible $\ol{\Theta}$ or $\wt{\Theta}$. Strictly speaking, the $T_w/T_r=1.0$ cases may have a small amount of heat transfer because an isothermal wall condition is being set; however, this value was found to be small and has loosely been called the adiabatic case throughout this paper. For the purposes of this study, for the $T_w/T_r=1.0$ cases, the wall heat flux vanishes and all values based on $q_w$ are omitted when reported in tables and figures. 

\begin{table}
    \centering
    \begin{tabular}{lcccccccc}
        Case & $x_{1,a}/\theta_i$ & $T_w/T_r$ & $\frac{T_w-T_e}{T_r-T_e}$ & $\delta_\Theta$, mm & $\delta_\Theta/\delta$&  $-\frac{\ln(\delta_\Theta/\delta)}{\ln(Pr)}$ & $-\Theta_\tau$, K & $-B_q\!\times\!10^3$ \\\hline
        TwTr1p0s & 394 & 1.0 & 1.00 & 15.04 & 1.06 & 0.200 & -- & --\\
        TwTr1p0r & 687 & 1.0 & 1.00 &  8.62 & 1.08 & 0.252 & -- & -- \\
        TwTr0p7s & 313 & 0.7 & 0.42 &  9.19 & 1.07 & 0.229 & 7.73  & 19.45 \\
        TwTr0p7r & 395 & 0.7 & 0.41 &  4.42 & 1.04 & 0.124 & 10.61 & 26.68 \\
        TwTr0p4s & 277 & 0.4 & -0.15 &  3.63 & 1.03 & 0.084 & 13.28 & 58.46 \\
        TwTr0p4r & 316 & 0.4 & -0.16 &  2.18 & 1.07 & 0.221 & 17.13 & 75.40 
    \end{tabular}
    \caption{Thermal boundary layer parameters at analysis location $(x_{1,a})$. The thermal boundary layer height has the thermal wall offset pre-subtracted.}
    \label{tab:thermal-bl-parameters}
\end{table}

The results from Table \ref{tab:thermal-bl-parameters} agree with those shown in \S \ref{sec:surface}, insomuch as the heat transfer into the wall increases with decreasing $T_w/T_r$. Another finding is the height of the thermal boundary layer is similar, yet slightly larger in magnitude to the momentum boundary layer, where the roughness and wall temperature conditions have limited effect on the $\delta_\Theta/\delta$ ratio. On average the thermal boundary layer height can be found as roughly $\delta_\Theta\approx\delta Pr^{-0.185}=\delta Pr^{-1/5.4}$.

\subsection{Inner Layer and Roughness Function Analysis}

When scaling the mean temperature profiles, like velocity, the idea is to scale the profiles by the mean property variations such that the resulting profile recovers to that of the incompressible flow. Before considering roughness, there are two notable challenges when forming these transformations. First, in cases with zero wall heat transfer, the friction temperature is undefined \citep{Chen2022_JFM}. Second, cases with cold walls have non-monotonic temperature profiles that lead to singularities in the transformations near the location of zero gradient $\pp{\wt{\Theta}}{x_2}=0$ at the peak temperature \citep{Liang2026, Zhang2026}. Due to these inherent challenges, only recently has the literature proposed temperature transformations that can accommodate either or both challenges \citep{Patel2017, Wan2020, Chen2022_JFM, Chen2022_PRF, Zhu2025, Liang2026, Zhang2026}. Even with the recent interest, the number of studies on $\Delta\Theta^+$ significantly lags those on $\Delta u_1^+$ \citep{Kadivar2025} and the present work aims to asses whether the modern temperature transformations are sufficient to reliably estimate the thermal roughness function. 

\begin{figure}[h!]
    \centering
    \begin{subfigure}[b]{0.495\textwidth}
        \centering
        \includegraphics[width=\textwidth]{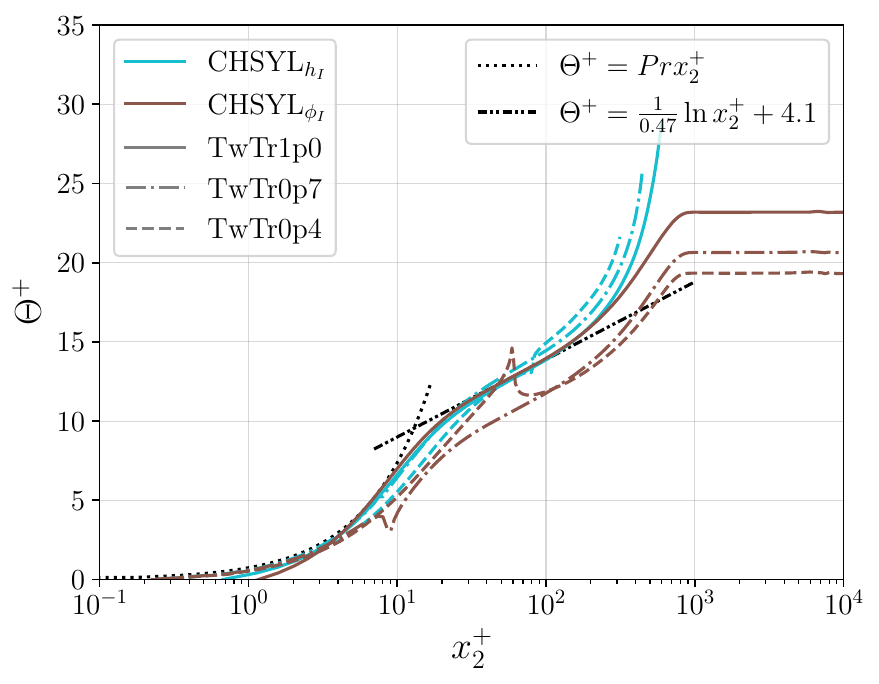}
        \caption{Wall scaling}
        \label{fig:TBL-local}
    \end{subfigure}
    \hfill
    \begin{subfigure}[b]{0.495\textwidth}
        \centering
        \includegraphics[width=\textwidth]{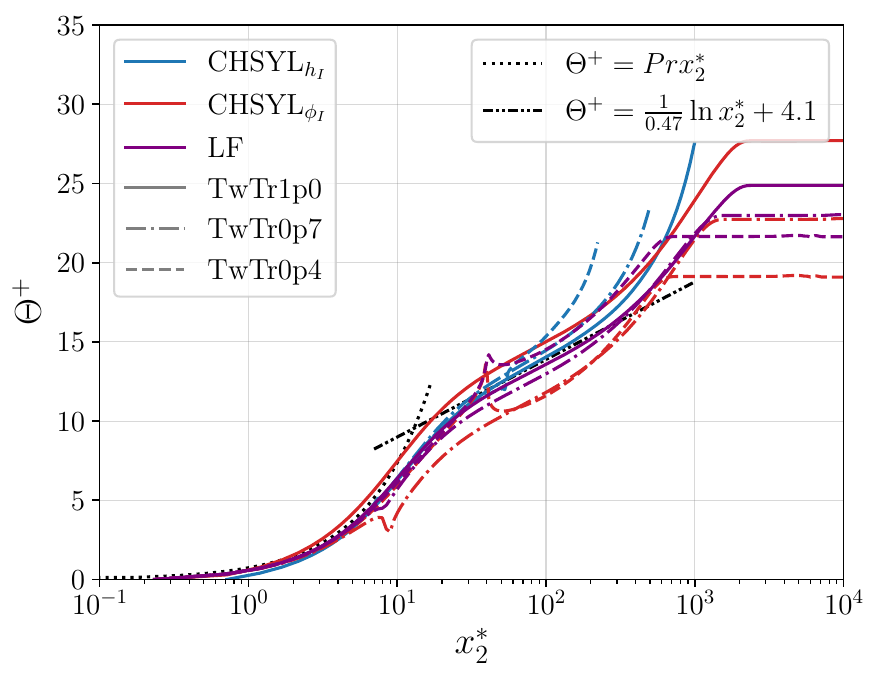}
        \caption{Semi-local scaling}
        \label{fig:TBL-semi-local}
    \end{subfigure}
    \caption{Smooth-wall mean temperature profiles in inner scaling at analysis location $(x_{1,a})$. The classical law of the wall is shown with dashed lines where the thermal sublayer and thermal log-layer are defined according to \cite{Kader1981}.}
    \label{fig:TBL-inner}
\end{figure}

Before attempting to compute a thermal roughness function, $\Delta\Theta^+$, and assessing the relative increase or decrease due to roughness, three recent temperature transformations from the literature are considered for the smooth-wall. Figure \ref{fig:TBL-inner} shows the smooth-wall mean temperature difference, $\Theta^+$, as a function of both the wall scaling and semi-local wall-normal coordinate. Both the van Driest and semi-local type transformations from \cite{Chen2022_JFM}, given the acronym CHSYL, and the composite-type transformation from \cite{Liang2026}, denoted LF are presently considered. The form of the transformations are provided in Appendix \ref{app:temp-trans}. The CHSYL transformations are evaluated taking both the form $\Theta^+=\int_0^{\wt{\Theta}} h_I d\wt{\Theta}$ as used by \cite{Chen2022_JFM} and the form $\Theta^+=\int_0^{x_2^*} \phi_I d x_2^*$ as used by \cite{Liang2026} because the choice of normalisation and simplifications present different numerical behaviour, especially surrounding the singularities. For a complete description of the temperature transformations herein, see Appendix \ref{app:temp-trans}.

\begin{figure}[h!]
    \centering
    \begin{subfigure}[b]{0.325\textwidth}
        \centering
        \includegraphics[width=\textwidth]{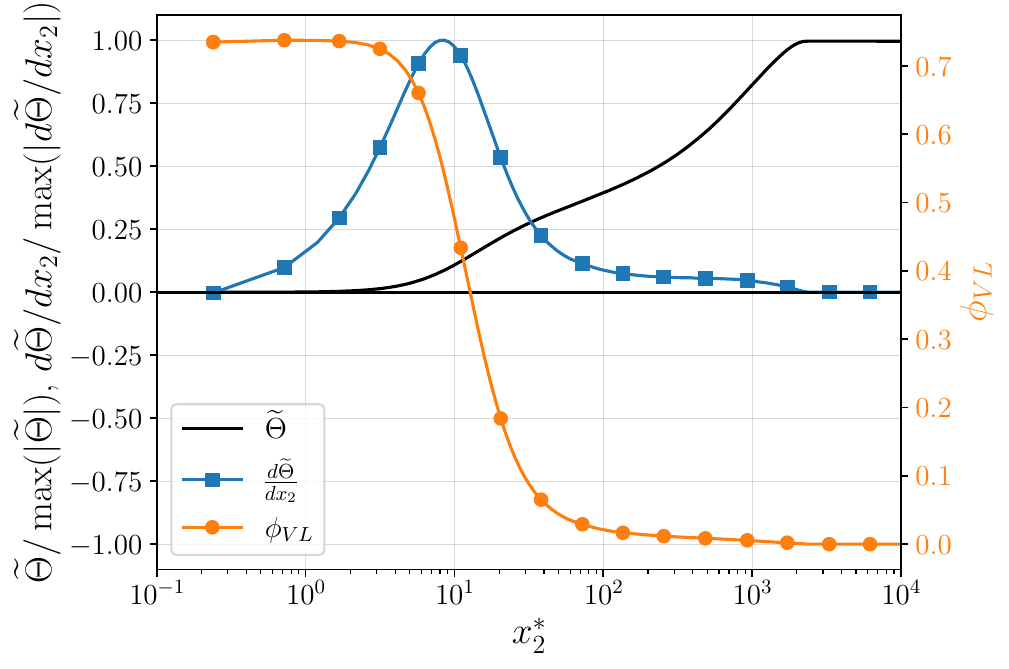}
        \caption{TwTr1p0s}
        \label{fig:singularity_1p0s}
    \end{subfigure}
    \hfill
    \begin{subfigure}[b]{0.325\textwidth}
        \centering
        \includegraphics[width=\textwidth]{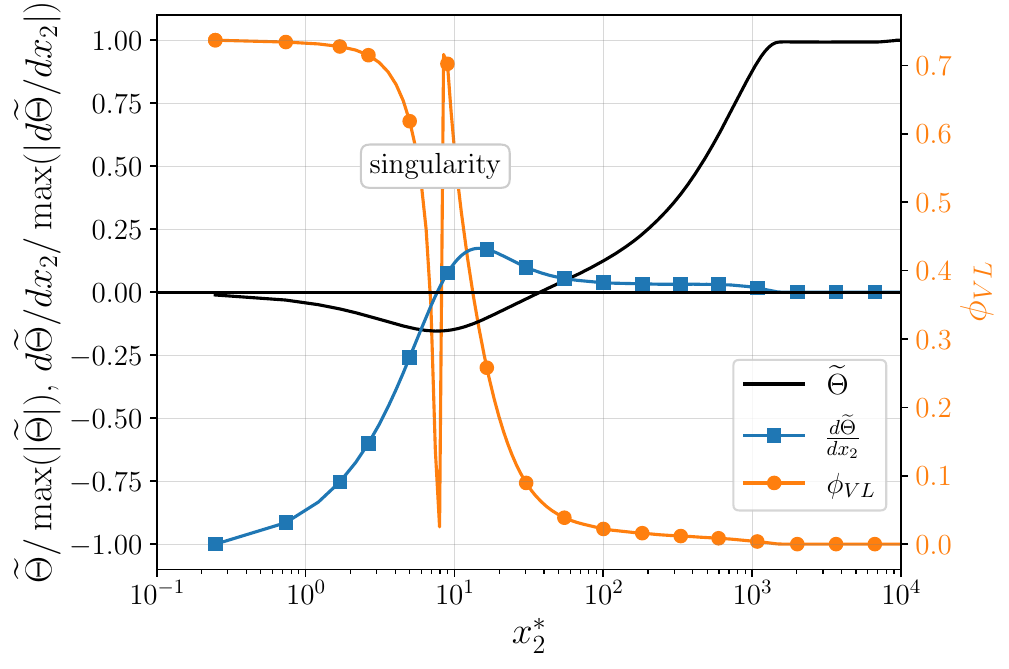}
        \caption{TwTr0p7s}
        \label{fig:singularity_0p7s}
    \end{subfigure}
    \hfill
    \begin{subfigure}[b]{0.325\textwidth}
        \centering
        \includegraphics[width=\textwidth]{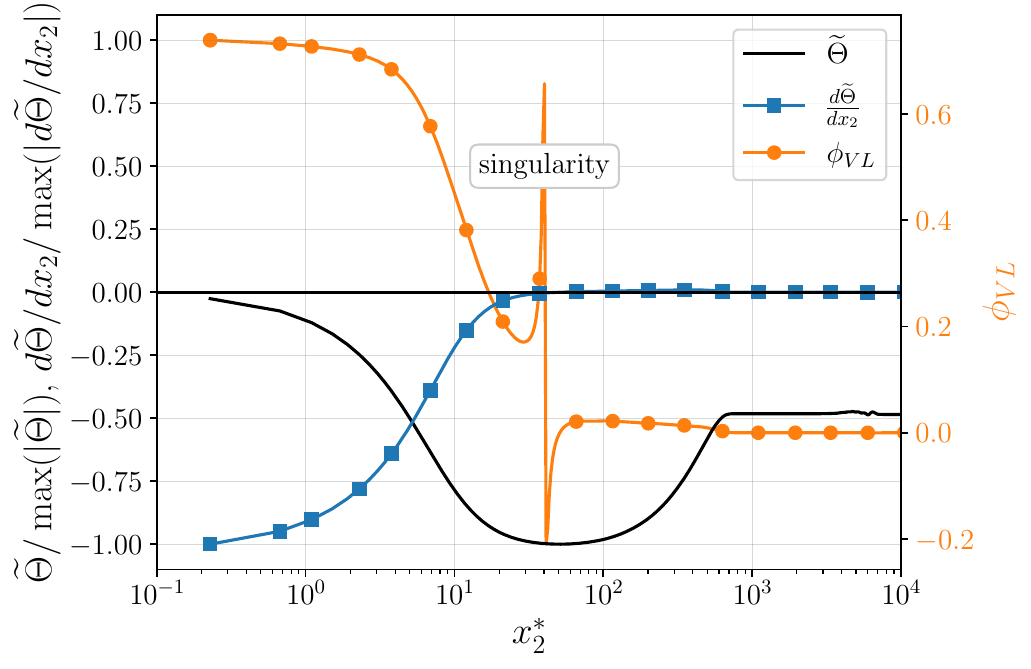}
        \caption{TwTr0p4s}
        \label{fig:singularity_0p4s}
    \end{subfigure}
    \caption{Singularities in the compressible temperature transformations at the peak temperature location. The temperature difference, $\widetilde{\Theta}$, and temperature gradient, $d\widetilde{\Theta}/dx_2$, are normalised to [-1, 1]. The Mach number and wall-temperature invariant function $\phi_{VL}$ from \cite{Liang2026}, from Appendix \ref{app:temp-trans}, is also shown to highlight the location of the singularity corresponding with the peak in temperature and zero in temperature gradient.}
    \label{fig:singularity}
\end{figure}

Both Figure \ref{fig:TBL-local} and \ref{fig:TBL-semi-local} have significant spread in the transformed temperature profiles. The CHSYL$_{h_I}$ has decent collapse in the sublayer, buffer layer, and start of the log-layer before ultimately becoming unbounded after $x_2^+\approx x_2^*>100$. The CHSYL$_{\phi_I}$ remain bounded throughout the entire boundary layer, but a spike exists near the beginning of the buffer layer $x_2^+\approx x_2^*\approx10$ for the TwTr0p7 and a spike near the end of the buffer layer $x_2^+\approx x_2^*\approx40$ for the TwTr0p4 case. Figure \ref{fig:singularity} overlays the mean temperature profile with temperature gradient, both normalised so that the schematic is between [-1,1]. This figure shows that the spikes in Figure \ref{fig:TBL-inner} correspond to singularities around the peak temperature. The primary difference between CHSYL$_{h_I}$ and CHSYL$_{\phi_I}$ is the choice of modelling $\ol{q}=\ol{t_{i2}u_i} - \ol{\rho u_2''u_i''}\wt{u_i} - \frac{\ol{\rho u_2'' u_i'' u_i''}}{2}$ as $\wt{u_1}\ol{\tau_w}$. Inspired by the \cite{Griffin2021} total stress based composite velocity transformation, \cite{Liang2026} attempt to improve upon the issue of the singularities by using a composite transformation that leverages two Mach-number and wall-temperature invariant functions that are blended based on the dominance of the molecular heat flux in the viscous sublayer where turbulent transport is negligible and the dominance of the turbulent heat flux in the logarithmic region, where molecular viscosity effects are minimal. Compared to the CHSYL the LF composite transformation performs better: at the scale of Figure \ref{fig:TBL-semi-local} the transformation almost collapses the TwTr0p7s and TwTr1p0s temperature profiles to the classical law of the wall. All cases remain bounded and in the log-layer the TwTr1p0s and TwTr0p7s follow close to $\Theta^+=\frac{1}{\kappa_\Theta}\ln x_2^*+A_\Theta$, where $\kappa_\Theta=\kappa/0.87=0.47$ and $A_\Theta=(3.85 Pr^{1/3} - 1.3)^2 + 2.12\ln Pr=4.09$ \citep{Kader1981}. However, it is not perfect, there is still a slight kink near $x_2^*=9$ for the TwTr0p7s case that shifts the profile downward which never fully recovers to the log-layer. This issue is even worse for the TwTr0p4s case with a spike at $x_2^*=40$, again affecting the remaining cumulative integral and logarithmic portion. We believe this limitation of the LF transformation was not seen in the original work by \cite{Liang2026} because their test matrix, although considering cold wall cases with $T_w/T_r<1$, did not have any cases with $\frac{T_w-T_e}{T_r-T_e}<1$ like the present TwTr0p4 simulation. This strongly cooled case with wall temperature colder than freestream temperature is an extreme condition. The composite method attempts to leverage different invariant functions in the viscous sublayer and the logarithmic region because they are deficient outside of their respective zone of applicability. Unfortunately for the present cold wall conditions, the singularities  happen between the viscous sublayer and logarithmic regions where both invariant functions are semi-activated and the deficiencies are not dominated by one or the other function. Figure \ref{fig:singularity} shows the invariant functions and the manifestation of the singularity. Nevertheless, the \cite{Liang2026} transformation performs best and will be used for the rough-wall cases. Unfortunately, the results for the cold wall cases TwTr0p7 and TwTr0p4 should be used with caution because of the presence of the singularity. 

\begin{figure}[h!]
    \centering
    \begin{subfigure}[b]{0.495\textwidth}
        \centering
        \includegraphics[width=\textwidth]{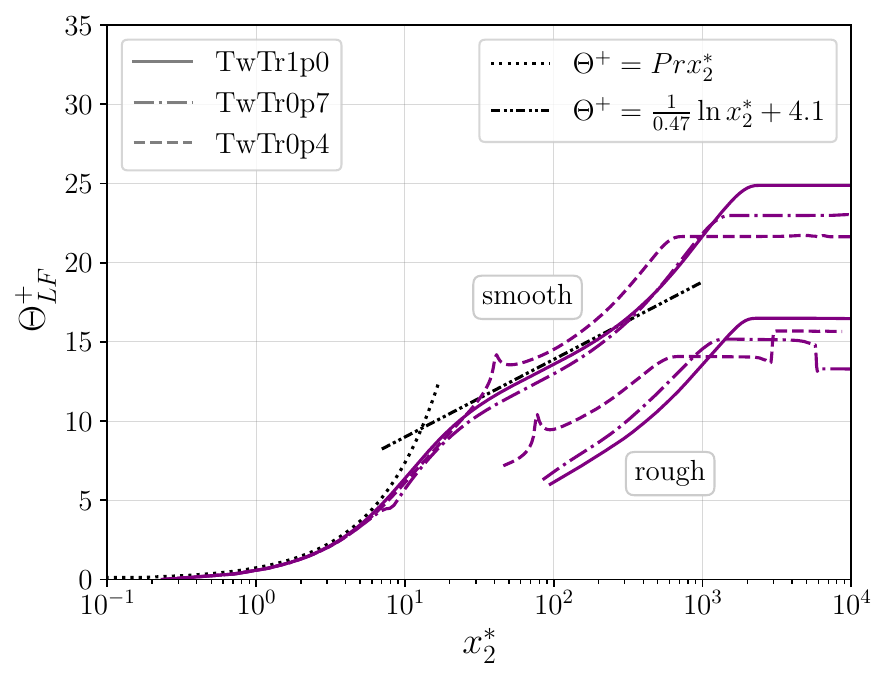}
        \caption{Mean temperature}
        \label{fig:TBL-inner-smooth&rough}
    \end{subfigure}
    \hfill
    \begin{subfigure}[b]{0.495\textwidth}
        \centering
        \includegraphics[width=\textwidth]{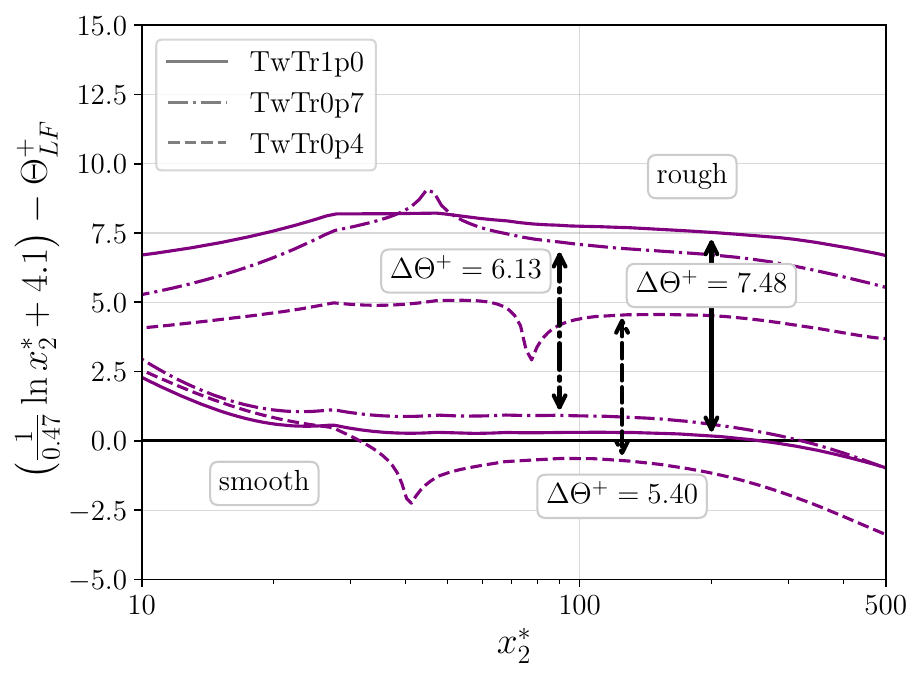}
        \caption{Roughness function}
        \label{fig:thermal-rough-func}
    \end{subfigure}
    \caption{Mean temperature profiles at the analysis location $(x_{1,a})$ with corresponding thermal roughness function $\Delta \Theta^+=\Theta^+_{sw} - \Theta^+_{rw}$ in the logarithmic layer based on the \cite{Liang2026} temperature transformation. Thermal roughness function $\Delta \Theta^+$ measured as average along viable logarithmic region, per \S \ref{sec:wall-offset}, with the exception of the TwTr0p4 case further limiting the lower bound to be $x_2^*>90$ instead of $k^*$ to circumvent the singularity.}
    \label{fig:BL-rough-func}
\end{figure}

Figure \ref{fig:TBL-inner-smooth&rough} shows the \cite{Liang2026} transformed mean temperature for all three smooth- and rough-wall cases and Figure \ref{fig:thermal-rough-func} shows the roughness functions. Table \ref{tab:thermal-roughness-function} compares the thermal roughness function, $\Delta\Theta^+$, to the roughness function $\Delta u_1^+$. On average $\Delta u_1^+/\Delta\Theta^+=1.59^{+7\%}_{-9\%}$ and a linear regression of the limited data results in $\Delta \Theta^+ = 0.81\Delta u_1^+ -1.83$, with coefficient of determination $R^2=0.747$. From the limited dataset it is not possible to conclude whether the thermal roughness functions exhibit linear mapping to the momentum roughness functions. Additionally, the thermal roughness function exhibits an approximately linear dependence on $T_w/T_r$, specifically $\Delta \Theta^+ = 3.47\frac{T_w}{T_r} + 3.91$, with coefficient of determination $R^2=0.972$. However, additional data points are necessary to confirm the exact relationship. We further caution the use of these thermal roughness functions due to the prevalence of the singularities.

\begin{table}
    \centering
    \begin{tabular}{lccccc}
       Case  & $k_s^+$ & $k_s^*$ & $\Delta u^+_{1,GMF}$ & $\Delta \Theta^+_{LF}$  & $\frac{\Delta u^+_{1,GMF}}{\Delta \Theta^+_{LF}}$ \\\hline
       TwTr1p0r & 331.84 & 359.55 & 10.85 & 7.48 & 1.45\\
       TwTr0p7r & 381.28 & 308.13 & 10.48 & 6.13 & 1.71\\
       TwTr0p4r & 240.12 & 152.04 & 8.75 & 5.40 & 1.62
    \end{tabular}
    \caption{Comparison of thermal roughness function, $\Delta\Theta^+$, and roughness function $\Delta u_1^+$, with respect to the equivalent sandgrain roughness height, $k_s$. The \cite{Liang2026} temperature transformation and the \cite{Griffin2021} velocity transformation are used for this comparison.}
    \label{tab:thermal-roughness-function}
\end{table}

\section{Conclusions}\label{sec:Conclusion}

Six DNS were performed for a perfect gas air, ZPG, Mach 2.5 TBL, at matched $Re_\tau=784$. Three wall temperatures, $T_w/T_r$ = 1.0, 0.7, and 0.4 were considered, each repeated with a smooth and rough-walled surface. The surface roughness followed a three-dimensional sinusoidal profile with effective slope 0.5 and matched $k^+=79.1$.   

An exploration of the mean wall shear and heat transfer detailed the augmentation in skin friction and heat transfer coefficient between the smooth- and rough-walled cases. This highlights the role of the combined viscous and pressure contributions to the total shear stress. We confirm that the classical Reynolds analogy fails for the rough walls because the heat transfer does not increase by the same amount as skin friction due to the lack of additional roughness-dominated component like pressure. Moreover, we find that  differences in the shear stress across wall temperature conditions is driven by the viscous component even though the pressure component is dominant. Additional near-wall and surface analysis quantified the extent of the roughness sublayer, finding the roughness sublayer to be $R_{RSL}=3k-6k$, for the present conditions, consistent with the commonly cited $R_{RSL}=2k-5k$ from the literature. Quantification and applicability of the wall offset for both the momentum and thermal boundary layers found the wall offset to be $d=d_\Theta\approx0.5k$ for both momentum and thermal boundary layers. We clarify the common statement in literature that the wall offset is roughly the ``mean roughness height'' should rather be half the peak-to-valley height. Physically this indicates that the virtual origin sits halfway between the peaks and valleys of the present sinusoidal roughness.

Investigating the mean momentum boundary layer, we conclude that supersonic TBLs over fully rough sinusoidal roughness at varying wall temperatures recover the roughness function $\Delta u_1^+$  from incompressible theory if you match semi-local roughness Reynolds number $k^*$ rather than the wall-based roughness Reynolds number $k^+$. We find the present compressibility transformations for mean velocity from the literature hold regardless of wall temperature and roughness. This finding contradicts recent publications from \cite{Wang2024} and \cite{Wang2026} which claim the existing transformations fail and propose a new transformation. Exploiting the roughness function results we suggest an equivalent sand-grain roughness prediction for this type of rough surface and flow conditions to be $k_s^*\approx3.7k^*$, which can then be used to get roughness functions from fully rough predictions \citep{Nikuradse1933,Schlichting1937_NACA,Nikuradse1933_NACA} .

Regarding the mean velocity-temperature relationship, we report the various Reynolds analogies (quadratic velocity-temperature relationships), finding the GRA from \cite{Zhang2014} works well for smooth-wall cases regardless of wall thermal boundary condition. However, the GRA fails for the rough-wall cases because $C_h$ and $C_f$ increase by different amounts -- recalling the previous Reynolds analogy results and discussion. We identified the difference in $\wt{T}/T_e$ as a function of $\wt{u_1}/u_e$ between the DNS and GRA was linear in $\wt{u_1}-u_e$, and we proposed a roughness correction for the generalized Reynolds analogy $\frac{\wt{T}}{T_e}=\left(\frac{\wt{T}}{T_e}\right)_{GRA}+\left(\frac{\wt{T}}{T_e}\right)_{rc}$, where $\left(\frac{\wt{T}}{T_e}\right)_{rc} = \frac{\Delta s Pr (T_w-T_r)}{T_e}\left(\frac{\wt{u_1}}{u_e}-1\right)$. The correction satisfies outer layer similarity and the coefficient is derived from existing GRA literature and depends on the difference between the smooth- and rough-walled Reynolds analogy factor, $\Delta s$.

Finally, we demonstrated that the current compressible mean temperature transformations fail for very cold wall conditions, namely $(T_w-T_e)/(T_r-T_e)<0$. Recognizing the limitations of the transformed mean temperature profiles, we nonetheless computed the thermal roughness function, $\Delta\Theta^+$, and found a potential dependence on $T_w/T_r$ and no clear mapping to $\Delta u_1^+$.  However, we caution the unreliability of the transformations due to singularities in the cold wall cases and no significant conclusions should be taken from the presently reported thermal roughness functions.
        
Overall, the theory developed for incompressible TBLs over roughness mostly holds when applied to compressible TBLs over sinusoidal roughness with varying wall temperature, when correctly accounting for compressibility effects and mean property variation. Building on these results, a detailed analysis of the turbulence dynamics driving these mean flow characteristics through examination of Reynolds stresses, turbulent kinetic energy budgets, higher-order thermal statistics, coherent structures, and spectral characteristics remains future work.

\begin{bmhead}[Acknowledgements.]
The authors would like to thank Graham V. Candler and the University of Minnesota for their computing resources. 
\end{bmhead}

\begin{bmhead}[Funding statement.]
This work has been supported under a NASA Space Technology Research Institute Award (ACCESS, grant number 80NSSC21K1117). 
\end{bmhead}

\begin{bmhead}[Declaration of interests.]
    The authors report no conflict of interest.
\end{bmhead}

\begin{appen}
\section{Velocity Transformations} \label{app:vel-trans}
Classically, fully turbulent, high-Reynolds number ($Re_\tau\gg 1$), no pressure gradient incompressible boundary layer flows collapse to the law of the wall when the mean velocity and wall-normal distance are normalised in inner scaling as follows:

\begin{align}
    x_2^+ &= \frac{x_2 u_\tau\rho_w}{\mu_w} \label{eq:yplus-defn}\\
    u_1^+ &= \frac{\ol{u_1}}{u_\tau}
\end{align}
where $Re_\tau=\delta u_\tau\rho_w/\mu_w$ is the friction Reynolds number, $\delta$ is the boundary layer height, $u_\tau=\sqrt{\tau_w/\rho_w}$ is the friction velocity, $\tau_w$ is the wall shear stress, and subscript $w$ denotes a `wall' quantity. Note that in incompressible flow $\rho$ and $\mu$ are assumed constant so $\rho_w=\rho=\ol{\rho}$ and $\mu=\mu_w=\ol{\mu}$. For compressible flow, the idea is to scale the velocity profile and wall coordinate by the mean property variation such as density and viscosity, such that you get an equivalent incompressible form ${x_2}_I$ and ${u_1}_I$, which then can be normalised and collapsed to the incompressible law of the wall (${x_2}_I^+ = \frac{{x_2}_I u_\tau\rho_w}{\mu_w}$ and ${u_1}_I^+=\frac{{u_1}_I}{u_\tau}$). The functional form for accounting for the mean property variation in terms of mapping functions $f_I$ and $g_I$ for wall distance and mean velocity, respectively, is taken from \cite{Modesti2016}:

\begin{align}
    {x_2}_I&=\int_0^{x_2} f_I\;dx_2\\
    {u_1}_I&=\int_0^{\wt{u_1}} g_I\;d\wt{u_1} \label{eq:velocity-mapping}
\end{align}

A number of velocity transformations are available in the literature. Here six are considered and the wall distance and mean velocity mappings are tabulated in Table \ref{tab:vel-trans} following a similar format to \citep{Modesti2016,Sciacovelli2024}. \cite{Griffin2021} and \cite{Hasan2023} only provide velocity transformations -- no unique wall-distance transformation like \cite{Trettel2016}. Therefore, in Table \ref{tab:vel-trans}, their wall distance function $f_I$ is given as 1, similarly presented in \citep{Sciacovelli2024}. For plotting results, the semi-local wall-normal coordinate $x_2^*=\frac{{x_2}_I\sqrt{\tau_w/\rho(x_2)}}{\nu(x_2)}={x_2}_{TL}^+=\frac{{x_2}_{I,TL} u_\tau\rho_w}{\mu_w}$ is used, consistent with the original publications. In contrast, \cite{vanDriest1951} and \cite{Zhang2012} wall-normal coordinates utilize the wall-based scaling $x_2^+=\frac{{x_2}_I u_\tau\rho_w}{\mu_w}$. 

\begin{table}
    \centering
    \begin{tabular}{lccc}
        \hline\hline
        Transformation & Acronym & Wall Distance $f_I$ & Mean Velocity $g_I$ \\\hline
        \cite{vanDriest1951} & VD & 1 & $\left(\rho^+\right)^{1/2}$\\
        \cite{Zhang2012} & ZBHLS & 1 & $\frac{g_Z}{\mu^+}$ \\
        \cite{Trettel2016} & TL & $\pp{}{x_2}\left[\frac{x_2\left(\rho^+\right)^{1/2}}{\mu^+}\right]$  & $\mu^+\pp{}{x_2}\left[\frac{x_2\left(\rho^+\right)^{1/2}}{\mu^+}\right]$ \\
        \cite{Volpiani2020} & VIPL & $\frac{\left(\rho^+\right)^{1/2}}{\left(\mu^+\right)^{3/2}}$  &  $\frac{\left(\rho^+\right)^{1/2}}{\left(\mu^+\right)^{1/2}}$\\
        \cite{Griffin2021} & GFM & 1  & $S_t^+\frac{dx_2^*}{d\wt{u_1}^+}$ \\
        \cite{Hasan2023} & HLPP &  1 & $\left(\frac{1+\kappa x_2^* D^c}{1+\kappa x_2^* D^i}\right)\left(1-\frac{x_2}{\delta_v^*}\frac{d \delta_v^*}{dx_2}\right)(\rho^+)^{1/2}$ \\
        \hline\hline
    \end{tabular}
    \caption{Velocity transformations}
    \label{tab:vel-trans}
\end{table}

For clarity, the full length terms shown in Table \ref{tab:vel-trans} are provided outside the table. Effort was made to preserve the notation from the original publications where appropriate without conflict or loss of clarity. Consistent with the body of the paper, $\gamma$ is the ratio of specific heats and $R_{gas}$ is the specific gas constant. The list of terms is as follows:

\begin{align}
    \rho^+ &= \frac{\ol{\rho}}{\ol{\rho_w}}\\
    \mu^+ &= \frac{\ol{\mu}}{\ol{\mu_w}}\\
    S_Z &= \frac{1}{\mu^+}\pp{\wt{u_1}^+}{x_2^+}\\
    g_Z &= \frac{-\frac{S_z}{2}+\left(\left(\frac{S_Z}{2}\right)^2+1-\left({\mu^+}\right)^2S_Z\right)^{1/2}}{1-\left({\mu^+}\right)^2S_Z}\\
    S_{eq}^+ &= \frac{1}{\mu^+}\pp{\wt{u_1}^+}{x_2^*}\\
    S_{TL}^+ &= \mu^+\pp{\wt{u_1}^+}{x_2^*}\\
    \tau^+ &= \tau_{visc}^+ + \tau_{turb}^+\\
    S_{t}^+ &= \frac{\tau^+ S_{eq}^+}{\tau^+ + S_{eq}^+ - S_{TL}^+}\\
    \kappa &= 0.41\;\;\;\;(\text{von K\'arm\'an constant})\\
    A^+ &= 17 \\
    f(M_\tau) &= 19.3 M_\tau\\
    M_\tau &= \frac{u_\tau}{\sqrt{\gamma R_{gas} T_w}} \\
    u_\tau^* &= \sqrt{\frac{\tau_w}{\ol{\rho}(x_2)}}\\
    \delta_v^* &= \frac{\ol{\mu}(x_2)}{\ol{\rho}(x_2)u_\tau^*}
\end{align}
\begin{align}
    D^i &= \left[1 - \exp\left(-\frac{x_2^*}{ A^+}\right)\right]^2\\
    D^c &= \left[1 - \exp\left(-\frac{x_2^*}{ A^+ + f(M_\tau)}\right)\right]^2
\end{align}

\section{Temperature Transformations} \label{app:temp-trans}
Stemming from the idea that momentum and heat transfer are transported by the same turbulent mechanisms, similarly to the momentum (velocity) boundary layer, the thermal (temperature) boundary layer can also be scaled in inner units and reduced to a common logarithmic form. Early work addressing temperature scalings were by \cite{Kader1981} and \cite{Bradshaw1995}. Necessary for the temperature analysis, a mean temperature difference is defined as follows: $\ol{\theta} = \ol{T_w}-\ol{T}$ or $\wt{\theta} = \wt{T_w}-\wt{T}$. A friction temperature analogous to the friction velocity is based on the wall heat flux $q_w$, $\theta_\tau = \frac{q_w}{\rho_w c_p u_\tau}$, where $c_p$ is the specific heat at constant pressure. Normalising the mean temperature difference by the friction temperature results in the conventional scaling, where the normalised wall-normal distance is the same as Eq. \ref{eq:yplus-defn}, $\theta^+=\ol{\theta}/\theta_\tau$. For incompressible flow, the density is constant and the wall subscript, $w$, is redundant. However, $\theta^+=\wt{\theta}/\theta_\tau$ presents a couple of challenges for compressible flows, namely: This scaling does not account for mean property variations and the friction temperature normalisation is undefined for adiabatic flow that has zero wall heat flux. The first issue has been addressed by mimicking the idea of an equivalent incompressible mapping, as seen in Eq. \ref{eq:velocity-mapping} for the velocity, but now in terms of the temperature difference $\theta^+=\theta_I/\theta_\tau$. van Driest type or semi-local type transformed temperatures are common place, and recent temperature transformations have been proposed and assessed \citep{Patel2017,Wan2020,Chen2022_JFM,Chen2022_PRF,Zhu2025,Liang2026,Zhang2026}. Equation \ref{eq:temp-map} shows the functional form to account for the mean property variation in terms of mapping functions $h_I$ for the mean temperature difference.

\begin{equation}
    \theta_I=\int_0^{\wt{\theta}} h_I \;d\wt{\theta} \;\;\;\;\;\text{or}\;\;\;\;\; \theta^+=\int_0^{\theta^+} h_I\; d\theta^+ \label{eq:temp-map}
\end{equation}

The second issue has only recently received attention. Recent efforts by \cite{Chen2022_JFM} have been to define a friction temperature that accounts for the diffusive flux from the Favre-averaged energy equation in addition to the wall heat flux, such that the temperature transformation applies for both isothermal and adiabatic walls. Akin to the composite velocity transformation from \cite{Griffin2021}, \cite{Zhang2026} and \cite{Liang2026} have proposed composite transformations that leverage different Mach-number and wall-temperature invariant functions for the mean temperature field that vary based on whether in the viscous sublayer or log-layer. These composite scalings attempt to reduce the singularity challenges near the temperature peak (where $\partial \wt{\theta}/\partial x_2 =0$) due to the pronounced non-monotonicity of mean temperature profiles in supersonic and hypersonic turbulent boundary layers with cold walls. A list of the temperature transformations presently considered is provided in Table \ref{tab:temp-trans}. Only the transformations adequate for both isothermal and adiabatic wall conditions are shown presently. Moreover, an additional column for the mean temperature transformation in terms of the invariant function models $\phi_I$ have also been included. This $\phi_I$ form is to be integrated with respect to $x_2^*$ to get $\theta^+$ directly, $\theta^+=\int_0^{x_2^*}\phi_I\;dx_2^*$ and follows the approximated normalisations from \cite{Liang2026}. Whereas the $h_I$ are to be integrated with respect to $\wt{\theta}$, $\theta^+ = \int_0^{\wt{\theta}} h_I \;d\wt{\theta}$, and follows the normalisations from \cite{Chen2022_JFM}. Note that both integrals are slightly different than the form shown in Eq. \ref{eq:temp-map} and present different numerical behaviour, especially surround the singularities. Lastly, in Table \ref{tab:temp-trans} no unique wall-distance transformation is provided, therefore the wall distance mapping function is set at one. Nevertheless, when plotting, the wall $(x_2^+)$ or semi-local $(x_2^*)$ wall-normal coordinate should be used appropriately. 

\begin{table}
    \centering
    \begin{tabular}{lcccc}
        \hline\hline
        Transformation & Acronym & Wall Distance $f_I$ & Mean Temperature $h_I$ & Mean Temperature $\phi_I$ \\\hline
        \makecell{van Driest - type \\\cite{Chen2022_JFM}} & CHSYL & 1 & $\left(\theta_{\tau,C}^*\right)^{-1}$ & $\left(\theta_{\tau,L}^*\right)^{-1}\pp{\wt{\theta}}{x_2^+}$\\
        \makecell{semi-local - type \\\cite{Chen2022_JFM}} & CHSYL & 1 & $\left(\theta_{\tau,C}^*\right)^{-1}\left[1 + \frac{x_2}{Re_\tau^*}\frac{d Re_\tau^*}{dx_2} \right]$ & $\frac{c_p \ol{\mu}}{\ol{q_w}+\wt{u_1}\ol{\tau_w}}\pp{\wt{\theta}}{x_2}$\\
        \makecell{composite - type \\\cite{Liang2026}} & LF & 1 & $\frac{\left(\ol{q_{x_2}}+\ol{q_{t,m}}\right)\phi_V\phi_L}{\ol{q_{x_2}}\phi_L+\ol{q_{t,m}}\phi_V}\pp{x_2^*}{\wt{\theta}}$ & $\frac{\left(\ol{q_{x_2}}+\ol{q_{t,m}}\right)\phi_V\phi_L}{\ol{q_{x_2}}\phi_L+\ol{q_{t,m}}\phi_V}$ \\
        \hline\hline
    \end{tabular}
    \caption{Temperature transformations}
    \label{tab:temp-trans}
\end{table}

For clarity, the full length terms shown in Table \ref{tab:temp-trans} are provided after the table. Effort was made to preserve the notation from the original publications where appropriate without conflict or loss of clarity. The full definition of the various components in Table \ref{tab:temp-trans} are as follows:

\begin{align}
    \rho^+ &= \frac{\ol{\rho}}{\ol{\rho_w}}\\
    Re_\tau^* &= Re_\tau\sqrt{\frac{\ol{\rho}}{\ol{\rho_w}}}\frac{\ol{\mu_w}}{\ol{\mu}}\\
    u_\tau^* &= \sqrt{\frac{\ol{\tau_w}}{\ol{\rho}}} \\
    \theta_{\tau,C}^* &= \frac{\ol{q_w}+\ol{q}}{\ol{\rho}c_pu_\tau^*}\\
    \theta_{\tau,L}^* &= \frac{\ol{q_w}+\wt{u_1}\ol{\tau_w}}{\ol{\rho}c_pu_\tau^*}\\
    \ol{q} &= \ol{t_{i2}u_i} - \ol{\rho u_2''u_i''}\wt{u_i} - \frac{\ol{\rho u_2'' u_i'' u_i''}}{2}\\
    \ol{q_{x_2}} & = \frac{c_p\ol{\mu}}{Pr}\pp{\wt{\theta}}{x_2}\\
    \ol{q_{t,m}} &= \frac{\ol{\tau_w}-\ol{\mu}\pp{\wt{u_1}}{x_2}}{\ol{\tau_w}}\left(\ol{q_w}+\wt{u_1}\ol{\tau_w}\right) \approx \ol{q_t}=-c_p\ol{\rho u_2'' T''}\\
    \phi_V &= \frac{c_p\ol{\mu}\pp{\wt{\theta}}{x_2}}{\frac{c_p\ol{\mu}}{Pr}\pp{\wt{\theta}}{x_2}-\frac{\ol{\mu}\pp{\wt{u_1}}{x_2}-\ol{\tau_w}}{\ol{\tau_w}}\left(\ol{q_w}+\wt{u_1}\ol{\tau_w}\right)}\\
    \phi_L & =  \frac{c_p\ol{\mu}\sqrt{\rho^+}}{\ol{q_w}+\wt{u_1}\ol{\tau_w}}\pp{\wt{\theta}}{\left(x_2\sqrt{\rho^+}\right)}
\end{align}

$\ol{q}$ collects all the diffusion (transport) terms of the kinetic energy, both mean and turbulent, from the Favre-averaged energy equation. Moreover, it makes use of the assumptions that the flow is steady and the boundary layer is thin, where the boundary layer thickness $\delta(x)\ll x$ such that $\ol{u_2}\ll\ol{u_1}$ and $\pp{}{x_1}\ll\pp{}{x_2}$. $\ol{q_{x_2}}$ is the wall-normal molecular heat flux and $\ol{q_{t,m}}$ is a model for the turbulent heat flux $\ol{q_{t}}$, shown to work best in the viscous sublayer and buffer layer. The composite transformation from \cite{Liang2026} has $\phi_V$ active in the viscous sublayer where $\ol{q_{x_2}}$ is dominant and then $\phi_L$ active in the log-layer where $\ol{q_{t}}$ is dominant. The $h_I$ form of the \cite{Liang2026} transformation shown in Table \ref{tab:temp-trans} maintains the modelling assumptions from \cite{Liang2026} but scaled by $\pp{x_2^*}{\wt{\theta}}$ so that it may be integrated with respect to $d\wt{\theta}$. It is observed that the choice of normalisation $\ol{q_w}+\ol{q}$ as opposed to $\ol{q_w}+\wt{u_1}\ol{\tau_w}$ makes a large difference in the behaviour of the subsequent integrated result around and after the peak temperature (singularity).

\end{appen}

\bibliographystyle{jfm}
\bibliography{jfm}

\end{document}